\documentclass[10pt,journal,compsoc,twocolumn]{IEEEtran}

\pdfoutput=1

\usepackage{bobbystyle_IEEEjournal}

\usepackage{tikz}
\usetikzlibrary{spy}

\usepackage[resetlabels]{multibib}
\newcites{Supp}{References}
\newcommand{\quotes}[1]{`#1'}
\newcommand{\dquotes}[1]{``#1''}

\usepackage{bbm}

\tikzstyle{only in spy node magn 1.75}=[%
   transform canvas={%
      shift={(tikzspyinnode)},
      scale=1.75,
   }
]

\newcommand{\specialcell}[2][c]{%
  \begin{tabular}[#1]{@{}c@{}}#2\end{tabular}}

\newcolumntype{L}[1]{>{\raggedright\let\newline\\\arraybackslash\hspace{0pt}}m{#1}}
\newcolumntype{C}[1]{>{\centering\let\newline\\\arraybackslash\hspace{0pt}}m{#1}}
\newcolumntype{R}[1]{>{\raggedleft\let\newline\\\arraybackslash\hspace{0pt}}m{#1}}

\newcommand{\conv}{\mathop{\scalebox{1.5}{\raisebox{-0.2ex}{$\ast$}}}}%

\renewcommand{\theequation}{\arabic{equation}}

\newenvironment{myquote}[1]%
  {\list{}{\leftmargin=#1\rightmargin=#1}\item[]}%
  {\endlist}

\begin{document}

\title{Momentum-Net: Fast and convergent \\ 
iterative neural network for inverse problems
}
%\title{\fontsize{23.4}{23.4}Momentum-Net: Fast and convergent recurrent neural network for inverse problems}

\author{Il Yong Chun,~\IEEEmembership{Member,~IEEE,} Zhengyu Huang$^\ast$, Hongki Lim$^\ast$,~\IEEEmembership{Student Member,~IEEE,} \\ and Jeffrey A. Fessler,~\IEEEmembership{Fellow,~IEEE}

%\vspace{-0.25\baselineskip}
%\vspace{-1.5\baselineskip}

\IEEEcompsocitemizethanks{
\IEEEcompsocthanksitem The authors indicated by asterisks ($^\ast$) contributed equally to this work.
\IEEEcompsocthanksitem This work is supported in part by the Keck Foundation, NIH U01 EB018753, NIH R01 EB023618, NIH R01 EB022075, and NSF IIS 1838179.
\IEEEcompsocthanksitem Il Yong Chun was with the Department of Electrical Engineering and Computer Science, The University of Michigan, Ann Arbor, MI 48019 USA,
and is now with the Department of Electrical Engineering, the University of Hawai'i at M\=anoa, Honolulu, HI 96822 USA (email: iychun@hawaii.edu).
Zhengyu Huang, Hongki Lim, and Jeffrey A. Fessler are with the Department of Electrical Engineering and Computer Science, The University of Michigan, Ann Arbor, MI 48019 USA (email: 
zyhuang@umich.edu; hongki@umich.edu; fessler@umich.edu).
\IEEEcompsocthanksitem This paper has appendices. The prefix ``A'' indicate the numbers in section, theorem, equation, figure, table, and footnote in the appendices.
\IEEEcompsocthanksitem This paper was presented in part at the Allerton Conference on Communication, Control, and Computing, Monticello, IL, USA, in Oct. 2018.
}
}

\IEEEtitleabstractindextext{%
\begin{abstract}
Iterative neural networks (INN) are rapidly gaining attention 
for solving inverse problems in imaging, image processing, and computer vision.
INNs combine regression NNs and an iterative model-based image reconstruction (MBIR) algorithm, 
often leading to both good generalization capability and 
outperforming reconstruction quality over existing MBIR optimization models.
This paper proposes the first fast and convergent INN architecture, \emph{Momentum-Net},
by generalizing a block-wise MBIR algorithm that uses momentum and majorizers with regression NNs.
For fast MBIR, Momentum-Net uses \emph{momentum} terms in extrapolation modules,
and noniterative MBIR modules at each iteration by using \emph{majorizers},
where each iteration of Momentum-Net consists of three core modules:
image refining, extrapolation, and MBIR.
Momentum-Net guarantees convergence to a fixed-point
for general differentiable (non)convex MBIR functions (or data-fit terms) and convex feasible sets, 
under two asymptomatic conditions.
To consider data-fit variations across training and testing samples, 
we also propose a regularization parameter selection scheme 
based on the \dquotes{spectral spread} of majorization matrices.
Numerical experiments for light-field photography using a focal stack and sparse-view computational tomography 
demonstrate that, given identical regression NN architectures,
Momentum-Net significantly improves MBIR speed and accuracy over several existing INNs;
it significantly improves reconstruction quality 
compared to a state-of-the-art MBIR method in each application.  
\end{abstract}

% Note that keywords are not normally used for peerreview papers.
}

\maketitle

% To allow for easy dual compilation without having to reenter the
% abstract/keywords data, the \IEEEtitleabstractindextext text will
% not be used in maketitle, but will appear (i.e., to be "transported")
% here as \IEEEdisplaynontitleabstractindextext when the compsoc 
% or transmag modes are not selected <OR> if conference mode is selected 
% - because all conference papers position the abstract like regular
% papers do.
\IEEEdisplaynontitleabstractindextext
% \IEEEdisplaynontitleabstractindextext has no effect when using
% compsoc or transmag under a non-conference mode.

\IEEEraisesectionheading{\section{Introduction}\label{sec:intro}}

\IEEEPARstart{D}{eep} regression neural network (NN) methods 
have been actively studied for solving diverse inverse problems,
due to their effectiveness at mapping noisy signals into clean signals. 
Examples include 
image denoising 
\cite{Vincent&etal:10JMLR, Xie&Xu&Chen:12NIPS, Mao&Shen&Yang:16NIPS, Zhang&etal:17TIP},
image deconvolution 
\cite{Xu&etal:14NIPS, Sun&etal:15CVPR},
image super-resolution 
\cite{Dong&etal:16TPAMI, Kim&Lee&Lee:16CVPR},
magnetic resonance imaging (MRI) 
\cite{Yang&etal:18TMI, Quan&Nguyen-Duc&Jeong:18TMI},
X-ray computational tomography (CT) 
\cite{Chen&etal:17TMI, Jin&etal:17TIP, Ye&Han&Cha:18SJIS},
and light-field (LF) photography \cite{Wu&etal:17CVPR, Gupta&etal:17CVPRW}.
However, regression NNs with a greater mapping capability
have increased overfitting/hallucination risks \cite{Chun&etal:19MICCAI, Zheng&etal:19TCI, Lim&etal:20TMI, Ye&Long&Chun:ICIP20}. 
An alternative approach to solving inverse problems
is an \emph{iterative NN} (INN) that combines 
regression NNs -- called \dquotes{refiners} or denoisers -- with an unrolled iterative model-based image reconstruction (MBIR) algorithm 
\cite{Yang&etal:16NIPS, Chen&Pock:17PAMI, Romano&Elad&Milanfar:17SJIS, Buzzard&etal:18SJIS, Chun&Fessler:20TIP, Chun&Fessler:18IVMSP, Mardani&etal:18NIPS, Chun&etal:18Allerton}.
This alternative approach can regulate overfitting of regression NNs,
by balancing physical data-fit of MBIR and prior information estimated by refining NNs \cite{Lim&etal:20TMI, Chun&etal:19MICCAI}.
This \dquotes{soft-refiner} approach has been successfully applied to several extreme imaging systems, 
e.g., highly undersampled MRI
\cite{Yang&etal:16NIPS, Hammernik&etal:17MRM, Mardani&etal:18TMI, Aggarwal&Mani&Jacobs:18TMI, Chun&Fessler:18IVMSP},
low-dose or sparse-view CT
\cite{Gupta&etal:18TMI, Chun&Fessler:20TIP, Chun&etal:18Allerton, Chun&etal:19MICCAI, Ye&Long&Chun:ICIP20}, 
and low-count emission tomography \cite{Kim&etal:18TMI, Lim&etal:18NSSMIC, Lim&etal:20TMI, Lim&etal:19SNMMI}.

\subsection{Notation} 
\label{sec:notation}

This section provides mathematical notations.
We use $f(x;y)$ to denote a function $f$ of $x$ given $y$.
We use $\nm{\cdot}_{p}$ to denote the $\ell^p$-norm and write $\ip{\cdot}{\cdot}$ for the standard inner product on $\bbC^N$.  
The weighted $\ell^2$-norm with a Hermitian positive definite matrix $A$ is denoted by $\| \cdot \|_{A} = \| A^{\frac{1}{2}} (\cdot) \|_2$.
The Frobenius norm of a matrix is denoted by $\| \cdot \|_\mathrm{F}$. 
$( \cdot )^T$, $( \cdot )^H$, and $( {\cdot} )^*$ indicate the transpose, complex conjugate transpose (Hermitian transpose), and complex conjugate, respectively. 
$\diag(\cdot)$ denotes the conversion of a vector into a diagonal matrix or diagonal elements of a matrix into a vector.
For (self-adjoint) matrices $A,B \in \bbC^{N \times N}$, the notation $B \preceq A$ denotes that $A-B$ is a positive semi-definite matrix.

\subsection{From block-wise optimization to INN} \label{sec:intro:INN}

To recover signals $x \in \bbC^N$ from measurements $y \in \bbC^m$, 
consider the following MBIR optimization problem:
\ea{
\label{sys:mbir:z}
\argmin_{x \in \cX}~F(x;y,z), \quad F(x;y,z) \triangleq f(x;y) + \frac{\gamma}{2} \| x - z \|_2^2,
\tag{P0}
}
where $\cX$ is a set of feasible points,
$f(x; y)$ is data-fit function,
$\gamma$ is a regularization parameter, 
and $z \!\in\! \bbC^N$ is some high-quality approximation to the true unknown signal $x$.
The data-fit $f(x; y)$ measures deviations of model-based predictions of $x$ from data $y$, 
considering models of imaging physics (or image formation) and noise statistics in $y$.
In \R{sys:mbir:z}, the signal recovery accuracy increases as the quality of $z$ improves \cite[Prop.~3]{Zheng&etal:19TCI}; 
however, it is difficult to obtain such $z$ in practice. 
Alternatively, there has been a growing trend
in learning sparsifying regularizers (e.g., convolutional regularizers \cite{Chun&Fessler:18TIP, Chun&Fessler:20TIP, Chun&etal:19SPL, Chun&Fessler:18Asilomar, Crockett&etal:19CAMSAP}) from training datasets
and applying the trained regularizers to the following block-wise MBIR problem:
$\argmin_{x \in \cX} f(x;y) + \min_\zeta r(x, \zeta; \cO)$.
Here, a learned regularizer $\min_\zeta r(x, \zeta; \cO)$
quantifies consistency between $x$ and refined sparse signal $\zeta$ via some learned operators $\cO$.
Recently, we have constructed INNs
by generalizing the corresponding block-wise MBIR updates with regression NNs without convergence analysis
\cite{Chun&Fessler:18IVMSP, Chun&etal:18Allerton}.
In existing INNs, two major challenges exist: convergence and acceleration.

\subsection{Challenges in existing INNs: Convergence} 
\label{sec:intro:convg}

Existing convergence analysis has some practical limitations.
The original form of plug-and-play (PnP \cite{Sreehari&etal:16TCI, Zhang&etal:17CVPR, Chan&Wang&Elgendy:17TCI, Buzzard&etal:18SJIS}) is motivated by the alternating direction method of multipliers (ADMM \cite{Boyd&Parikh&Chu&Peleato&Eckstein:11FTML}),
and its fixed-point convergence has been analyzed with consensus equilibrium perspectives \cite{Buzzard&etal:18SJIS}.
However, similar to ADMM, its practical convergence depends on how one selects ADMM penalty parameters.
For example, \cite{Romano&Elad&Milanfar:17SJIS} reported unstable convergence behaviors
of PnP-ADMM with fixed ADMM parameters.
To moderate this problem,
\cite{Chan&Wang&Elgendy:17TCI} proposed a scheme that adaptively controls the ADMM parameters based on relative residuals.
Similar to the residual balancing technique \cite[\S3.4.1]{Boyd&Parikh&Chu&Peleato&Eckstein:11FTML},
the scheme in \cite{Chan&Wang&Elgendy:17TCI} requires tuning initial parameters.
Regularization by Denoising (RED \cite{Romano&Elad&Milanfar:17SJIS})
is an alternative
that moderates some such limitations.
In particular, RED aims to make a clear connection between optimization and a denoiser $\cD$, by defining its prior term by (scaled) $x^T (x - \cD(x))$.
Nonetheless, \cite{Reehorst&Schniter:19TCI} showed that many practical denoisers do not satisfy the Jacobian symmetry in \cite{Romano&Elad&Milanfar:17SJIS}, and proposed a less restrictive method, score-matching by denoising.

The convergence analysis of the INN inspired by the relaxed projected gradient descent (RPGD) method in \cite{Gupta&etal:18TMI} has the least restrictive conditions on the regression NN among the existing INNs. 
This method replaces the projector of a projected gradient descent method with an image refining NN.
However, the RPGD-inspired INN directly applies an image refining NN to gradient descent updates of data-fit;
thus, this INN relies heavily on the mapping performance of a refining NN and can have overfitting risks, similar to non-MBIR regression NNs, e.g., FBPConvNet \cite{Jin&etal:17TIP}.
In addition, it exploits the data-fit term only for the first few iterations \cite[Fig.~5(c)]{Gupta&etal:18TMI}.
We refer the perspective used in RPGD-inspired INN and its related works \cite{Kamilov&Mansour&Wohlberg:17SPL, Mardani&etal:18NIPS} as \dquotes{hard-refiner}: different from soft-refiners, these methods do not use a refining NN as a regularizer. 
More recently, 
\cite{Mardani&etal:18NIPS} presented convergence analysis
for an INN inspired by a proximal gradient descent method.
However, their analysis is based on noiseless measurements,
which is typically impractical.

Broadly speaking, existing convergence analysis largely depends
on the (firmly) nonexpansive property
of image refining NNs \cite{Buzzard&etal:18SJIS, Romano&Elad&Milanfar:17SJIS, Reehorst&Schniter:19TCI}, \cite[PGD]{Gupta&etal:18TMI}, \cite{Mardani&etal:18NIPS}. 
However, except for a single-hidden layer convolutional NN (CNN),
it is yet unclear which analytical conditions guarantee the non-expansiveness of general refining NNs \cite{Chun&etal:18Allerton}.
To guarantee convergence of INNs even when using possibly \textit{expansive} image refining NNs,
we proposed a method that normalizes the output signal of image refining NNs
by their Lipschitz constants
\cite{Chun&etal:18Allerton}.
However, if one uses expansive NNs that are identical across iterations,
it is difficult to obtain \dquotes{best} image recovery
with that normalization scheme.
The spectral normalization based training \cite{Miyato&etal:18ICLR, Ryu&etal:19ICML} can ensure the non-expansiveness of refining NNs
by single-step power iteration. 
However, similar to the normalization method in \cite{Chun&etal:18Allerton}, 
refining NNs trained with the spectral normalization method \cite{Ryu&etal:19ICML} 
degraded the image reconstruction accuracy for an INN using iteration-wise refining NNs \cite{Ye&Long&Chun:ICIP20}.
In addition, there does not yet exist theoretical convergence results
when refining NNs change across iterations,
yet iteration-wise refining NNs are widely studied 
\cite{Yang&etal:16NIPS, Chen&Pock:17PAMI, Hammernik&etal:17MRM, Chun&Fessler:18IVMSP}. 
Finally, existing analysis considers only a narrow class of data-fit terms:
most analyses consider a quadratic function with a linear imaging model \cite{Gupta&etal:18TMI, Mardani&etal:18NIPS}
or more generally, a convex cost function \cite{Reehorst&Schniter:19TCI, Buzzard&etal:18SJIS, Ryu&etal:19ICML} that can be minimized with a practical closed-form solution. 
No theoretical convergence results exist for \textit{general} (non)convex data-fit terms, iteration-wise NN denoisers, and a general set of feasible points.

\subsection{Challenges in existing INNs: Acceleration}
\label{sec:intro:acc}

Compared to non-MBIR regression NNs that do not exploit the data-fit $f(x;y)$ in \R{sys:mbir:z},
INNs require more computation
because they consider the imaging physics.
Computation increases as the imaging system or image formation model becomes larger-scale,
e.g., LF photography from a focal stack, 3D CT, parallel MRI using many receive coils, and image super-resolution.
Thus, acceleration becomes crucial for INNs.

First,
consider the existing methods motivated by ADMM or block coordinate descent (BCD) method: 
examples include PnP-ADMM~\cite{Chan&Wang&Elgendy:17TCI, Buzzard&etal:18SJIS}, 
RED-ADMM \cite{Romano&Elad&Milanfar:17SJIS, Reehorst&Schniter:19TCI}, 
MoDL \cite{Aggarwal&Mani&Jacobs:18TMI}, 
BCD-Net \cite{Chun&Fessler:18IVMSP, Lim&etal:20TMI, Chun&etal:19MICCAI}, etc.
These methods can require multiple inner iterations
to balance data-fit and prior information estimated by trained refining NNs,
increasing total MBIR time.
For example, in solving such problems, each outer iteration involves
$
x^{(i+1)} \!=\! \argmin_{x} F(x;y,z^{(i+1)})
$,
where $F$ is given as in \R{sys:mbir:z}
and $z^{(i+1)}$ is the output from the $i$th image refining NN.
For LF imaging system using a focal stack data \cite{Lien&etal:20NP}, 
solving the above problem requires multiple iterations,
and the total computational cost scale with the numbers of photosensors and sub-aperture images.
In addition,
nonconvexity of the data-fit term $f(x;y)$ can break
convergence guarantees of these methods,
because in general, the proximal mapping $\argmin_{x} f(x;y) + \gamma \| x - z^{(i+1)} \|_2^2$ is no longer nonexpansive.

Second,
consider the existing works motivated by gradient descent methods
\cite{Chen&Pock:17PAMI, Hammernik&etal:17MRM, Gupta&etal:18TMI, Mardani&etal:18NIPS}.
These methods resolve the inner iteration issue; 
however, they lack a sophisticated step-size control or backtracking scheme that influences convergence guarantee and acceleration.
Accelerated proximal gradient (APG) methods using \emph{momentum} terms can significantly accelerate convergence rates
for solving composite convex problems \cite{Nesterov:13MP, Beck&Teboulle:09SIAM},
so we expect that INN methods in the second class have yet to be maximally accelerated. 
The work in \cite{Kamilov&Mansour&Wohlberg:17SPL} applied PnP to the APG method \cite{Beck&Teboulle:09SIAM}; 
\cite{Ono:17SPL} applied PnP to the primal-dual splitting (PDS) algorithm \cite{Chambolle&Pock:11JMIV}.
However, similar to RPGD \cite{Gupta&etal:18TMI}, 
these are hard-refiner methods using some state-of-the-art denoisiers (e.g., BM3D \cite{Dabov&etal:07TIP}) but not trained NNs.
Those methods lack convergence analyses and guarantees may be limited to convex data-fit function.

\subsection{Contributions and organization of the paper} \label{sec:contribute}

This paper proposes \textit{Momentum-Net}, the first INN architecture that aims for fast and convergent MBIR.
The architecture of Momentum-Net is motivated by applying the Block Proximal Extrapolated Gradient method using a Majorizer (BPEG-M) \cite{Chun&Fessler:18TIP, Chun&Fessler:20TIP} to MBIR using trainable convolutional autoencoders \cite{Chun&Fessler:20TIP, Chun&Fessler:18IVMSP, Chun&Fessler:18Asilomar}.
Specifically, each iteration of Momentum-Net consists of three core modules: image refining, extrapolation, and MBIR. 
At each Momentum-Net iteration, an extrapolation module uses \emph{momentum} from previous updates to amplify the changes in subsequent iterations and accelerate convergence,
and an MBIR module is \emph{noniterative}.
In addition, Momentum-Net resolves the convergence issues mentioned in \S\ref{sec:intro:convg}: for general differentiable (non)convex data-fit terms and convex feasible sets, it guarantees convergence to a point that satisfies fixed-point and critical point conditions, under some mild conditions and two asymptotic conditions, i.e., \emph{asymptotically nonexpansive paired refining NNs} and \emph{asymptotically block-coordinate minimizer}.

The remainder of this paper is organized as follows.
\S\ref{sec:momnet} constructs the Momentum-Net architecture 
motivated by BPEG-M algorithm that solves MBIR problem using a learnable convolutional regularizer, 
describes its relation to existing works, 
analyzes its convergence, and summarizes the benefits of Momentum-Net over existing INNs.
\S\ref{sec:train} provides details of training INNs, including image refining NN architectures, single-hidden layer or \dquotes{shallow} CNN (sCNN)
and multi-hidden layer or \dquotes{deep} CNN (dCNN), and training loss function,
and proposes a regularization parameter selection scheme to consider data-fit variations across training and testing samples.
\S\ref{sec:exp} considers two extreme imaging applications: sparse-view CT and LF photography using a focal stack. 
\S\ref{sec:exp} reports numerical experiments of applications where
the proposed Momentum-Net using extrapolation significantly improves MBIR speed and accuracy, 
over the existing INNs, 
BCD-Net \cite{Chun&Fessler:18IVMSP, Aggarwal&Mani&Jacobs:18TMI, Romano&Elad&Milanfar:17SJIS},
Momentum-Net using \textit{no extrapolation} \cite{Chen&Pock:17PAMI, Hammernik&etal:17MRM},
ADMM-Net \cite{Yang&etal:16NIPS, Chan&Wang&Elgendy:17TCI, Buzzard&etal:18SJIS},
and PnP-PDS \cite{Ono:17SPL} using refining NNs.
Furthermore, \S\ref{sec:exp} reports numerical experiments where 
Momentum-Net significantly improves reconstruction quality compared to a state-of-the-art MBIR method in each application.

\section{Momentum-Net: Where BPEG-M meets NNs for inverse problems}
\label{sec:momnet}

\subsection{Motivation: BPEG-M algorithm for MBIR using learnable convolutional regularizer} 
\label{sec:mbir:learn_reg}

This section motivates the proposed Momentum-Net architecture, based on our previous works \cite{Chun&Fessler:20TIP, Chun&Fessler:18Asilomar}.
Consider the following approach for recovering signal $x$ from measurements $y$ (see the setup of block multi-(non)convex problems in \S\ref{sec:bpgm:block}):
\begingroup
\allowdisplaybreaks
\ea{
\label{sys:recov&caol}
& \argmin_{x \in \cX} ~ f(x; y) + \gamma \bigg( \min_{\{ \zeta_k \}} r(x, \{ \zeta_k \}; \{ h_k \}) \bigg),
\nn \\
& r(x, \{ \zeta_k \}; \{ h_k \}) \triangleq \sum_{k=1}^K \frac{1}{2} \nm{ h_k \conv x - \zeta_k  }_2^2 + \beta_k \nm{ \zeta_k }_1,
}
\endgroup
where $\cX$ is a closed set, 
$f(x; y) + \gamma r(x, \{ \zeta_k \}; \{ h_k \})$ is a (continuosly) differentiable (non)convex function in $x$, 
$\min_{\{ \zeta_k \}} r(x, \{ \zeta_k \} ; \{ h_k \})$ is a learnable convolutional regularizer \cite{Chun&Fessler:20TIP, Chun&etal:19SPL}, 
$\{ \zeta_k \!:\! k \!=\! 1,\ldots,K \}$ is a set of sparse features that correspond to $\{ h_k \conv x  \}$, 
$\{ h_k \!\in\! \bbC^R \!:\! k \!=\! 1,\ldots,K \}$ is a set of trainable filters, 
and $R$ and $K$ denote the size and number of trained filters, respectively.

\begin{figure*}[!pt]
% \vspace{-1em}
\centering
\small\addtolength{\tabcolsep}{-7.5pt}

\begin{tabular}{c}
\includegraphics[trim={0 6.15cm 0 6.4cm},clip,scale=0.48]{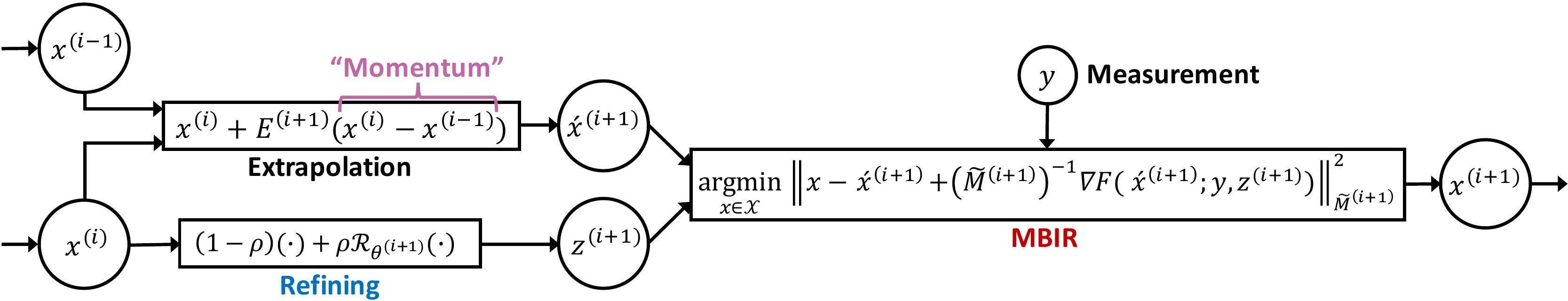}
\\
\small{(a) Momentum-Net}
\\
\includegraphics[trim={5.1cm 8.6cm 4.8cm 6.3cm},clip,scale=0.48]{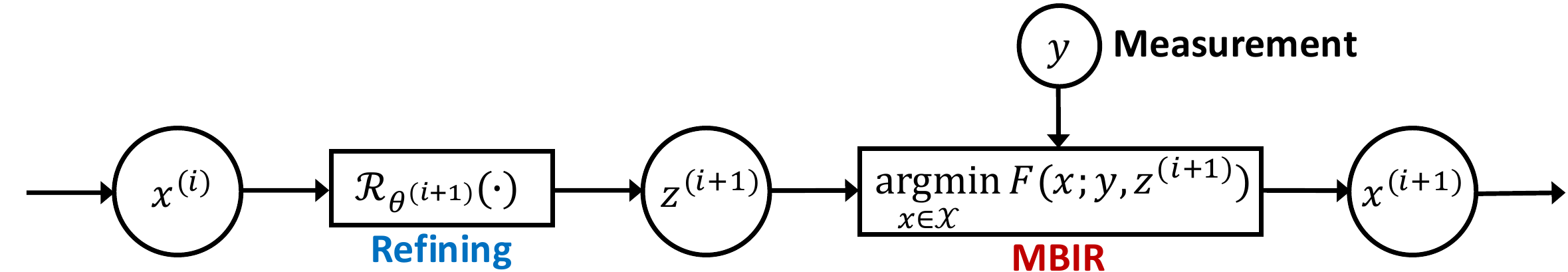} 
\\ 
\small{(b) BCD-Net \cite{Chun&Fessler:18IVMSP}} 
\end{tabular}

%\vspace{-0.5em}
\caption{Architectures of different INNs for MBIR.
(a--b)~The architectures of Momentum-Net and BCD-Net \cite{Chun&Fessler:18IVMSP} are constructed by generalizing BPEG-M and BCD algorithms that solve MBIR problem using a convolutional regularizer trained via \emph{convolutional analysis operator learning} (CAOL) \cite{Chun&Fessler:20TIP, Chun&etal:19SPL}, respectively.
(a)~Removing extrapolation modules (i.e., setting the extrapolation matrices $\{ E^{(i+1)} \!:\! \forall i \}$ as a zero matrix),
Momentum-Net specializes to the existing gradient-descent-inspired INNs \cite{Chen&Pock:17PAMI, Hammernik&etal:17MRM}.
When the MBIR cost function $F(x;y,z^{(i+1)})$ in \R{sys:mbir} has a sharp majorizer $\widetilde{M}^{(i+1)}$, $\forall i$, 
Momentum-Net (using $\rho \!\approx\! 1$) specializes to BCD-Net; see Examples~\ref{eg:bcdnet:denoising}--\ref{eg:bcdnet:mri}.
(b)~BCD-Net is a general version of the existing INNs in \cite{Yang&etal:16NIPS, Sreehari&etal:16TCI, Zhang&etal:17CVPR, Chan&Wang&Elgendy:17TCI, Romano&Elad&Milanfar:17SJIS, Aggarwal&Mani&Jacobs:18TMI, Buzzard&etal:18SJIS} by using iteration-wise image refining NNs, i.e., $\{ \cR_{\theta^{(i+1)}} : \forall i \}$, or considering general convex data-fit $f(x;y)$.
}
\label{fig:arch}
% \vspace{-1em}
\end{figure*}

Problem \R{sys:recov&caol} can be viewed as a two-block optimization problem in terms of the image $x$ and the features $\{ \zeta_k \}$.
We solve \R{sys:recov&caol} using the recent BPEG-M optimization framework \cite{Chun&Fessler:18TIP, Chun&Fessler:20TIP}
that has attractive convergence guarantee and rapidly solved several block optimization problems \cite{Xu&Yin:13SIAM, Xu&Yin:16IPI, Xu&Yin:17JSC, Chun&Fessler:18TIP, Chun&Fessler:20TIP}.
BPEG-M has the following key ideas for each block optimization problem
(see details in \S\ref{sec:bpgm}):
\bulls{
%[\setlength{\topsep}{1pt}]

\item $M_b$-Lipschitz continuity for the gradient of the $b\rth$ block optimization problem, $\forall b$:

\defn{[\mbox{$M$-Lipschitz continuity \cite{Chun&Fessler:20TIP}}] \label{d:QM}
A function $g: \bbR^n \rightarrow \bbR^{n}$ is \emph{$M$-Lipschitz continuous} on $\bbR^n$ if there exists a (symmetric) positive definite matrix $M$ such that
\bes{
\nm{g(u) - g(v)}_{M^{-1}} \leq \nm{u - v}_{M}, \quad \forall u,v \in \bbR^n.
}
}
Definition~\ref{d:QM} is a more general concept than the classical Lipschitz continuity.

\item A sharper majorization matrix $M$ that gives a tighter bound in Definition~\ref{d:QM} 
leads to a tighter quadratic majorization bound in the following lemma: 

\lem{[Quadratic majorization via $M$-Lipschitz continuous gradients~\!\mbox{\cite{Chun&Fessler:20TIP}}] \label{l:QM}
Let $f(u) : \bbR^n \rightarrow \bbR$. If $\nabla f$ is $M$-Lipschitz continuous, then 
\begingroup
\setlength{\thinmuskip}{1.5mu}
\setlength{\medmuskip}{2mu plus 1mu minus 2mu}
\setlength{\thickmuskip}{2.5mu plus 2.5mu}
%\fontsize{9.5pt}{11.4pt}\selectfont
\bes{
f(u) \leq f(v) + \ip{\nabla_u f(v)}{u-v} + \frac{1}{2} \nm{u - v}_M^2, \quad \forall u,v \in \bbR^n.
}
\endgroup
}

Having tighter majorization bounds, 
sharper majorization matrices tend to accelerate BPEG-M convergence.

\item The majorized block problems are \dquotes{proximable}, i.e., proximal mapping of majorized function is \dquotes{easily} computable depending on the properties of $b\rth$ block majorizer and regularizer, $M_b$ and $r_b$, 
where the proximal mapping operator is defined by
\be{
\label{eq:d:prox}
\mathrm{Prox}_{r_b}^{M_b} (z) \triangleq \argmin_{u} \, \frac{1}{2} \nm{u - z}_{M_b}^2 + r_b(u), \quad \forall b.
}

\item Block-wise extrapolation and momentum terms to accelerate convergence.
}

Suppose that \textit{1)} gradient of $f(x; y) + \gamma r(x, \{ \zeta_k \}; \{ h_k \})$ is $M$-Lipschitz continuous at an extrapolated point $\acute{x}^{(i+1)}$, $\forall i$; \textit{2)} filters in \R{sys:recov&caol} satisfy the tight-frame (TF) condition, 
$\sum_{k=1}^K \nm{ h_k \conv u }_2^2 = \nm{ u }_2^2$, $\forall u$,
for some boundary conditions \cite{Chun&Fessler:20TIP}.
Applying the BPEG-M framework (see Algorithm~\ref{alg:bpgm}) to solving \R{sys:recov&caol} leads to the following block updates:
\begingroup
\setlength{\thinmuskip}{1.5mu}
\setlength{\medmuskip}{2mu plus 1mu minus 2mu}
\setlength{\thickmuskip}{2.5mu plus 2.5mu}
\fontsize{9.5pt}{11.4pt}\selectfont
\ea{
\label{eq:bpgm:zk}
z^{(i+1)} &= \sum_{k=1}^K \mathrm{flip}(h_k^*) \conv \cT_{\beta_k} ( h_k \conv x^{(i)} ),
\\
\label{eq:bpgm:xacute}
\acute{x}^{(i+1)} &= x^{(i)} + E^{(i+1)} \big( x^{(i)} - x^{(i-1)} \big),
\\
\label{eq:bpgm:x}
x^{(i+1)} & = \mathrm{Prox}_{\bbI_{\cX}}^{\widetilde{M}^{(i+1)}} \!\!\! 
\Big( \! \acute{x}^{(i+1)} - \big( \widetilde{M}^{(i+1)} \big)^{\!\!-1} \nabla F (\acute{x}^{(i+1)}; y, z^{(i+1)})  \! \Big),
}
\endgroup
where $E^{(i+1)}$ is an extrapolation matrix that is given in \R{up:Ex:cvx} or \R{up:Ex:ncvx} below,
$\widetilde{M}^{(i+1)}$ is a (scaled) majorization matrix for $\nabla F(x; y, z^{(i+1)})$ that is given in \R{up:MFtilde} below, $\forall i$,
the proximal operator $\mathrm{Prox}_{\bbI_{\cX}}^{\widetilde{M}^{(i+1)}} \! (\cdot)$ in \R{eq:bpgm:x} is given by \R{eq:d:prox},
and $\bbI_{\cX} (x)$ is the characteristic function of set $\cX$ (i.e., $\bbI_{\cX}$ equals to $0$ if $x \in \cX$, and $\infty$ otherwise).

Proximal mapping update \R{eq:bpgm:zk} has a \emph{single-hidden layer convolutional autoencoder} architecture 
that consists of encoding convolution, nonlinear thresholding, and decoding convolution,
where $\mathrm{flip}(\cdot)$ flips a filter along each dimension,
and the soft-thresholding operator $\cT_{\alpha} (u): \bbC^N \rightarrow \bbC^N$ is defined by 
\be{
\label{eq:soft-threshold}
( \cT_{\alpha} (u) )_n \triangleq \left\{ \begin{array}{cc} u_n - \alpha \cdot \sgn (u_n), & | u_n | > \alpha, \\ 0, & \mbox{otherwise}, \end{array} \right.
}
for $n = 1,\ldots,N$, in which $\sgn(\cdot)$ is the sign function.
See details of deriving BPEG-M updates \R{eq:bpgm:zk}--\R{eq:bpgm:x} in \S\ref{sec:bpgm:caol}.
The BPEG-M updates in \R{eq:bpgm:zk}--\R{eq:bpgm:x} guarantee convergence to a critical point, 
when MBIR problem~\R{sys:recov&caol} satisfies some mild conditions, e.g., lower-boundedness and existence of critical points; see
Assumption~S.1 in \S\ref{sec:bpgm:analysis}. 

The following section generalizes the BPEG-M updates in \R{eq:bpgm:zk}--\R{eq:bpgm:x} and constructs the Momentum-Net architecture.

\subsection{Architecture} \label{sec:momnet:setup}

This section establishes the INN architecture of \emph{Momentum-Net} by generalizing BPEG-M updates \R{eq:bpgm:zk}--\R{eq:bpgm:x} that solve \R{sys:recov&caol}.
Specifically, we replace the proximal mapping in \R{eq:bpgm:zk} with a general image refining NN $\cR_{\theta} (\cdot)$, 
where $\theta$ denotes the trainable parameters.
To effectively remove iteration-wise artifacts and give \dquotes{best} signal estimates at each iteration,
we further generalize a refining NN $\cR_{\theta} (\cdot)$ to 
\emph{iteration-wise} image refining NNs $\{ \cR_{\theta^{(i+1)}} (\cdot) \!:\! i \!=\! 0,\ldots,N_{\text{iter}}\!-\!1 \}$, 
where $\theta^{(i+1)}$ denotes the parameters for the $i\rth$ iteration refining NN $\cR_{\theta^{(i+1)}}$, 
and $N_{\text{iter}}$ is the number of Momentum-Net iterations.
The iteration-wise NNs are particularly useful for reducing overfitting risks in regression,
because $\cR_{\theta^{(i+1)}}$ is responsible for removing noise features only at the $i\rth$ iteration, 
and thus one does not need to greatly increase dimensions of its parameter $\theta^{(i+1)}$ \cite{Lim&etal:20TMI, Chun&etal:19MICCAI}.
In low-dose CT reconstruction, for example, the refining NNs at the early and later iterations remove streak artifacts and Gaussian-like noise, respectively \cite{Chun&etal:19MICCAI}.

\begin{algorithm}[!pt]
\caption{Momentum-Net}
\label{alg:momnet}

\begin{algorithmic}
\REQUIRE $\{ \cR_{\theta^{(i)}} : i= 1,\ldots,N_{\text{iter}} \}$, $\rho \in (0,1)$, $\gamma >  0$, $x^{(0)} = x^{(-1)}$, $y$

\FOR{$i = 0,\ldots,N_{\text{iter}}\!-\!1$}

\STATE Calculate $\widetilde{M}^{(i+1)}$ by \R{up:MFtilde}, and $E^{(i+1)}$ by \R{up:Ex:cvx} or \R{up:Ex:ncvx}

\STATE {\em Image refining}:
\begingroup
\setlength\abovedisplayskip{0.2\baselineskip}
\setlength\belowdisplayskip{0.2\baselineskip}
\be{
\label{eq:momnet:map}
z^{(i+1)} = (1-\rho) x^{(i)} + \rho \cR_{\theta^{(i+1)}} \big( x^{(i)} \big) \tag{Alg.1.1}
}
\endgroup

\STATE {\em Extrapolation}:
\begingroup
\setlength\abovedisplayskip{0.2\baselineskip}
\setlength\belowdisplayskip{0.2\baselineskip}
\be{
\label{eq:momnet:exp}
\acute{x}^{(i+1)} = x^{(i)} + E^{(i+1)} \big( x^{(i)} - x^{(i-1)} \big) \tag{Alg.1.2}
}
\endgroup

\STATE {\em MBIR}: 
\begingroup
\setlength{\thinmuskip}{1.5mu}
\setlength{\medmuskip}{2mu plus 1mu minus 2mu}
\setlength{\thickmuskip}{2.5mu plus 2.5mu}
%\fontsize{9.5pt}{11.4pt}\selectfont
\setlength\abovedisplayskip{0.2\baselineskip}
\setlength\belowdisplayskip{0.2\baselineskip}
\ea{
\label{eq:momnet:mbir}
\nn
%\tag{Alg.1.3}
&~ x^{(i+1)} 
 \\[-0.2\baselineskip]
&= \mathrm{Prox}_{\bbI_{\cX}}^{\widetilde{M}^{(i+1)}} \!\!\! 
\Big( \! \acute{x}^{(i+1)} - \big( \widetilde{M}^{(i+1)} \big)^{\!\!-1} \nabla F (\acute{x}^{(i+1)}; y, z^{(i+1)}) \! \Big)
%\nn
\tag{Alg.1.3}
}
\endgroup

\ENDFOR

\end{algorithmic}
\end{algorithm}

Each iteration of Momentum-Net consists of \textit{1)} image refining, \textit{2)} extrapolation, and \textit{3)} MBIR modules, corresponding to the BPEG-M updates \R{eq:bpgm:zk}, \R{eq:bpgm:xacute}, and \R{eq:bpgm:x}, respectively. 
See the architecture of Momentum-Net in Fig.~\ref{fig:arch}(a) and Algorithm~\ref{alg:momnet}.
At the $i\rth$ iteration, Momentum-Net performs the following three processes:

\bulls{
%[\setlength{\topsep}{1pt}]

\item {\em Refining}: 
The $i\rth$ image refining module gives the \dquotes{refined} image $z^{(i+1)}$, by applying the $i\rth$ refining NN, $\cR_{\theta^{(i+1)}}$, to an input image at the $i\rth$ iteration, $x^{(i)}$ (i.e., image estimate from the $(i-1)\rth$ iteration).
Different from existing INNs, e.g., ADMM-Net \cite{Yang&etal:16NIPS}, PnP-ADMM \cite{Chan&Wang&Elgendy:17TCI, Buzzard&etal:18SJIS}, RED \cite{Romano&Elad&Milanfar:17SJIS}, 
MoDL \cite{Aggarwal&Mani&Jacobs:18TMI}, BCD-Net \cite{Chun&Fessler:18IVMSP} (see Fig.~\ref{fig:arch}(b)), TNRD \cite{Chen&Pock:17PAMI, Hammernik&etal:17MRM}, 
we apply $\rho$-relaxation with $\rho \in (0,1)$; see \R{eq:momnet:map}.
The parameter $\rho$ controls the strength of inference from refining NNs, 
but does not affect the convergence guarantee of Momentum-Net.
Proper selection of $\rho$ can improve MBIR accuracy (see \S\ref{sec:discuss:param}).

\item {\em Extrapolation}: 
The $i\rth$ extrapolation module gives the  extrapolated point $\acute{x}^{(i+1)}$, based on \emph{momentum} terms $x^{(i)} - x^{(i-1)}$; see \R{eq:momnet:exp}.
Intuitively speaking, momentum is information from previous updates to amplify the changes in subsequent iterations. 
Its effectiveness has been shown in diverse optimization literature, e.g., convex optimization \cite{Nesterov:13MP, Beck&Teboulle:09SIAM} and block optimization \cite{Chun&Fessler:18TIP, Chun&Fessler:20TIP}.

\item {\em MBIR}: 
Given a refined image $z^{(i+1)}$ and a measurement vector $y$,
the $i\rth$ MBIR module \R{eq:momnet:mbir} applies the proximal operator $\mathrm{Prox}_{\bbI_{\cX}}^{\widetilde{M}^{(i+1)}} \! (\cdot)$
to the \emph{extrapolated gradient update using a quadratic majorizer} of $F(x;y,z^{(i+1)})$, 
where $F$ is defined in \R{sys:mbir:z}.
Intuitively speaking, this step solves a \emph{majorized} version of the following MBIR problem at the extrapolated point $\acute{x}^{(i+1)}$:
\be{
\label{sys:mbir}
\min_{x \in \cX} F(x;y,z^{(i+1)}),
\tag{P1}
}
and gives a reconstructed image $x^{(i+1)}$.
In Momentum-Net, we consider (non)convex differentiable MBIR cost functions $F$ with $M$-Lipschitz continuous gradients,
and a convex and closed set $\cX$.
For a wide range of large-scale inverse imaging problems, 
the majorized MBIR problem \R{eq:momnet:mbir} has a practical closed-form solution and thus, does not require an iterative solver,
depending on the properties of practically invertible majorization matrices $M^{(i+1)}$ and constraints.
Examples of $M^{(i+1)}$-$\cX$ combinations that give a noniterative solution for \R{eq:momnet:mbir} include 
scaled identity and diagonal matrices with a box constraint and the non-negativity constraint,
and matrices decomposable by unitary transforms, e.g., a circulant matrix \cite{Muckley&etal:16ISMRM, Mcgaffin&Fessler:15arXiv}, with $\cX = \bbC^N$.
The updated image $x^{(i+1)}$ is the input to the next Momentum-Net iteration.
}

The followings are details of Momentum-Net in Algorithm~\ref{alg:momnet}.
A scaled majorization matrix is
\be{
\label{up:MFtilde}
\widetilde{M}^{(i+1)} = \lambda \cdot M^{(i+1)} \succ 0, \qquad \lambda \geq 1,
}
where $M^{(i+1)} \!\in\! \bbR^{N \times N}$ is 
a symmetric positive definite majorization matrix of $\nabla F(x;y,z^{(i+1)})$
in the sense of $M$-Lipschitz continuity (see Definition~\ref{d:QM}).
In \R{up:MFtilde}, $\lambda = 1$ and $\lambda > 1$ for convex and nonconvex $F(x;y,z^{(i+1)})$ (or convex and nonconvex $f(x;y)$), respectively.
We design the extrapolation matrices as follows:
\begingroup
\setlength{\thinmuskip}{1.5mu}
\setlength{\medmuskip}{2mu plus 1mu minus 2mu}
\setlength{\thickmuskip}{2.5mu plus 2.5mu}
%\fontsize{9.5pt}{11.4pt}\selectfont
\allowdisplaybreaks
\setlength\abovedisplayskip{0.2\baselineskip}
\setlength\belowdisplayskip{0.2\baselineskip}
\ea{
& \mbox{for \emph{convex}~$F$,}
\nn \\
& E^{(i+1)} =
\delta^2 m^{(i)} \cdot \big( M^{(i+1)} \big)^{\!-\frac{1}{2}} 
\big( M^{(i)} \big)^{\frac{1}{2}};
\label{up:Ex:cvx} 
\\
& \mbox{for \emph{nonconvex}~$F$,}
\nn \\
& E^{(i+1)} = 
\delta^2 m^{(i)} \cdot \frac{\lambda-1}{2 (\lambda+1)} \cdot \big( M^{(i+1)} \big)^{\!-\frac{1}{2}} 
\big( M^{(i)} \big)^{\!\frac{1}{2}}, 
\label{up:Ex:ncvx}
}
\endgroup
for some $\delta \!<\! 1$ and $\{ 0 \!\leq\! m^{(i)} \!\leq\! 1 \!:\! \forall i \}$.
We update the momentum coefficients $\{ m^{(i+1)} \!:\! \forall i \}$ by the following formula \cite{Chun&Fessler:18TIP, Chun&Fessler:20TIP}:
\be{
\label{eq:mom_coeff}
m^{(i+1)} = \frac{\theta^{(i)}  - 1}{\theta^{(i+1)}}, \qquad \theta^{(i+1)} = \frac{1 + \sqrt{1 + 4 (\theta^{(i)})^2}}{2};
}
if $F(x;y,z^{(i+1)})$ has a sharp majorizer, i.e., $\nabla F(x;y,z^{(i+1)})$ has $M^{(i+1)}$ 
such that the corresponding bound in Definition~\ref{d:QM} is tight, then we set $m^{(i+1)} = 0$, $\forall i$.
\S\ref{sec:param} lists parameters of Momentum-Net, 
and summarizes selection guidelines or gives default values.

\subsection{Relations to previous works} \label{sec:relation}

Several existing MBIR methods can be viewed as a special case of Momentum-Net:

\examp{\label{eg:caol}
(MBIR model \R{sys:recov&caol} using convolutional autoencoders that satisfy the TF condition \cite{Chun&Fessler:20TIP}). 
The BPEG-M updates in \R{eq:bpgm:zk}--\R{eq:bpgm:x} are special cases of the modules in Momentum-Net (Algorithm~\ref{alg:momnet}), 
with $\{ \cR_{\theta^{(i+1)}} ( \cdot ) \!=\! \sum_{k=1}^K \mathrm{flip}(h_k^*) \conv \cT_{\beta_k} (h_k \conv (\cdot) ) \!:\! \forall i \}$ 
(i.e., single hidden-layer convolutional autoencoder \cite{Chun&Fessler:20TIP}) and $\rho \approx 1$.
These give a clear mathematical connection between a denoiser \R{eq:bpgm:zk} and cost function \R{sys:recov&caol}.
One can find a similar relation between a multi-hidden layer CNN and a multi-layer convolutional regularizer \cite[Appx.]{Chun&Fessler:20TIP}.
}

\examp{\label{eg:tnrd}
(INNs inspired by gradient descent method, e.g., TNRD \cite{Chen&Pock:17PAMI, Hammernik&etal:17MRM}).
Removing extrapolation modules, i.e., setting $\{ E^{(i+1)} = 0 : \forall i \}$ in \R{eq:momnet:exp}, and setting $\rho \approx 1$, Momentum-Net becomes the existing INN in \cite{Chen&Pock:17PAMI, Hammernik&etal:17MRM}.
}

\example{\label{eg:bcdnet:denoising}
(BCD-Net for image denoising \cite{Chun&Fessler:18IVMSP}).
To obtain a clean image $x \in \bbR^N$ from a noisy image $y \in \bbR^N$ corrupted by an additive white Gaussian noise (AWGN),
MBIR problem \R{sys:mbir} considers the data-fit $f(x;y) = \frac{1}{2} \| y -  x \|_W^2$ 
with the inverse covariance matrix $W = \frac{1}{\sigma^2} I$, where $\sigma^2$ is a variance of AWGN, 
and the box constraint $\cX = [0, U]^N$ with an upper bound $U > 0$.
For this $f(x;y)$, the MBIR module \R{eq:momnet:mbir} can use the exact majorizer 
$\{ \widetilde{M}^{(i+1)} = ( \frac{1}{\sigma^2} + \gamma) I  \}$ and one does not need to use the extrapolation module \R{eq:momnet:exp},
i.e., $\{ E^{(i+1)} = 0  \}$.
Thus, Momentum-Net (with $\rho \approx 1$) becomes BCD-Net.
}

\examp{\label{eg:bcdnet:mri}
(BCD-Net for undersampled single-coil MRI \cite{Chun&Fessler:18IVMSP}).
To obtain an object magnetization $x \in \bbR^N$ from a \textit{k}-space data $y \in \bbC^m$ obtained by undersampling (e.g., compressed sensing \cite{Chun&Adcock:17TIT}) MRI,
MBIR problem \R{sys:mbir} considers the data-fit $f(x;y) = \frac{1}{2} \| y -  A x \|_W^2$ 
with an undersampling Fourier operator $A$
(disregarding relaxation effects and considering Cartesian \textit{k}-space),
the inverse covariance matrix $W = \frac{1}{\sigma^2} I$, where $\sigma^2$ is a variance of complex AWGN \cite{AjaFernandez&VegasSanchez&TristanVega:14MRI},
and $\cX = \bbC^N$.
For this $f(x;y)$, the MBIR module \R{eq:momnet:mbir} can use the exact majorizer $\{ \widetilde{M}^{(i+1)} = F_{\textmd{disc}}^H ( \frac{1}{\sigma^2} P + \gamma I) F_{\textmd{disc}}  \}$ that is practically invertible,
where $F_{\textmd{disc}}$ is the discrete Fourier transform and 
$P$ is a diagonal matrix with either $0$ or $1$ 
(their positions correspond to sampling pattern in $k$-space), 
and the extrapolation module \R{eq:momnet:exp} uses the zero extrapolation matrices $\{ E^{(i+1)} = 0 \}$.
Thus, Momentum-Net (with $\rho \approx 1$) becomes BCD-Net.
}

The following section analyzes the convergence of Momentum-Net.

\begin{figure}[!pt]
\centering
\begin{tabular}{c}
\includegraphics[scale=0.6, trim=0em 0.3em 1.9em 1.0em, clip]{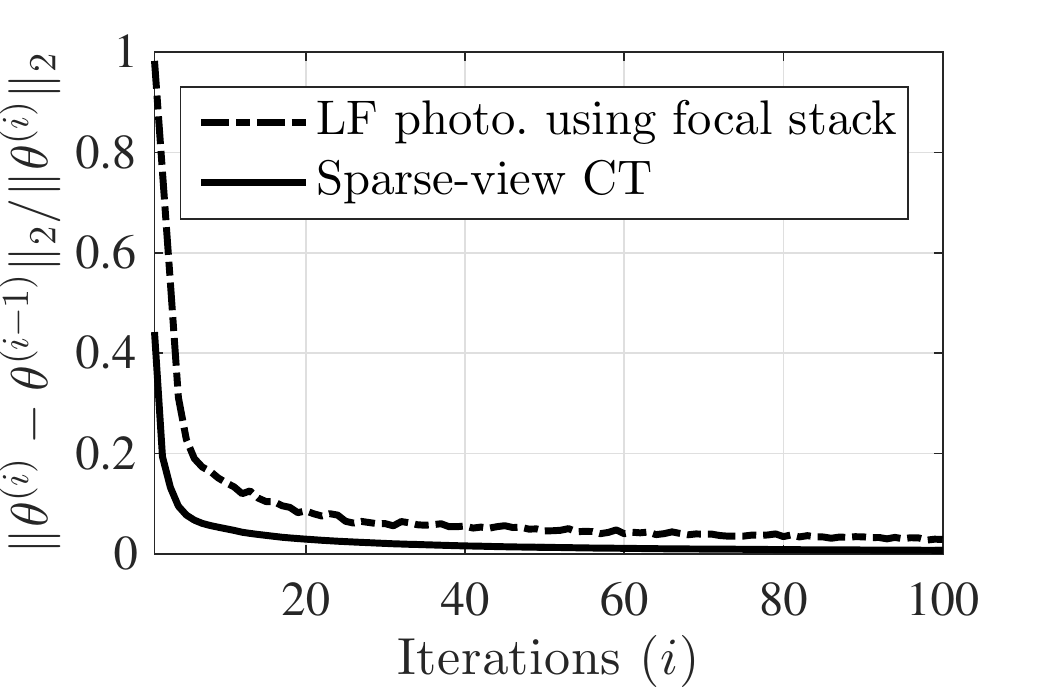} 
\end{tabular}

\vspace{-0.75em}
\caption{
Convergence behavior of Momentum-Net's dCNNs refiners $\{ \cR_{\theta^{(i)}} \}$
in different applications
($\theta^{(i)}$ denotes the parameter vector of the $i\rth$ iteration refiner $\cR_{\theta^{(i)}}$, for $i \!=\! 1,\ldots,N_{\text{iter}}$;
see details of $\{ \cR_{\theta^{(i)}} \}$ in \R{sys:dcnn} and \S\ref{sec:exp:INN:param}; $N_{\text{iter}} \!=\! 100$).
\underline{Sparse-view CT} (fan-beam geometry with $12.5$\% projections views):
$\cR_{\theta^{(i)}}$ quickly converges, where majorization matrices of training data-fits have similar condition numbers.
\underline{LF photography using a focal stack} (five detectors and reconstructed LFs consists of $9 \!\times\! 9$ sub-aperture images): 
$\cR_{\theta^{(i)}}$ has slower convergence, where majorization matrices of training data-fits have largely different condition numbers.
}
\label{fig:momnet:convg}
\vspace{-0.25em}
\end{figure}

\subsection{Convergence analysis} \label{sec:momnet:convg}

In practice, INNs, i.e., \dquotes{unrolled} or PnP methods using refining NNs,
are trained and used with a specific number of iterations.
Nevertheless, similar to optimization algorithms, studying convergence properties of INNs with $N_{\text{iter}} \rightarrow \infty$ \cite{Gupta&etal:18TMI, Buzzard&etal:18SJIS, Ryu&etal:19ICML} is important; 
in particular, it is crucial to know if a given INN tends to converge as $N_{\text{iter}}$ increases.
For INNs using iteration-wise refining NNs, e.g., BCD-Net \cite{Chun&Fessler:18IVMSP} and proposed Momentum-Net, 
we expect that refiners converge,
i.e., their image refining capacity converges,
because information provided by data-fit function $f(x;y)$ in MBIR (e.g., likelihood) reaches some \dquotes{bound} after a certain number of iterations.
Fig.~\ref{fig:momnet:convg} illustrates that dCNN parameters of Momentum-Net tend to converge for different applications.
(The similar behavior was reported for sCNN refiners in BCD-Net \cite{Chun&etal:19MICCAI}.)
Although refiners do not \emph{completely} converge, in practice,  
one could use a refining NN at a sufficiently large iteration number, e.g., $N_{\text{iter}} \!=\! 100$ in Momentum-Net,  
for the later iterations.

There are two key challenges in analyzing the convergence of Momentum-Net in Algorithm~\ref{alg:momnet}: 
both challenges relate to its image refining modules \R{eq:momnet:map}.
First, image refining NNs $\cR_{\theta^{(i+1)}}$ change across iterations; 
even if they are identical across iterations, they are not necessarily nonexpansive operators \cite{Rockafellar:76SIAMCO, Ryu&Boyd:16ACM} in practice.
Second, the iteration-wise refining NNs are not necessarily proximal mapping operators, 
i.e., they are not written explicitly in the form of \R{eq:d:prox}.
This section proposes two new asymptotic definitions to overcome these challenges, 
and then uses those conditions to analyze convergence properties of Momentum-Net in Algorithm~\ref{alg:momnet}.

\subsubsection{Preliminaries} \label{sec:prelim}

To resolve the challenge of iteration-wise refining NNs and the practical difficulty in guaranteeing their non-expansiveness, 
we introduce the following generalized definition of the non-expansiveness \cite{Rockafellar:76SIAMCO, Ryu&Boyd:16ACM}.

\defn{[Asymptotically nonexpansive paired operators] \label{d:pair-map}
A sequence of paired operators $( \cR_{\theta^{(i)}}, \cR_{\theta^{(i+1)}} )$ is asymptotically nonexpansive if there exist a summable nonnegative sequence $\{ \epsilon^{(i+1)} \geq 0 : \sum_{i=0}^\infty \epsilon^{(i+1)} < \infty \}$ such that\footnote{
One could replace the bound in \R{cond:eps} with $\|  \cR_{\theta^{(i+1)}}(u) - \cR_{\theta^{(i)}} (v) \|_2^2 \leq (1+\epsilon^{(i+1)}) \nm{u - v}_2^2$ (and summable $\{ \epsilon^{(i+1)} : \forall i \}$), and the proofs for our main arguments go through. 
}
\be{
\label{cond:eps}
\left\| \cR_{\theta^{(i+1)}} (u) - \cR_{\theta^{(i)}} (v) \right\|_2^2 \leq  \nm{u - v}_2^2 + \epsilon^{(i+1)}, \qquad \forall u,v, i.
}
}

When $\cR_{\theta^{(i+1)}} \!=\! \cR_{\theta}$ and  $\epsilon^{(i+1)} \!=\! 0$, $\forall i$, 
Definition~\ref{d:pair-map} becomes the standard non-expansiveness of a mapping operator $ \cR_{\theta}$.
If we replace the inequality ($\leq$) with the strict inequality ($<$) in \R{cond:eps}, then we say that the sequence of paired operators $( \cR_{\theta^{(i+1)}}, \cR_{\theta^{(i+1)}} )$ is asymptotically contractive. (This stronger assumption is used to prove convergence of BCD-Net in Proposition~\ref{p:bcdnet:cauchy}.) 
Definition~\ref{d:pair-map} also implies that mapping operators $\cR_{\theta^{(i+1)}}$ converge to some nonexpansive operator, 
if the corresponding parameters ${\theta^{(i+1)}}$ converge.

\begin{figure*}[!pt]
% \vspace{-0.75em}
 \centering
 \small\addtolength{\tabcolsep}{-7.5pt}
 \renewcommand{\arraystretch}{1}

     \begin{tabular}{ccc}
     \multicolumn{3}{c}{\small (a) Sparse-view CT: Condition numbers of data-fit majorizers have \textit{mild} variations.} 
     \\
     \small{(a1) $\{ \Delta^{(i)} : i \geq 2 \}$} &
     \small{(a2) $\{ \epsilon^{(i)} : i \geq 2 \}$} &
     \small{(a3) $\{ \kappa^{(i)} : i \geq 1 \}$} 
     \\
     \includegraphics[scale=0.55, trim=0.2em 0.2em 1.1em 1em, clip]{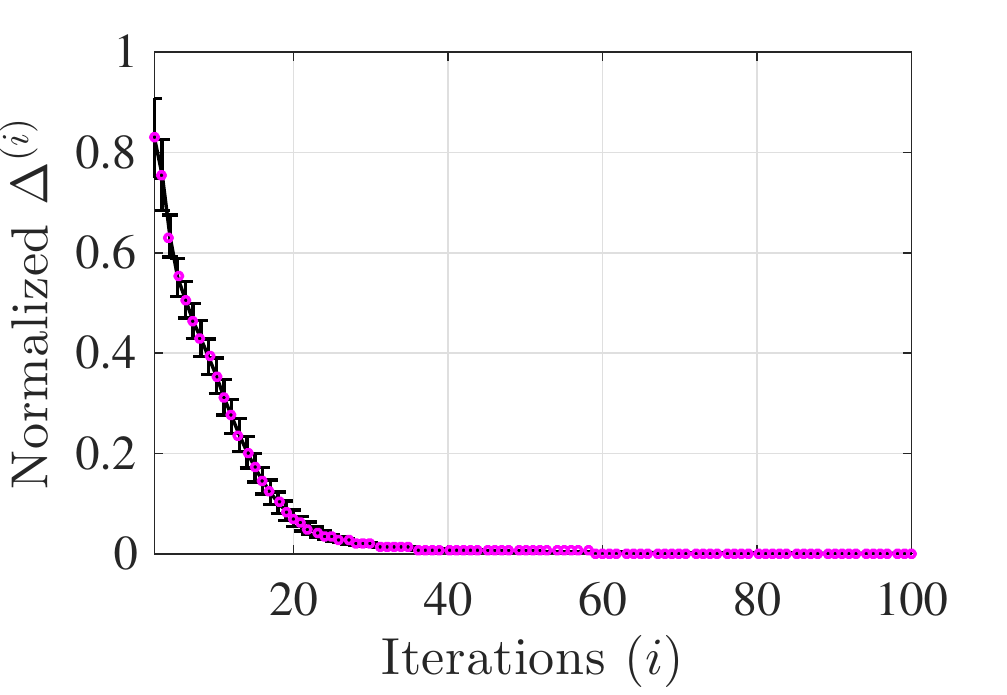} &
     \includegraphics[scale=0.55, trim=0.2em 0.2em 1.1em 1em, clip]{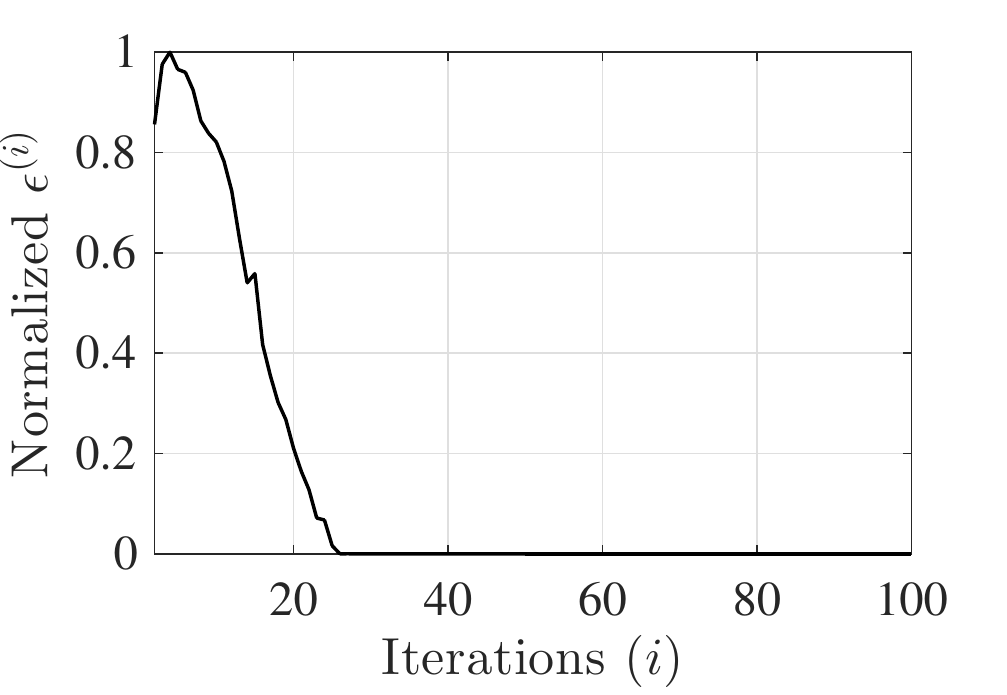} &
     \includegraphics[scale=0.55, trim=0.2em 0.2em 1.1em 1em, clip]{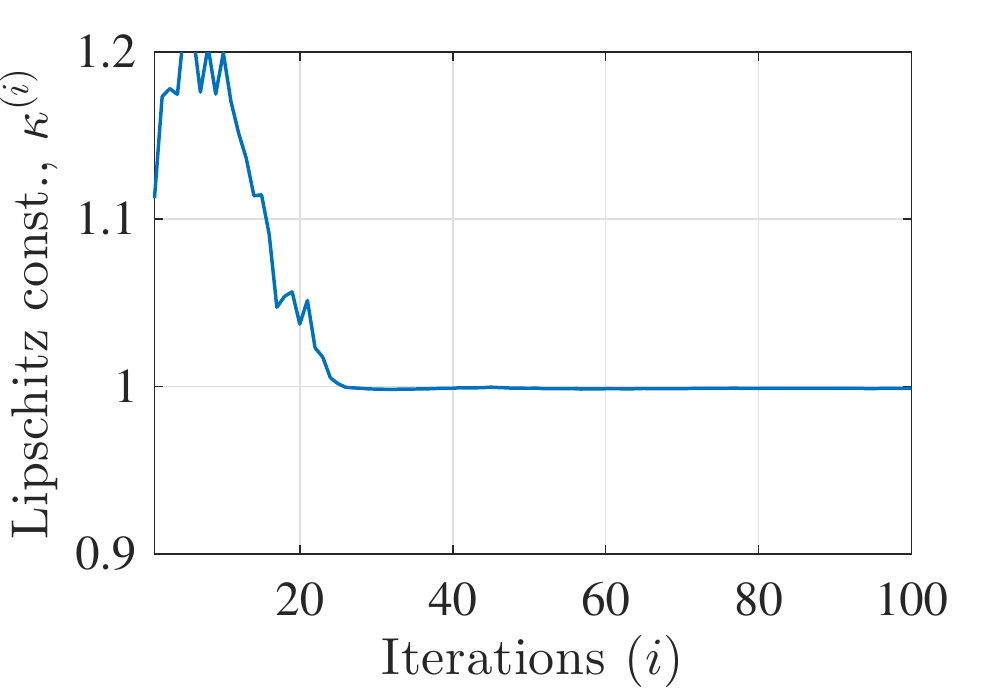} 
     \\
     \multicolumn{3}{c}{\small (b) LF photography using a focal stack: Condition numbers of data-fit majorizers have \textit{large} variations.}
     \\
     \small{(b1) $\{ \Delta^{(i)} : i \geq 2 \}$} &
     \small{(b2) $\{ \epsilon^{(i)} : i \geq 2 \}$} &
     \small{(b3) $\{ \kappa^{(i)} : i \geq 1 \}$} 
     \\
     \includegraphics[scale=0.55, trim=0.2em 0.2em 1.1em 1em, clip]{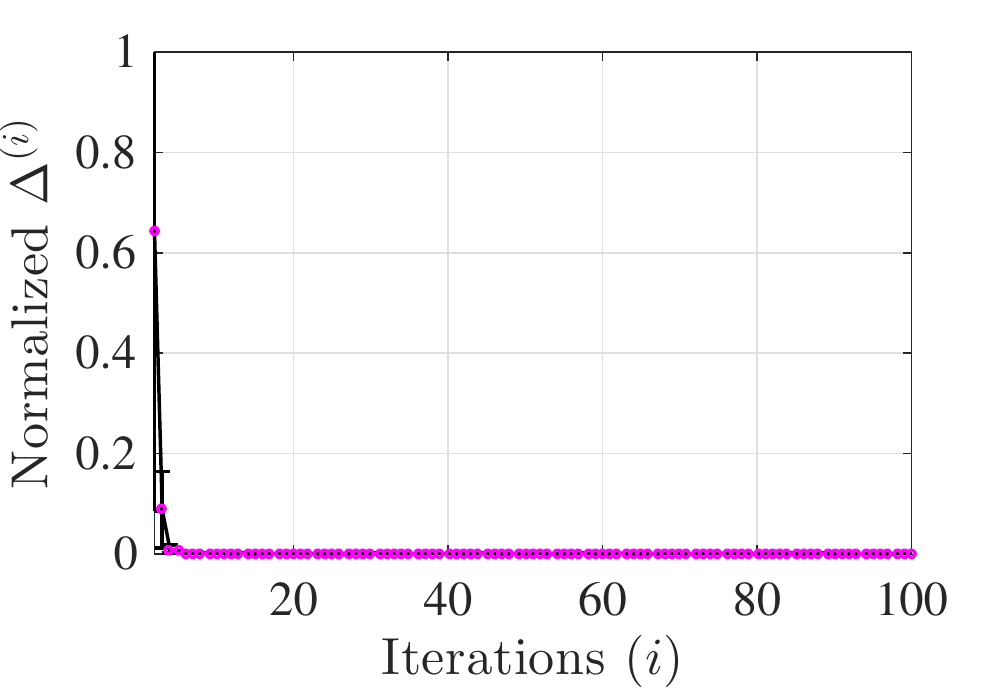} &
     \includegraphics[scale=0.55, trim=0.2em 0.2em 1.1em 1em, clip]{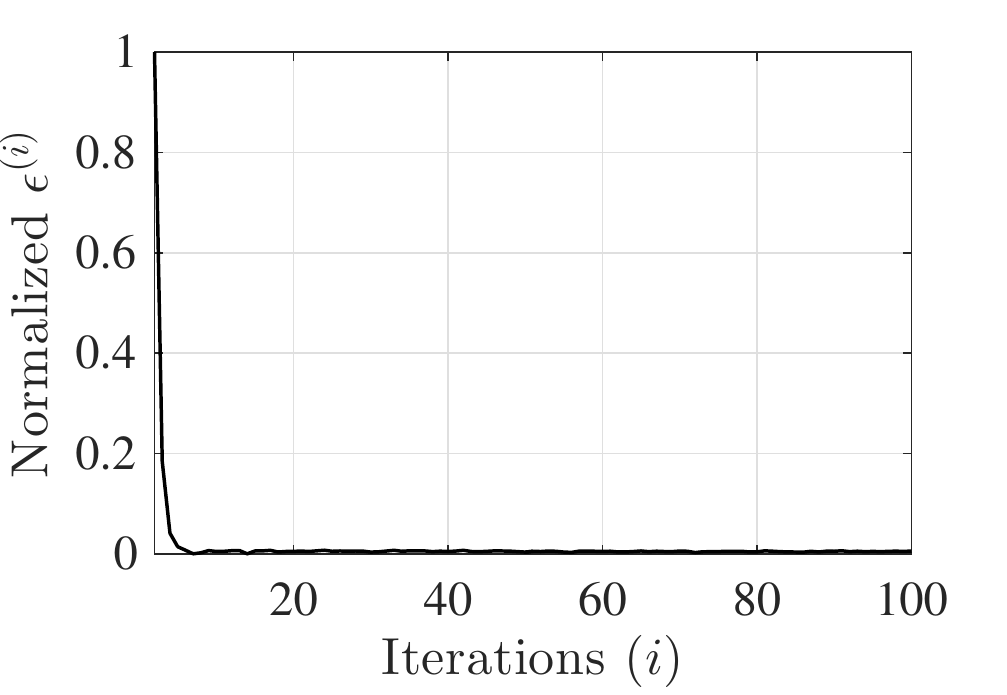} &
	\includegraphics[scale=0.55, trim=0.2em 0.2em 1.1em 1em, clip]{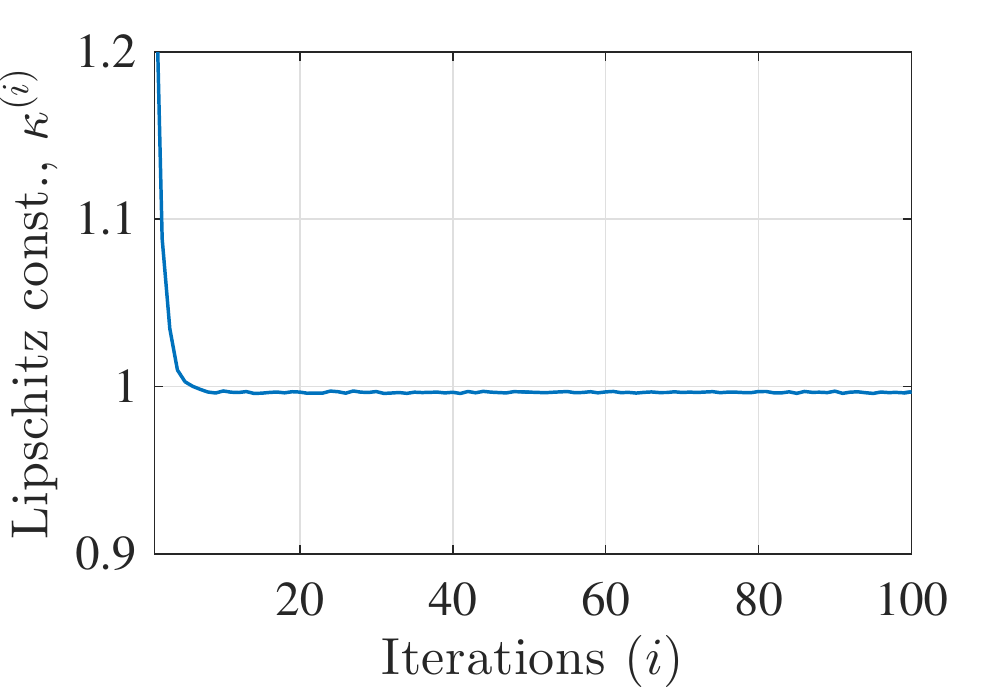}
     \end{tabular}
     
\vspace{-0.75em}
 \caption{
 Empirical measures related to Assumption~4 for guaranteeing convergence of Momentum-Net using \emph{dCNNs} refiners 
 (for details, see \R{sys:dcnn} and \S\ref{sec:exp:INN:param}), in different applications.
 See estimation procedures in \S\ref{sec:momnet:assume:scnn}.
 (a)~The sparse-view CT reconstruction experiment used fan-beam geometry with $12.5$\% projections views. 
 (b)~The LF photography experiment used five detectors and reconstructed LFs consisting of $9 \!\times\! 9$ sub-aperture images.
 (a1,~b1)~For both the applications, we observed that $\Delta^{(i)} \rightarrow 0$. 
 This implies that the $z^{(i+1)}$-updates in \R{eq:momnet:map} satisfy the asymptotic block-coordinate minimizer condition in Assumption~4.
 (Magenta dots denote the mean values and black vertical error bars denote standard deviations.)
 (a2)~Momentum-Net trained from training data-fits, where their majorization matrices have \emph{mild} condition number variations,
 shows that $\epsilon^{(i)} \rightarrow 0$.
 This implies that paired NNs $( \cR_{\theta^{(i+1)}},  \cR_{\theta^{(i)}} )$ in \R{eq:momnet:map} are asymptotically nonexpansive.
 (b2)~Momentum-Net trained from training training data-fits, where their majorization matrices have \emph{mild} condition number variations, 
 shows that $\epsilon^{(i)}$ becomes close to zero, but does not converge to zero in one hundred iterations.
 (a3,~b3)~The NNs, $\cR_{\theta^{(i+1)}}$ in \R{eq:momnet:map}, become nonexpansive, 
 i.e., its Lipschitz constant $\kappa^{(i)}$ becomes less than $1$, as $i$ increases.
 }
 \label{fig:momnet:assume:dcnn}
 \end{figure*}

Definition~\ref{d:pair-map} incorporates a pairing property because Momentum-Net uses iteration-wise image refining NNs. 
Specifically, the pairing property helps prove convergence of Momentum-Net,  
by connecting image refining NNs at adjacent iterations.
Furthermore, the asymptotic property in Definition~\ref{d:pair-map} allows Momentum-Net to use \textit{expansive} refining NNs (i.e., mapping operators having a Lipschitz constant larger than $1$) for some iterations, while guaranteeing convergence; see Figs.~\ref{fig:momnet:assume:dcnn}(a3) and \ref{fig:momnet:assume:dcnn}(b3).
Suppose that refining NNs are identical across iterations, i.e., $\cR_{\theta^{(i+1)}} \!=\! \cR_{\theta}$, $\forall i$, 
similar to some existing INNs, e.g., PnP \cite{Buzzard&etal:18SJIS}, RED \cite{Romano&Elad&Milanfar:17SJIS}, and other methods in \S\ref{sec:intro:convg}.
In such cases, if $\cR_{\theta}$ is expansive, Momentum-Net may diverge;
this property corresponds to the limitation of existing methods described in \S\ref{sec:intro:convg}.
Momentum-Net moderates this issue by using iteration-wise refining NNs 
that satisfy the asymptotic paired non-expansiveness in Definition~\ref{d:pair-map}.

Because the sequence $\{ z^{(i+1)} : \forall i \}$ in \R{eq:momnet:map} is not necessarily updated with a proximal mapping, 
we introduce a generalized definition of block-coordinate minimizers \cite[(2.3)]{Xu&Yin:13SIAM} for $z^{(i+1)}$-updates:

\defn{[Asymptotic block-coordinate minimizer]  \label{d:bcm}
The update $z^{(i+1)}$ is an asymptotic block-coordinate minimizer if there exists a summable nonnegative sequence $\{ \Delta^{(i+1)} \geq 0 : \sum_{i=0}^\infty \Delta^{(i+1)} < \infty \}$ such that
\be{
\label{cond:delta}
\left\| z^{(i+1)} - x^{(i)} \right\|_2^2 \leq \left\| z^{(i)} - x^{(i)} \right\|_2^2 + \Delta^{(i+1)}, \qquad \forall i.
}
}

Definition~\ref{d:bcm} implies that as $i \rightarrow \infty$, the updates $\{ z^{(i+1)} : i \geq 0 \}$ approach a block-coordinate minimizer trajectory that satisfies \R{cond:delta} with $\{  \Delta^{(i+1)} \!=\! 0 : i \!\geq\! 0 \}$.
In particular, $\Delta^{(i+1)}$ quantifies how much the update $z^{(i+1)}$ in \R{eq:momnet:map} perturbs a block-coordinate minimizer trajectory.
The bound $\| z^{(i+1)}  - x^{(i)} \|_2^2 \leq \| z^{(i)}  - x^{(i)} \|_2^2$ always holds, $\forall i$, when one uses the proximal mapping in \R{eq:bpgm:zk} within the BPEG-M framework.
Note that while applying trained Momentum-Net, \R{cond:delta} is easy to examine empirically, whereas \R{cond:eps} is harder to check.

\subsubsection{Assumptions} \label{sec:momnet:convg:assume}

This section introduces and interprets the assumptions for convergence analysis of Momentum-Net in Algorithm~\ref{alg:momnet}:

\bulls{
%[\setlength{\topsep}{1pt}]

\item {\em Assumption~1)} In MBIR problems \R{sys:mbir},
(non)convex $F(x;y,z^{(i+1)})$ is (continuously) differentiable, proper, and lower-bounded in $\dom(F)$,\footnote{
$F : \bbR^n \rightarrow (- \infty, + \infty]$ is proper if $\dom F \neq \emptyset$. 
$F$ is lower bounded in $\dom(F) \triangleq \{ u : F(u) < \infty \}$ if $\inf_{u \in \dom(F)} F(u) > -\infty$.
}
$\forall i$, and $\cX$ is convex and closed.
Algorithm~\ref{alg:momnet} has a fixed-point.

\item {\em Assumption~2)} $\nabla F(x;y,z^{(i+1)})$ is $M^{(i+1)}$-Lipschitz continuous with respect to $x$ (see Definition~\ref{d:QM}),
%specifically,
%\begingroup
%\setlength\abovedisplayskip{0.5\baselineskip}
%\setlength\belowdisplayskip{0.5\baselineskip}
%\eas{
%&~\nm{ \nabla F (u;y,z^{(i+1)}) - \nabla F (v;y,z^{(i+1)})  }_{\big( \! M^{(i+1)} \! \big)^{\!-1}} 
%\\[-0.2\baselineskip]
%&\leq \nm{u - v}_{M^{(i+1)}},
%}
%\endgroup
%$\forall u,v \in \bbR^{N}$, 
where $M^{(i+1)}$ is a iteration-wise majorization matrix that satisfies $m_{F,\min} I_{N} \!\preceq\! M^{(i+1)} \!\preceq\! m_{F,\max} I_{N}$ with $0 < m_{F,\min} \leq m_{F,\max} < \infty$, $\forall i$.

\item {\em Assumption~3)} The extrapolation matrices $E^{(i+1)} \succeq 0$ in \R{up:Ex:cvx}--\R{up:Ex:ncvx} satisfy the following conditions:
\begingroup
\setlength\abovedisplayskip{0.2\baselineskip}
\setlength\belowdisplayskip{0.2\baselineskip}
\allowdisplaybreaks
\ea{
\label{cond:Ex:cvx}
& \mbox{for \emph{convex}~$F$,} 
\nn \\
& \big( E^{(i+1)} \big)^{\!T} M^{(i+1)} E^{(i+1)}
\preceq \delta^2 \cdot M^{(i)}, \qquad \delta < 1;
\\
\label{cond:Eu:ncvx}
& \mbox{for \emph{nonconvex}~$F$,} 
\nn \\
& \big( E^{(i+1)} \big)^{\!T} M^{(i+1)} E^{(i+1)}
\preceq \frac{ \delta^2 (\lambda-1)^2}{4 (\lambda+1)^2} \cdot M^{(i)}, \qquad \delta < 1.
}
\endgroup

\item {\em Assumption~4)}  
The sequence of paired operators $( \cR_{\theta^{(i+1)}}, \cR_{\theta^{(i)}} )$ is asymptotically nonexpansive with a summable sequence $\{ \epsilon^{(i+i)} \!\geq\! 0 \}$;
the update $z^{(i+1)}$ is an asymptotic block-coordinate minimizer with a summable sequence $\{ \Delta^{(i+i)} \!\geq\! 0 \}$.
The mapping functions $\{ \cR_{\theta^{(i+1)}} : \forall i \}$ are continuous with respect to input points and the corresponding parameters $\{ \theta^{(i+1)} : \forall i \}$ are bounded.

}

Assumption~1 is a slight modification of Assumption~S.1 of BPEG-M, and Assumptions~2--3 are identical to Assumptions~S.2--S.3 of BPEG-M;
see Assumptions~S.1--S.3 in \S\ref{sec:bpgm:analysis}.
The extrapolation matrix designs \R{up:Ex:cvx} and \R{up:Ex:ncvx} satisfy conditions \R{cond:Ex:cvx} and \R{cond:Eu:ncvx} in Assumption~3, respectively.

We provide empirical justifications for the first two conditions in Assumption~4.
First, Figs.~\ref{fig:momnet:assume:dcnn}(a2) and \ref{fig:momnet:assume:scnn}(a2) illustrate that 
paired refining NNs $( \cR_{\theta^{(i+1)}}, \cR_{\theta^{(i)}} )$ of Momentum-Net 
appear to be asymptotically nonexpansive 
in an application that has mild condition number variations across training data-fit majorization matrices.
Figs.~\ref{fig:momnet:assume:dcnn}(a3), \ref{fig:momnet:assume:dcnn}(b3), \ref{fig:momnet:assume:scnn}(a3), and \ref{fig:momnet:assume:scnn}(b3)
illustrate for different applications that 
refining NNs $\{ \cR_{\theta^{(i+1)}} \}$ become nonexpansive:
their Lipschitz constants at the first several iterations are larger than $1$, 
and their Lipschitz constants in later iterations become less than $1$.
Alternatively, the asymptotic non-expansiveness of paired operators $( \cR_{\theta^{(i+1)}}, \cR_{\theta^{(i)}} )$ 
can be satisfied by a stronger assumption that the sequence $\{ \cR_{\theta^{(i+1)}} \}$ 
converges to some nonexpansive operator.
(Fig.~\ref{fig:momnet:convg} illustrates that dCNN parameters of Momentum-Net appear to converge.)

Figs.~\ref{fig:momnet:assume:dcnn}(a3), \ref{fig:momnet:assume:dcnn}(b3), \ref{fig:momnet:assume:scnn}(a3), and \ref{fig:momnet:assume:scnn}(b3)
illustrate for different applications that the $z^{(i+1)}$-updates are asymptotic block-coordinate minimizers.
Lemma~\ref{l:delta:prob} and \S\ref{sec:l:delta:prob} in the appendices provide 
a \emph{probabilistic} justification for the asymptotic block-coordinate minimizer condition.

%Finally, we remark that Example~\ref{eg:caol} in \S\ref{sec:relation} satisfies the two asympototic conditions in Assumption~4.
%In particular, the convolutional autoencoder in \R{eq:bpgm:zk} that satisfies the TF condition in \S\ref{sec:mbir:learn_reg} is nonexpansive \cite{Chun&etal:18Allerton}; 
%update \R{eq:bpgm:zk} leads to a block coordinate minimizer $z^{(i+1)}$ because the convolutional autoencoder in \R{eq:bpgm:zk} is a proximal mapping operator.

\subsubsection{Main convergence results} \label{sec:momnet:convg:result}

This section analyzes fixed-point and critical point convergence of Momentum-Net in Algorithm~\ref{alg:momnet}, under the assumptions in the previous section. 
We first show that differences between two consecutive iterates generated by Momentum-Net converge to zero:

\prop{[Convergence properties]
\label{p:momnet:sum}
Under Assumptions 1--4, let $\{ x^{(i+1)}, z^{(i+1)} : i \geq 0 \}$ be the sequence generated by Algorithm~\ref{alg:momnet}. 
Then, the sequence satisfies 
\be{
\label{p:momnet:convg:sqSum}
\sum_{i=0}^{\infty} \nm{ \left[ \arraycolsep=1.5pt \begin{array}{c} x^{(i+1)} \\  z^{(i+1)} \end{array} \right] -  \left[ \arraycolsep=1.5pt \begin{array}{c} x^{(i)} \\  z^{(i)} \end{array} \right] }_2^2 < \infty,
}
and hence $\nm{ \left[ \arraycolsep=1.5pt \begin{array}{c} x^{(i+1)} \\  z^{(i+1)} \end{array} \right] -  \left[ \arraycolsep=1.5pt \begin{array}{c} x^{(i)} \\  z^{(i)} \end{array} \right] }_2 \rightarrow 0$.
}
\prf{\renewcommand{\qedsymbol}{}
See \S\ref{sec:p:momnet:sum} in the appendices.
}

Using Proposition~\ref{p:momnet:sum}, our main theorem provides that any limit points of the sequence generated by Momentum-Net satisfy critical point and fixed-point conditions:

\thm{[A limit point satisfies both critical point and fixed-point conditions]
\label{t:momnet:convg}
Under Assumptions 1--4 above, let $\{ x^{(i+1)}, z^{(i+1)} : i \geq 0 \}$ be the sequence generated by Algorithm~\ref{alg:momnet}.
Consider either a fixed majorization matrix with general structure, i.e., $M^{(i+1)} \!=\! M$ for $i \!\geq\! 0$, 
or a sequence of diagonal majorization matrices, i.e., $\{ M^{(i+1)} : i \!\geq\! 0 \}$.
Then, any limit point $\bar{x}$ of $\{ x^{(i+1)} \}$ satisfies both the critical point condition:
\be{
\label{t:momnet:convg:crit}
\ip{\nabla F (\bar{x};y,\bar{z}) }{ x - \bar{x} } \geq 0, \qquad \forall x \in \cX,
}
where $\bar{z}$ is a limit point of $\{ z^{(i+1)} \}$, and the fixed-point condition:
\be{
\label{t:momnet:convg:fix}
\left[ \begin{array}{c} \bar{x} \\ \bar{x} \end{array} \right]  = \cA_{\cR_{\bar{\theta}}}^{\bar{M}} \! \left( \left[ \begin{array}{c} \bar{x} \\ \bar{x} \end{array} \right] \right),
}
where $\left[ \begin{smallmatrix} x^{(i+1)} \\ x^{(i)} \end{smallmatrix} \right] \!=\!  \cA_{\cR_{\theta^{(i+1)}}}^{M^{(i+1)}} \!\! \left( \left[ \begin{smallmatrix} x^{(i)} \\ x^{(i-1)} \end{smallmatrix} \right] \right)$,
$\cA_{\cR_{\theta^{(i+1)}}}^{M^{(i+1)}} \! (\cdot)$ denotes performing the $i\rth$ updates in Algorithm~\ref{alg:momnet}, 
and $\bar{\theta}$ and $\bar{M}$ is a limit point of $\{ \theta^{(i+1)} \}$ and $\{ M^{(i+1)} \}$, respectively.
}
\prf{\renewcommand{\qedsymbol}{}
See \S\ref{sec:t:momnet:convg} in the appendices.
}

Observe that, if $\cX = \bbR^N$ or $\bar{x}$ is an interior point of $\cX$, 
\R{t:momnet:convg:crit} reduces to the first-order optimality condition $0 \in \partial F(\bar{x};y,\bar{z})$,
where $\partial F(x)$ denotes the limiting subdifferential of $F$ at $x$.
With additional isolation and boundedness assumptions for the points satisfying \R{t:momnet:convg:crit} and \R{t:momnet:convg:fix},
we obtain whole sequence guarantees:

\cor{[Whole sequence convergence]
\label{c:momnet:convg:whole}
Consider the construction in Theorem~\ref{t:momnet:convg}.
Let $\cS$ be the set of points satisfying the critical point condition in \R{t:momnet:convg:crit} and the fixed-point condition in \R{t:momnet:convg:fix}.
If $\{ x^{(i+1)} : i \geq 0 \}$ is bounded, then $\mathrm{dist} (x^{(i+1)}, \cS) \rightarrow 0$, where $\mathrm{dist} (u, \cV) \triangleq \inf \{ \| u - v \| : v \in \cV \}$ denotes the distance from $u$ to $\cV$, for any point $u \in \bbR^N$ and any subset $\cV \subset \bbR^N$.
If $\cS$ contains uniformly isolated points, i.e., there exists $\eta > 0$ such that $\| u - v \| \geq \eta$ for any distinct points $u,v \in \cS$, then $\{ x^{(i+1)} \}$ converges to a point in $\cS$.
}
\prf{\renewcommand{\qedsymbol}{}
See \S\ref{sec:c:momnet:convg:whole} in the appendices.
}

The boundedness assumption for $\{ x^{(i+1)} \}$ in Corollary~\ref{c:momnet:convg:whole} is standard in block-wise optimization, e.g., \cite{Bolte&Sabach&Teboulle:14MA, Xu&Yin:13SIAM, Xu&Yin:17JSC, Chun&Fessler:18TIP, Chun&Fessler:20TIP}. The assumption can be satisfied if 
the set $\cX$ is bounded (e.g., box constraints),
one chooses appropriate regularization parameters in Algorithm~\ref{alg:momnet} \cite{Xu&Yin:17JSC, Chun&Fessler:18TIP, Chun&Fessler:20TIP}, 
the function $F(x;y,z)$ is coercive \cite{Bolte&Sabach&Teboulle:14MA}, 
or the level set is bounded \cite{Xu&Yin:13SIAM}.
However, for general $F(x;y,z)$, it is hard to verify the isolation condition for the points in $\cS$ in practice.
Instead, one may use Kurdyka-{\L}ojasiewicz property \cite{Bolte&Sabach&Teboulle:14MA, Xu&Yin:13SIAM} to analyze the whole sequence convergence with some appropriate modifications.

For simplicity, we focused our discussion to noniterative MBIR module \R{eq:momnet:mbir}.
However, Momentum-Net practically converges with \emph{any} proximable MRIR function \R{eq:momnet:mbir} that may need an iterative solver,
if sufficient inner iterations are used. 
To maximize the computational benefit of Momentum-Net, 
one needs to make sure that majorized MBIR function \R{eq:momnet:mbir} is better proximable
over its original form \R{sys:mbir}.

\subsection{Benefits of Momentum-Net} \label{sec:benefit}

Momentum-Net has several benefits over existing INNs:
\bulls{[\setlength{\topsep}{1pt}]

\item {\em Benefits from refining module}:
The image refining module~\R{eq:momnet:map} can use iteration-wise image refining NNs $\{ \cR_{\theta^{(i+1)}} \!:\! i \!\geq\! 0 \}$:
those are particularly useful to reduce overfitting risks by reducing dimensions of their parameters $\theta^{(i+1)}$ at each iteration \cite{Lim&etal:20TMI, Chun&etal:19MICCAI, Ye&Long&Chun:ICIP20}.
Iteration-wise refining NNs 
require less memory for training, compared to methods that use a single refining NN for all iterations, e.g., \cite{Gilton&Ongie&Rebecca:19TCI}.
Different from the existing methods mentioned in \S\ref{sec:intro:convg},
Momentum-Net does not require (firmly) nonexpansive mapping operators $\{ \cR_{\theta^{(i+1)}} \}$
to guarantee convergence.
Instead, $\{ \cR_{\theta^{(i+1)}} \}$ in \R{eq:momnet:map} assumes a generalized notion of
the (firm) non-expansiveness condition assumed for convergence of the existing methods 
that use identical refining NNs across iterations, including 
PnP \cite{Yang&etal:16NIPS, Sreehari&etal:16TCI, Zhang&etal:17CVPR, Chan&Wang&Elgendy:17TCI, Buzzard&etal:18SJIS, Ryu&etal:19ICML}, 
RED \cite{Romano&Elad&Milanfar:17SJIS, Reehorst&Schniter:19TCI}, etc.
The generalized concept is the first practical condition to 
guarantee convergence of INNs using iteration-wise refining NNs; 
see Definition~\ref{d:pair-map}.

\item {\em Benefits from extrapolation module}:
The extrapolation module~\R{eq:momnet:exp} uses the momentum terms $x^{(i)} - x^{(i-1)}$
that accelerate the convergence of Momentum-Net.
In particular, compared to the existing gradient-descent-inspired INNs,
e.g., TNRD \cite{Chen&Pock:17PAMI, Hammernik&etal:17MRM},
Momentum-Net converges faster.
(Note that the way the authors of \cite{Reehorst&Schniter:19TCI} used momentum is less conventional.
The corresponding method, RED-APG \cite[Alg.~6]{Reehorst&Schniter:19TCI},
still can require multiple inner iterations to solve its quadratic MBIR problem, similar to BCD-Net-type methods.)

\item  {\em Benefits from MBIR module}: 
The MBIR module~\R{eq:momnet:mbir} does not require multiple inner iterations for a wide range of imaging problems and has both theoretical and practical benefits.
Note first that convergence analysis of INNs (including Momentum-Net) assumes that their MBIR operators are \emph{noniterative}.
In other words, related convergence theory (e.g., Proposition~\ref{p:bcdnet:cauchy}) is inapplicable 
if iterative methods, particularly with insufficient number of iterations, are applied to MBIR modules.
Different from the existing BCD-Net-type methods \cite{Chun&Fessler:18IVMSP, Yang&etal:16NIPS, Sreehari&etal:16TCI, Zhang&etal:17CVPR, Chan&Wang&Elgendy:17TCI, Romano&Elad&Milanfar:17SJIS, Aggarwal&Mani&Jacobs:18TMI, Reehorst&Schniter:19TCI, Buzzard&etal:18SJIS} that can require iterative solvers for their MBIR modules,
MBIR module~\R{eq:momnet:mbir} of Momentum-Net can have practical close-form solution (see examples in \S\ref{sec:momnet:setup}), 
and its corresponding convergence analysis (see \S\ref{sec:momnet:convg}) can hold stably for a wide range of imaging applications.
Second, combined with extrapolation module~\R{eq:momnet:exp}, 
noniterative MBIR modules~\R{eq:momnet:mbir} lead to faster MBIR, compared to the existing BCD-Net-type methods that can require multiple inner iterations for their MBIR modules for convergence.
Third, Momentum-Net guarantees convergence even for nonconvex MBIR cost function $F(x;y,z)$ or nonconvex data-fit $f(x;y)$ of which the gradient is $M$-Lipschitz continuous (see Definition~\ref{d:QM}), while existing INNs overlooked nonconvex $F(x;y,z)$ or $f(x;y)$.
}

Furthermore, \S\ref{sec:momnet-vs-bcdnet} analyzes the sequence convergence of BCD-Net \cite{Chun&Fessler:18IVMSP}, 
and describes the convergence benefits of Momentum-Net over BCD-Net.

\section{Training INNs} \label{sec:train}

This section describes training of all the INNs compared in this paper.

\subsection{Architecture of refining NNs and their training} \label{sec:train:NN}

For all INNs in this paper, we train the refining NN at each iteration
to remove artifacts from the input image $x^{(i)}$ that is fed from the previous iteration. 
For the $i\rth$ iteration NN, we first consider the following sCNN architecture, residual single-hidden layer convolutional autoencoder:
\be{
\label{sys:auto:res}
\cR_{\theta^{(i+1)}} (u) = \sum_{k=1}^K d_k^{(i+1)} \conv \cT_{\exp(\alpha_k^{(i+1)})} \! \big( e_k^{(i+1)} \conv u \big) + u, 
}
where $\theta^{(i+1)} \!=\! \{ d_k^{(i+1)}, \alpha_k^{(i+1)}, e_k^{(i+1)} \!:\! \forall k  \}$ is the parameter set of the $i\rth$ image refining NN, 
$\{ d_k^{(i+1)}, e_k^{(i+1)} \in \bbC^{R} \!:\! k \!=\! 1,\ldots,K \}$ is a set of $K$ decoding and encoding filters of size $R$, 
$\{ \exp( \alpha_k^{(i+1)} ) \!:\! k \!=\! 1,\ldots,K \}$ is a set of $K$ thresholding values, 
and $\cT_{\alpha} (u)$ is the soft-thresholding operator with parameter $\alpha$ defined in \R{eq:soft-threshold},
for $i \!=\! 0,\ldots,N_{\text{iter}}\!-\!1$.
We use the exponential function $\exp(\cdot)$ to prevent the thresholding parameters $\{ \alpha_k \}$ from becoming negative during training.
We observed that the residual convolutional autoencoder in \R{sys:auto:res} gives better results compared to
the convolutional autoencoder, i.e., \R{sys:auto:res} without the second term \cite{Chun&Fessler:18IVMSP, Lim&etal:20TMI}. 
This corresponds to the empirical result in \cite{He&etal:15CVPR, Li&etal:18NIPS} that 
having skip connections (e.g., the second term in \R{sys:auto:res})
can improve generalization.
The sequence of paired sCNN refiners \R{sys:auto:res} can satisfy the asymptotic non-expansiveness, 
if its convergent refiner satisfies that 
\bes{
\sigma_{\max} (\widebar{D}^H \widebar{D}) \leq 1/R, \qquad \sigma_{\max} (\widebar{E}^H \widebar{E}) \leq 1/R,
}
where 
$\sigma_{\max}(\cdot)$ is the largest eigenvalue of a matrix,
$\widebar{D}  \triangleq [\bar{d}_1, \ldots, \bar{d}_K, \delta_R]$, 
$\widebar{E}  \triangleq [\bar{e}_1, \ldots, \bar{e}_K, \delta_R]$, 
$\{ \bar{d}_k, \bar{e}_k : \forall k \}$ are limit point filters, and $\delta_R$ is the Kronecker delta filter of size $R$.

\begingroup
\setlength{\thinmuskip}{1.5mu}
\setlength{\medmuskip}{2mu plus 1mu minus 2mu}
\setlength{\thickmuskip}{2.5mu plus 2.5mu}
%\allowdisplaybreaks
For dCNN refiners, we use the following residual multi-hidden layer CNN architecture, 
a simplified DnCNN \cite{Zhang&etal:17TIP} using fewer layers, no pooling, and no batch normalization \cite{Ryu&etal:19ICML}
(we drop superscript indices $(\cdot)^{(i)}$ for simplicity):
%\fontsize{9.5pt}{11.4pt}\selectfont
\ea{
\label{sys:dcnn}
\cR_{\theta} (u) &= u - \sum_{k=1}^K e_k^{[L]} \conv u_k^{[L-1]},
\\
u_k^{[1]} &= \mathrm{ReLU} \Big( e_k^{[1]} \conv u \Big),
\quad
u_k^{[l]} = \mathrm{ReLU} \Bigg( \sum_{k'=1}^K e_{k,k'}^{[l]} \conv z_{k'}^{[l-1]} \Bigg), 
\nn
}
for $k \!=\! 1,\ldots, K$ and $l \!=\! 2,\ldots, L\!-\!1$,
where $\theta \!=\! \{ e_{k}^{[l]}, e_{k,k'}^{[l]} \!:\! \forall k,k',l \}$ is the parameter set of each refining NN,
$K$ is the number of feature maps,
$L$ is the number of layers,
$\{ e_{k}^{[l]} \in \bbR^R \!:\! k \!=\! 1,\ldots,K, l \!=\! 1,L \}$ is a set of filters at the first and last dCNN layer,
$\{ e_{k,k'}^{[l]} \in \bbR^R \!:\! k,k' \!=\! 1,\ldots,K, l \!=\! 2,\ldots,L\!-\!1 \}$ is a set of filters for remaining dCNN layer,
%$\{ u_k^{[l]} \in \bbR^N \!:\! k \!=\! 1,\ldots,K, l \!=\! 1,\ldots,L\!-\!1 \}$ is the $k\rth$ feature map at the $l\rth$ dCNN layer, 
and the rectified linear unit activation function is defined by
$\mathrm{ReLU} (u) \triangleq \max (0, u)$.
\endgroup

\begingroup
\setlength{\thinmuskip}{1.5mu}
\setlength{\medmuskip}{2mu plus 1mu minus 2mu}
\setlength{\thickmuskip}{2.5mu plus 2.5mu}
The training process of Momentum-Net requires $S$ high-quality training images, $\{ x_{s} \!:\! s \!=\! 1,\ldots,S \}$, 
$S$ training measurements simulated via imaging physics, $\{ y_{s} \!:\! s \!=\! 1,\ldots, S \}$, 
and $S$ data-fits $\{ f_{s} (x;y_{s}) \!:\! s \!=\! 1,\ldots,S \}$ 
and the corresponding majorization matrices $\{ M_{s}^{(i)} \!, \widetilde{M}_{s}^{(i)} \!:\! s \!=\! 1,\ldots,S, i \!=\! 1,\ldots,N_{\text{iter}} \}$.
Different from \cite{Chun&Fessler:18IVMSP, Chun&etal:19MICCAI, Li&Chun&Long:20ISBI} that train convolutional autoencoders from the patch perspective, 
we train the image refining NNs in \R{sys:auto:res}--\R{sys:dcnn} from the convolution perspective (that does not store many overlapping patches, e.g., see \cite{Chun&Fessler:20TIP}). 
From $S$ training pairs $( x_{s}, x_{s}^{(i)} )$, 
where $\{ x_{s}^{(i)} \!:\! s \!=\! 1,\ldots,S \}$ is a set of $S$ reconstructed images at the $(i-1)\rth$ Momentum-Net iteration,
we train the $i\rth$ iteration image refining NN in \R{sys:auto:res} by solving the following optimization problem: 
\endgroup
\be{
\label{sys:auto:res:train}
\theta^{(i+1)} 
= \argmin_{ \theta } \frac{1}{2S} \sum_{s=1}^S \nm{ x_{s} - \cR_{\theta} ( x_{s}^{(i)} ) }_2^2,
\tag{P2}
}
where $\theta^{(i+1)}$ is given as in \R{sys:auto:res}, for $i \!=\! 0,\ldots,N_{\text{iter}} \!-\! 1$
(see some related properties in \S\ref{sec:l:train}).
We solve the training optimization problems \R{sys:auto:res:train} by mini-batch stochastic optimization 
with the subdifferentials computed by the PyTorch \texttt{Autograd} package.

\subsection{Regularization parameter selection based on \dquotes{spectral spread}} \label{sec:reg:sel}

When majorization matrices of training data-fits $\{ f_{s} (x;y_{s}) \!:\! s \!=\! 1,\ldots,S \}$ have similar spectral properties, e.g., condition numbers,
the regularization parameter $\gamma$ in \R{sys:mbir} is trainable by substituting \R{eq:momnet:map}
into \R{eq:momnet:mbir} and modifying the training cost \R{sys:auto:res:train}.
However, the condition numbers of data-fit majorizers can greatly differ due a variety of
imaging geometries or image formation systems, 
or noise levels in training measurements, etc.
See such examples in \S\ref{sec:exp:imaging}--\ref{sec:exp:INN}.

To train Momentum-Net with diverse training data-fits,
we propose a parameter selection scheme based on the \dquotes{spectral spread} of
their majorization matrices $\{ M_{f_{s}}^{(i)} \}$.
For simplicity, consider majorization matrices of the form
$\widetilde{M}_{s}^{(i)} \!=\! \widetilde{M}_{s}  \!=\! \lambda ( M_{f_{s}}  + \gamma_{s} I )$ $\forall i$,
where the factor $\lambda$ is selected by \R{up:MFtilde} and 
$M_{f_{s}}$ is a symmetric positive semidefinite majorization matrix for $f_{s} (x;y_{s})$, $\forall s$.
We select the regularization parameter $\gamma_{s}$ for the $s\rth$ training sample as
\be{
\label{eq:reg:select}
\gamma_{s} = \frac{\sigma_{\text{spread}} (M_{f_{s}})}{\chi},
}
where the spectral spread of a symmetric positive definite matrix is defined by $\sigma_{\text{spread}}(\cdot) \triangleq \sigma_{\max}(\cdot) - \sigma_{\min} (\cdot)$
for $\sigma_{\max} (M_{f_{s}}) \!>\! \sigma_{\min} (M_{f_{s}}) \!\geq\! 0$,
and $\sigma_{\min} (\cdot)$ is the smallest eigenvalue of a matrix.
For the $s\rth$ training sample, 
a tunable factor $\chi$ controls $\gamma_{s}$ in \R{eq:reg:select}
according to $\sigma_{\text{spread}} (M_{f_{s}})$, $\forall s$.
The proposed parameter selection scheme also applies to testing Momentum-Net,
based on the tuned factor $\chi^\star$ in its training.
We observed that the proposed parameter selection scheme \R{eq:reg:select}
gives better MBIR accuracy than the condition number based selection scheme 
that is similarly used in selecting ADMM parameters \cite{Zheng&etal:19TCI} (for the two applications in \S\ref{sec:exp}).
One may further apply this scheme to iteration-wise majorization matrices $\widetilde{M}_{s}^{(i)}$
and select iteration-wise regularization parameters $\gamma_{s}^{(i)}$ accordingly.
For comparing different INNs, we apply \R{eq:reg:select} to all INNs.

\section{Experimental results and discussion} \label{sec:exp}

We investigated two extreme imaging applications: 
sparse-view CT and LF photography using a focal stack.
In particular, these two applications lack a practical closed-form solution 
for the MBIR modules of BCD-Net and ADMM-Net \cite{Yang&etal:16NIPS}, e.g., solving \R{eq:bcdnet:recon}.
For these applications, we compared the performances of the following five INNs: 
BCD-Net \cite{Chun&Fessler:18IVMSP} 
(i.e., generalization of RED \cite{Romano&Elad&Milanfar:17SJIS} and MoDL \cite{Aggarwal&Mani&Jacobs:18TMI}), 
ADMM-Net \cite{Yang&etal:16NIPS}, i.e., PnP-ADMM \cite{Chan&Wang&Elgendy:17TCI, Buzzard&etal:18SJIS} using iteration-wise refining NNs,
Momentum-Net \textit{without extrapolation} 
(i.e., generalization of TNRD \cite{Chen&Pock:17PAMI, Hammernik&etal:17MRM}), 
PDS-Net, i.e., PnP-PDS \cite{Ono:17SPL} using iteration-wise refining NNs,
and the proposed Momentum-Net using extrapolation.

\subsection{Experimental setup: Imaging} \label{sec:exp:imaging}

\subsubsection{Sparse-view CT} 
\label{sec:appl:ct}

To reconstruct a linear attenuation coefficient image $x \in \bbR^N$ 
from post-log sinogram $y \in \bbR^m$ in sparse-view CT, 
the MBIR problem \R{sys:mbir} considers a data-fit $f(x;y) = \frac{1}{2} \| y -  A x \|_W^2$
and the non-negativity constraint $\cX = [0, \infty)^N$,
where $A \in \bbR^{m \times N}$ is an undersampled CT system matrix,
$W \in \bbR^{m \times m}$ is a diagonal weighting matrix with elements 
$\{ W_{m',m'} = p_m'^2 / ( p_m' + \sigma^2 ) : \forall m'  \}$ based on 
a Poisson-Gaussian model \cite{Chun&Talavage:13Fully3D, Zheng&etal:19TCI} 
for the pre-log raw measurements $p \in \bbR^m$ 
with electronic readout noise variance $\sigma^2$.

We simulated 2D sparse-view sinograms of size $m \!=\! 888 \!\times\! 123$ 
-- \quotes{detectors or rays}~$\times$~\quotes{regularly spaced projection views or angles}, 
where $984$ is the number of full views -- with GE LightSpeed fan-beam geometry corresponding to a monoenergetic source with 
$10^5$ incident photons per ray and no background events, and electronic noise variance $\sigma^2 \!=\! 5^2$. 
We avoided an inverse crime in imaging simulation and 
reconstructed images of size $N \!=\! 420 \!\times\! 420$ with a coarser grid $\Delta_x \!=\! \Delta_y \!=\! 0.9766$~mm; 
see details in \cite[\S\Romnum{5}-A2]{Chun&Fessler:18Asilomar}.

\subsubsection{LF photography using a focal stack}
\label{sec:appl:lf}

To reconstruct a LF $x \!=\! [x_1^T, \ldots, x_C' ]^T \!\in\! \bbR^{SN'}$ 
that consists of $C'$ sub-aperture images
from focal stack measurements $y \!=\! [y_1^T, \ldots, y_C^T]^T \!\in\! \bbR^{CN'}$ 
that are collected by $C$ photosensors, 
the MBIR problem \R{sys:mbir} considers a data-fit $f(x;y) \!=\! \frac{1}{2} \| y -  A x \|_2^2$
and a box constraint $\cX \!=\! [0, U]^{C'N'}$ with $U \!=\! 1$ (or $255$ without rescaling),
where $A \!\in\! \bbR^{CN' \times C'N'}$ is a system matrix of LF imaging system using a focal stack
that is constructed blockwise with 
$C \cdot C'$ different convolution matrices $\{ \tau_c A'_{c,c'} \!\in\! \bbR^{N' \times N'} \!:\! c \!=\! 1,\ldots,C, c' \!=\! 1,\ldots,C' \}$~\cite{Lien&etal:20NP, Block&Chun&Fessler:18IVMSP},
$\tau_c \!\in\! (0,1]$ is a transparency coefficient for the $c\rth$ detector,
and $N'$ is the size of sub-aperture images, $x_{c'}$, $\forall c'$.\footnote{
Traditionally, one obtains focal stacks by physically moving imaging sensors and taking separate exposures across time.
Transparent photodetector arrays \cite{Lien&etal:20NP, Zhang&etal:19CLEO} allow one to collect focal stack data in a single exposure, 
making a practical LF camera using a focal stack.
If some photodetectors are not perfectly transparent, one can use $\tau_c < 1$, for some $c$.
}
In general, a LF photography system using a focal stack is extremely under-determined, 
because $C \ll C'$.

To avoid an inverse crime, our imaging simulation used higher-resolution synthetic LF dataset \cite{Honauer&etal:16ACCV}
(we converted the original RGB sub-aperture images to grayscale ones 
by the \dquotes{\texttt{rgb2gray.m}} function in MATLAB,
for simplicity and smaller memory requirements in training).
We simulated $C \!=\! 5$ focal stack images of size $N' \!=\! 255 \!\times\! 255$ 
with $40$~dB AWGN that models electronic noise at sensors, and setting transparency coefficients $\tau_c$ as $1$, for $c \!=\! 1,\ldots,C$.
The sensor positions were chosen such that five sensors focus at equally spaced depths;
specifically, the closest sensor to scenes and farthest sensor from scenes focus 
at two different depths that correspond to \quotes{$\texttt{disp}_{\min} \!+ 0.2$} and \quotes{$\texttt{disp}_{\max} \!- 0.2$}, respectively,
where $\texttt{disp}_{\max}$ and $\texttt{disp}_{\min}$ are the approximate maximum and minimum disparity values specified in \cite{Honauer&etal:16ACCV}.
We reconstructed 4D LFs that consist of $S \!=\! 9 \times 9$ sub-aperture images of size $N' \!=\! 255 \!\times\! 255$, 
with a coarser grid $\Delta_x \!=\! \Delta_y \!=\! 0.13572$~mm.

\begin{algorithm}[pt!]
\caption{BCD-Net \cite{Chun&Fessler:18IVMSP}}
\label{alg:bcdnet}

\begin{algorithmic}
\REQUIRE $\{ \cR_{\theta^{(i)}}: i= 1,\ldots,N_{\text{iter}} \}$, $\gamma >  0$, $x^{(0)} = x^{(-1)}$, $y$

\FOR{$i = 0,\ldots,N_{\text{iter}}\!-\!1$}

\STATE {\em Image refining}: 
\begingroup
\setlength\abovedisplayskip{0.2\baselineskip}
\setlength\belowdisplayskip{0.2\baselineskip}
\be{
\label{eq:bcdnet:map}
z^{(i+1)} =  \cR_{\theta^{(i+1)}} \big( x^{(i)} \big) 
\tag{Alg.2.1}
}
\endgroup

\STATE {\em MBIR}:
\begingroup
\setlength\abovedisplayskip{0.2\baselineskip}
\setlength\belowdisplayskip{0.2\baselineskip}
\be{
\label{eq:bcdnet:recon}
x^{(i+1)} = \argmin_{ x \in \cX } F(x;y,z^{(i+1)})
\tag{Alg.2.2}
}
\endgroup

\ENDFOR

\end{algorithmic}
\end{algorithm}

\subsection{Experimental setup: INNs} \label{sec:exp:INN}

\subsubsection{Parameters of INNs} \label{sec:exp:INN:param}

The parameters for the INNs compared in sparse-view CT experiments were defined as follows.
We considered two BCD-Nets (see Algorithm~\ref{alg:bcdnet}):
for one BCD-Net, we applied the APG method \cite{Beck&Teboulle:09SIAM} with $10$ inner iterations to \R{eq:bcdnet:recon},
and set $N_{\text{iter}} \!=\! 30$; 
for the other BCD-Net, we applied the APG method with $3$ inner iterations to \R{eq:bcdnet:recon},
and set $N_{\text{iter}} \!=\! 45$.
For ADMM-Net,
we used the identical configurations as BCD-Net and set the ADMM penalty parameter to $\gamma$ in \R{sys:mbir}, similar to \cite{Chun&etal:19MICCAI}.
For Momentum-Net without extrapolation,
we chose $N_{\text{iter}} \!=\! 100$ and $\rho \!=\! 1 - \varepsilon$.
For the proposed Momentum-Net,
we chose $N_{\text{iter}} \!=\! 100$ and $\rho \!=\! 0.5$.
For PDS-Net, 
we set the first step size to $\gamma_1 \!=\! \gamma^{-1}$ and the second step size to $\gamma_2 \!=\! \gamma_1^{-1} \sigma_{\max}^{-1} (M)$, 
per \cite{Ono:17SPL}.
For performance comparisons between different INNs,
all the INNs used sCNN refiners \R{sys:auto:res} with $\{ R , K \!=\! 7^2 \}$ to avoid the overfitting/hallucination risks.
For Momentum-Net using dCNN refiners, 
we chose $L\!=\!4$ layer dCNN \R{sys:dcnn} using $R \!=\! 3^2$ filters and $K \!=\! 64$ feature maps, following \cite{Ryu&etal:19ICML}.
(The chosen parameters gave lower RMSE values than $\{ L\!=\! 6, R \!=\! 3^3 , K \!=\! 64 \}$, 
for identical regularization parameters.)
For comparing different MBIR methods,
Momentum used 
extrapolation, i.e., \R{eq:momnet:exp} with \R{up:Ex:cvx} and \R{eq:mom_coeff}, 
and $\{ R \!\!=\!\! 7^2, K \!\!=\!\! 9^2 \}$ for \R{sys:auto:res}.
We designed the majorization matrices as $\{ \widetilde{M}^{(i+1)} \!=\! \diag( A^T W A 1 ) + \gamma I : i \!\geq\! 0 \}$,
using Lemma~\ref{l:diag(|At|W|A|1)} ($A$ and $W$ have nonnegative entries)
and setting $\lambda \!=\! 1$ by \R{up:MFtilde}.
We set an initial point of INNs, $x^{(0)}$, to filtered-back projection (FBP) using a Hanning window.
The regularization parameters of all INNs
were selected by the scheme in \S\ref{sec:reg:sel}
with $\chi^\star \!=\! 167.64$.
(This factor was estimated from the carefully chosen regularization parameter 
for sparse-view CT MBIR experiments using learned convolutional regularizers in \cite{Chun&Fessler:20TIP}.)

The parameters for the INNs compared in experiments of LF photography using a focal stack were defined as follows.
We considered two BCD-Nets and two ADMM-Nets with the identical parameters listed above.
For Momentum-Net without extrapolation and the proposed Momentum-Net,
we set $N_{\text{iter}} \!=\! 100$ and $\rho \!=\! 1 - \varepsilon$.
For PDS-Net, we used the identical parameter setup described above.
For performance comparisons between different INNs,
all the INNs used sCNN refiners \R{sys:auto:res} with $\{ R \!=\! 5^2, K \!=\! 32 \}$
(to avoid the overfitting risks)
in the epipolar domain.
For Momentum-Net using dCNN refiners, we chose $L\!=\!6$ layer dCNN \R{sys:dcnn} using $R \!=\! 3^2$ filters and $K \!=\! 16$ feature maps.
(The chosen parameters gave most accurate performances over the following setups, $\{ L\!=\!4, R \!=\! 3^2, K \!=\! 16, 32, 64 \}$, 
given the identical regularization parameters.)
To generate $\cR_{\theta^{(i+1)}} ( x^{(i)} )$ in \R{eq:momnet:map}, 
we applied a sCNN \R{sys:auto:res} with $\{ R \!=\! 5^2, K \!=\! 32 \}$ or 
a dCNN \R{sys:dcnn} with $\{ L\!=\!6, R \!=\! 3^2, K \!=\! 16 \}$
to two sets of horizontal and vertical epipolar plane images, 
and took the average of two LFs that were permuted back from 
refined horizontal and vertical epipolar plane image sets, $\forall i$.\footnote{
Epipolar images are 2D slices of a 4D LF $L_F ( c_x, c_y, c_u, c_v )$,
where $( c_x, c_y )$ and $( c_u, c_v )$ are spatial and angular coordinates, respectively.
Specifically, each horizontal epipolar plane image are obtained by
fixing $c_y$ and $c_v$,
and varying $c_x$ and $c_u$;
and each vertical epipolar image are obtained by
fixing $c_x$ and $c_u$,
and varying $c_y$ and $c_v$.
}
%For comparing different MBIR methods,
%%(i.e., Fig.~\ref{fig:recon:lf} and Table~\ref{tab:recon:lf}),
%Momentum-Net used extrapolation and $\{ R \!=\! 5^2, K \!=\! 32 \}$ for the epipolar-domain refiners \R{sys:auto:res} at each layer.
We designed the majorization matrices as $\{ \widetilde{M}^{(i+1)} \!=\! \diag( A^T A 1 ) + \gamma I : i \!\geq\! 0 \}$, 
using Lemma~\ref{l:diag(|At|W|A|1)} and setting $\lambda \!=\! 1$ by \R{up:MFtilde}.
We set an initial point of INNs, $x^{(0)}$, to $A^T y$ rescaled in the interval $[0, 1]$ (i.e., dividing by its max value).
The regularization parameters 
(i.e., $\gamma$ in BCD-Net/Momentum-Net, the ADMM penalty parameter in ADMM-Net, and the first step size in PDS-Net)
were selected by the proposed scheme in \S\ref{sec:reg:sel}
with $\chi^\star \!=\! 1.5$.
(We tuned the factor to achieve the best performances).

For different combinations of INNs and sCNN refiner \R{sys:auto:res}/dCNN refiner \R{sys:dcnn}, we use the following naming convention: \quotes{the INN name}-\quotes{sCNN} or \quotes{dCNN}.

\subsubsection{Training INNs} \label{sec:exp:INN:train}

For sparse-view CT experiments,
we trained all the INNs from the chest CT dataset with 
$\{ x_{s}, y_{s}, f_{s}(x;y_{s}) \!=\! \frac{1}{2} \| y_{s} - A x \|_{W_{s}}^2, \widetilde{M}_{s} \!:\! s \!=\! 1,\ldots,S, S \!=\! 142 \}$; we constructed the dataset by using XCAT phantom slices \cite{Segars&etal:08MP}. 
The CT experiment has mild data-fit variations across training samples: the standard deviation of the condition numbers ($\triangleq\! \sigma_{\max}(\cdot) / \sigma_{\min} (\cdot)$) of $\{ M_{f_{s}} \!=\! \diag( A^T W_s A 1 ) \!:\! \forall s \}$ is $1.1$.
For experiments of LF photography using a focal stack,
we trained all the INNs from the LF photography dataset with
$\{ y_{s}, f_{s}(x;y_{s}) \!=\! \frac{1}{2} \| y_{s} - A_{s} x \|_2^2, \widetilde{M}_{s} \!:\! s \!=\! 1,\ldots,S, S \!=\! 21 \}$ and 
two sets of ground truth epipolar images, 
$\{ x_{s,\text{epi-h}} ,  x_{s,\text{epi-v}}: s \!=\! 1,\ldots,S, S \!=\! 21 \!\cdot\! (255 \!\cdot\! 9) \}$;
we constructed the dataset by excluding four unrealistic \dquotes{stratified} scenes from the original LF dataset in \cite{Honauer&etal:16ACCV} 
that consists of $28$ LFs with diverse scene parameter and camera settings.
The LF experiment has large data-fit variations across training samples: the standard deviation of the condition numbers of $\{ M_{f_{s}} \!=\! \diag( A_s^T A_s 1 ) \!:\! \forall s \}$ is $2245.5$.

In training INNs for both the applications,
if not specified, we used identical training setups.
At each iteration of INNs, we solved \R{sys:auto:res:train} with the mini-batch version of Adam \cite{Kingma&Ba:15ICLR}
and trained iteration-wise sCNNs \R{sys:auto:res} or dCNNs \R{sys:dcnn}.
We selected the batch size and the number of epochs as follows:
for sparse-view CT reconstruction, we chose them as $20$ \& $300$, and $20$ \& $200$ for sCNN and dCNN refiners, respectively;
for LF photography using a focal stack, we chose them as $200$ \& $200$, and $200$ \& $100$, for sCNN and dCNN refiners, respectively.
We chose the learning rates for filters in sCNNs and dCNNs, 
and thresholding values $\{ \alpha_k^{(i+1)} \!:\! \forall k, i \}$ in sCNNs \R{sys:auto:res},
as $10^{-3}$ and $10^{-1}$, respectively;
we reduced the learning rates by $10$\% every $10$ epochs.
At the first iteration, we initialized filter coefficients with Kaiming uniform initialization \cite{He&etal:15ICCV};
in the later iteration, i.e., at the $i\rth$ INN iteration, for $i \geq 2$, we initialized filter coefficients 
from those learned from the previous iteration, i.e., $(i-1)\rth$ iteration 
(this also applies to initializing thresholding values).

\subsubsection{Testing trained INNs} \label{sec:exp:INN:test}

In sparse-view CT reconstruction experiments, we tested trained INNs to two samples where 
ground truth images and the corresponding inverse covariance matrices (i.e., $W$ in \S\ref{sec:appl:ct})
sufficiently differ from those in training samples
(i.e., they are a few cm away from training images).
We evaluated the reconstruction quality by the most conventional error metric in CT application,
RMSE (in HU), in a region of interest (ROI), 
where RMSE and HU stand for root-mean-square error and (modified) Hounsfield unit,
respectively,
and the ROI was a circular region that includes all the phantom tissues.
The $\textmd{RMSE}$ is defined by $\mathrm{RMSE} (x^\star \!, x^\text{true}) \!\triangleq\! ( \sum_{j=1}^{N_{\text{ROI}}}(x_j^{\star}-x^\text{true}_j)^2/{N_{\mathrm{ROI}}} )^{1/2}$, 
where $x^\star$ is a reconstructed image, 
$x^\text{true}$ is a ground truth image, 
and $N_{\text{ROI}}$ is the number of pixels in a ROI.
In addition, we compared the trained Momentum-Net (using extrapolation) 
to a standard MBIR method using a hand-crafted EP regularizer,
and an MBIR model using a learned convolutional regularizer \cite{Chun&Fessler:20TIP, Chun&Fessler:18Asilomar} 
which is the state-of-the-art MBIR method within an unsupervised learning setup.
We finely tuned their regularization parameters to achieve the lowest RMSE.
See details of these two MBIR models in \S\ref{sec:param:mbir:ct}.

In experiments of LF photography using a focal stack, we tested trained INNs to three samples 
of which scene parameter and camera settings are different from those in training samples 
(all training and testing samples have different camera and scene parameters).
We evaluated the reconstruction quality by the most conventional error metric in LF photography application,
PSNR (in dB), where PSNR stands for peak signal-to-noise.
In addition, we compared the trained Momentum-Net (using extrapolation) 
to MBIR method using the state-of-the-art non-trained regularizer, 4D EP introduced in \cite{Lien&etal:20NP}.
(The low-rank plus sparse tensor decomposition model \cite{Block&Chun&Fessler:18IVMSP, Kamal&etal:16CVIU} failed 
when inverse crimes and measurement noise are considered.)
We finely tuned its regularization parameter to achieve the lowest RMSE values.
See details of this MBIR model in \S\ref{sec:param:mbir:lf}.
We further investigated impacts of the LF MBIR quality on a higher-level depth estimation application,
by applying the robust Spinning Parallelogram Operator (SPO) depth estimation method \cite{Zhang&etal:16CVIU} to reconstructed LFs.

For comparing Momentum-Net with PDS-Net, we measured quality of refined images, $z^{(i+1)}$ in \R{eq:momnet:map},
because PDS-Net is a hard-refiner.

The imaging simulation and reconstruction experiments were based on 
the Michigan image reconstruction toolbox~\cite{fessler:16:irt},
and training INNs, i.e., solving \R{sys:auto:res:train}, was based on PyTorch 
(for sparse-view CT, we used ver.~1.2.0; for LF photography using a focal stack, we used ver.~0.3.1).
For sparse-view CT, 
single-precision MATLAB  and PyTorch implementations were tested 
on $2.6$ GHz Intel Core i7 CPU with $16$ GB RAM, 
and 1405 MHz Nvidia Titan Xp GPU with 12 GB RAM, respectively.
For LF photography using a focal stack, 
they were tested on $3.5$ GHz AMD Threadripper 1920X CPU with $32$ GB RAM,
and $1531$ MHz Nvidia GTX 1080 Ti GPU with $11$ GB RAM, respectively.

\begin{figure}[!pt]
\centering
\small\addtolength{\tabcolsep}{-6.5pt}
\begin{tabular}{ll}
{\small \hspace{0.6cm} (a) Momentum-Net vs.} & {\small (b) BCD-Net vs.} 
\\
{\small \hspace{1.06cm} PDS-Net \cite{Ono:17SPL}} & {\small \hspace{0.40cm} ADMM-Net \cite{Yang&etal:16NIPS, Chan&Wang&Elgendy:17TCI, Buzzard&etal:18SJIS}} 
\\
\includegraphics[scale=0.498, trim=0.2em 0.2em 2.8em 0.9em, clip]{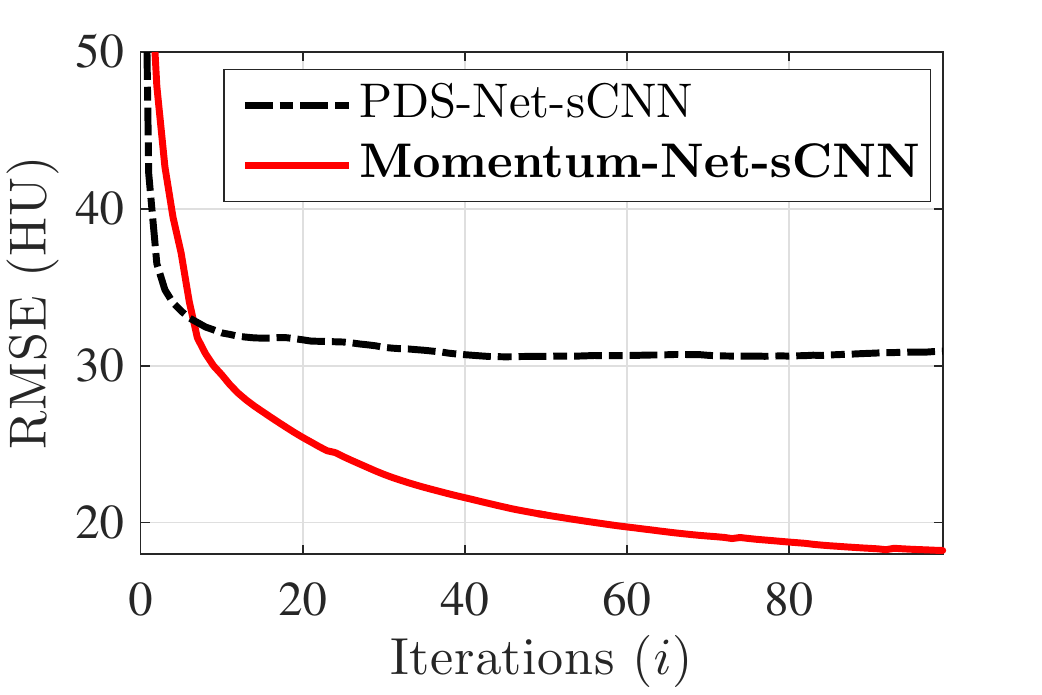} 
&
\includegraphics[scale=0.498, trim=4em 0.2em 1.5em 0.9em, clip]{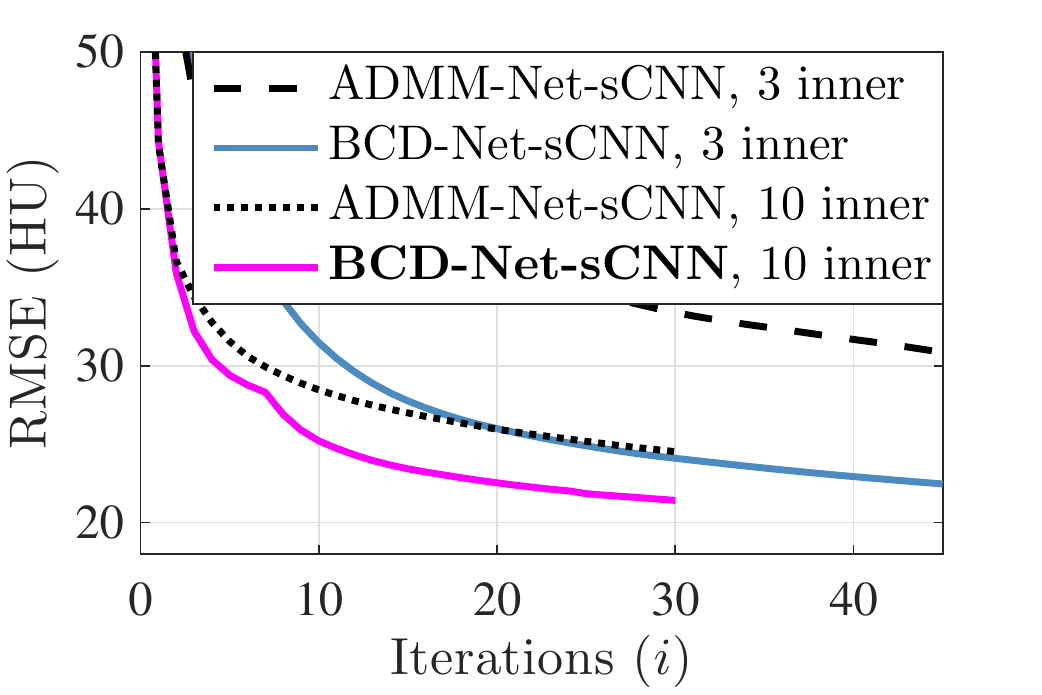} 
\end{tabular}

\vspace{-0.75em}
\caption{RMSE minimization comparisons
between different INNs for sparse-view CT  
(fan-beam geometry with $12.5$\% projections views
and $10^5$ incident photons; 
(a) averaged RMSE values across two test refined images;
(b) averaged RMSE values across two test reconstructed images).}
\label{fig:itercomp:ct}
\vspace{-0.5em}
\end{figure}

\begin{figure}[!pt]
\centering
\begin{tabular}{c}
\includegraphics[scale=0.55, trim=0.2em 0.2em 0.9em 1.3em, clip]{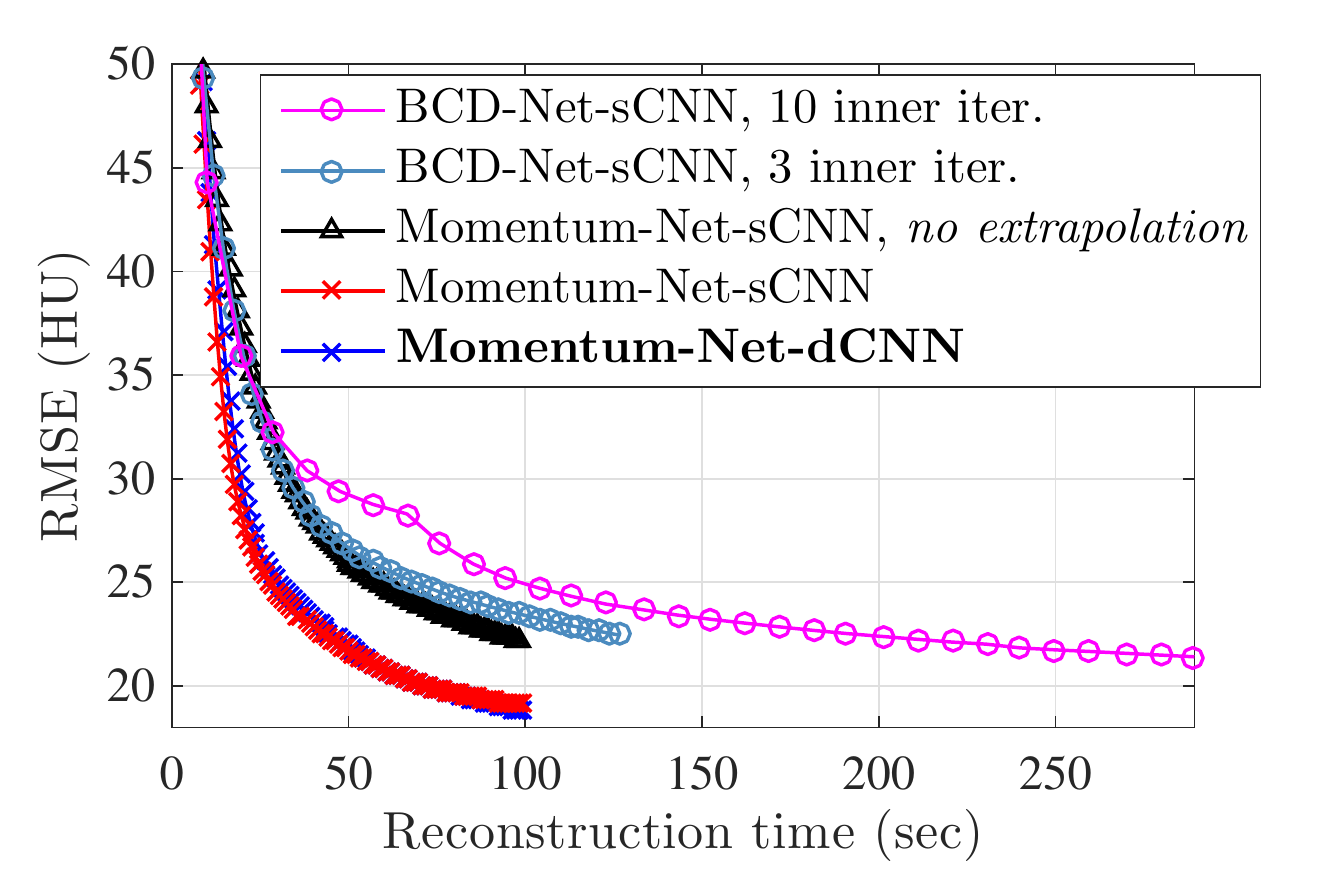} 
\end{tabular}

\vspace{-0.75em}
\caption{RMSE minimization comparisons
between different INNs for sparse-view CT
(fan-beam geometry with $12.5$\% projections views
and $10^5$ incident photons; 
averaged RMSE values across two test reconstructed images).}
\label{fig:timecomp:ct}
\vspace{-0.25em}
\end{figure}

\subsection{Comparisons between different INNs} \label{sec:comp:INN}

\begin{figure}[!pt]
\centering
\small\addtolength{\tabcolsep}{-6.5pt}
\begin{tabular}{ll}
{\small \hspace{0.6cm} (a) Momentum-Net vs.} & {\small (b) BCD-Net vs.} 
\\
{\small \hspace{1.06cm} PDS-Net \cite{Ono:17SPL}} & {\small \hspace{0.40cm} ADMM-Net \cite{Yang&etal:16NIPS, Chan&Wang&Elgendy:17TCI, Buzzard&etal:18SJIS}} 
\\
\includegraphics[scale=0.498, trim=0.2em 0.2em 2.8em 0.9em, clip]{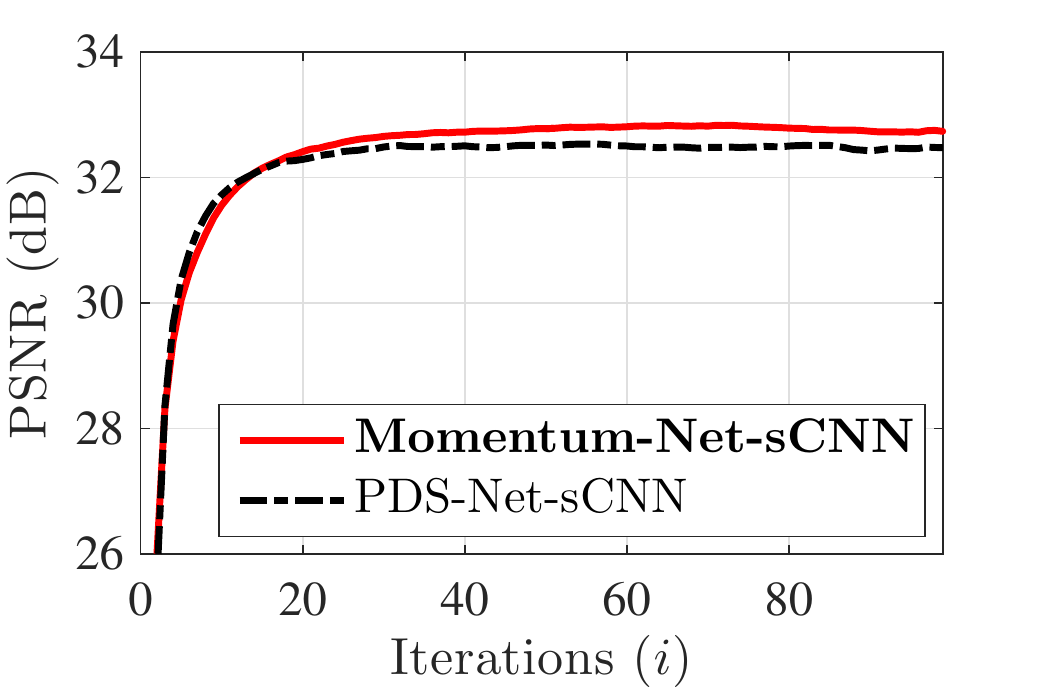} 
&
\includegraphics[scale=0.498, trim=4em 0.2em 1.5em 0.9em, clip]{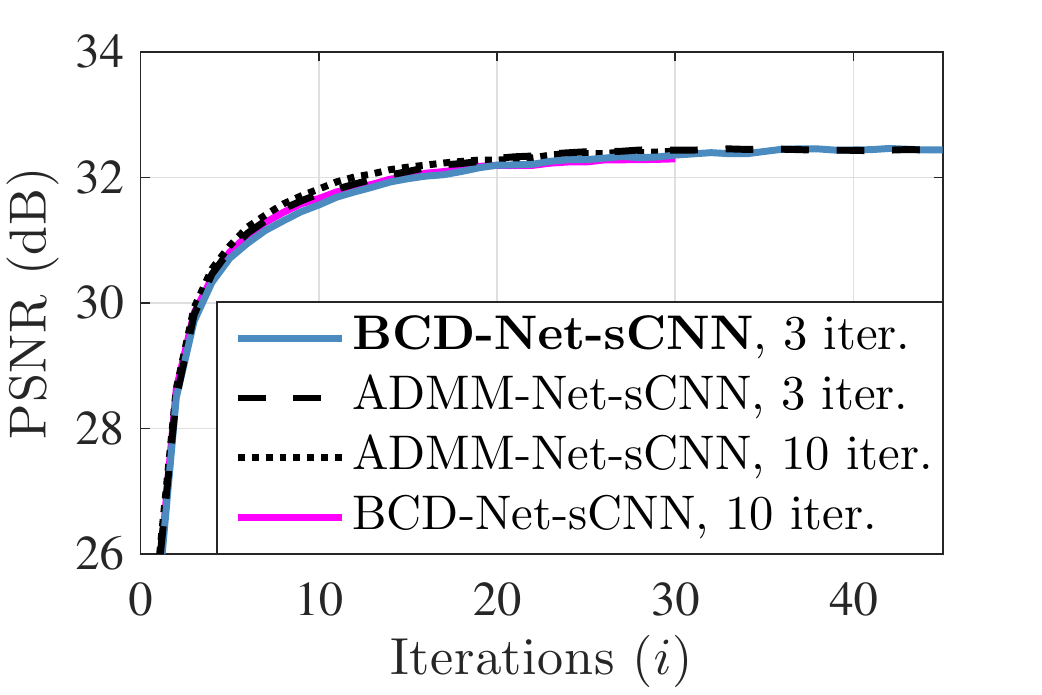} 
\end{tabular}

\vspace{-0.75em}
\caption{PSNR maximization comparisons
between different INNs in LF photography using a focal stack 
(LF photography systems with $C \!=\! 5$ detectors obtain a focal stack of
LFs consisting of $S \!=\! 81$ sub-aperture images; 
(a) averaged RMSE values across two test refined images);
(b) averaged RMSE values across two test reconstructed images.}
\label{fig:itercomp:lf}
\vspace{-0.5em}
\end{figure}

\begin{figure}[!pt]
\centering
\begin{tabular}{c}
\includegraphics[scale=0.55, trim=0.2em 0.2em 0.9em 1.3em, clip]{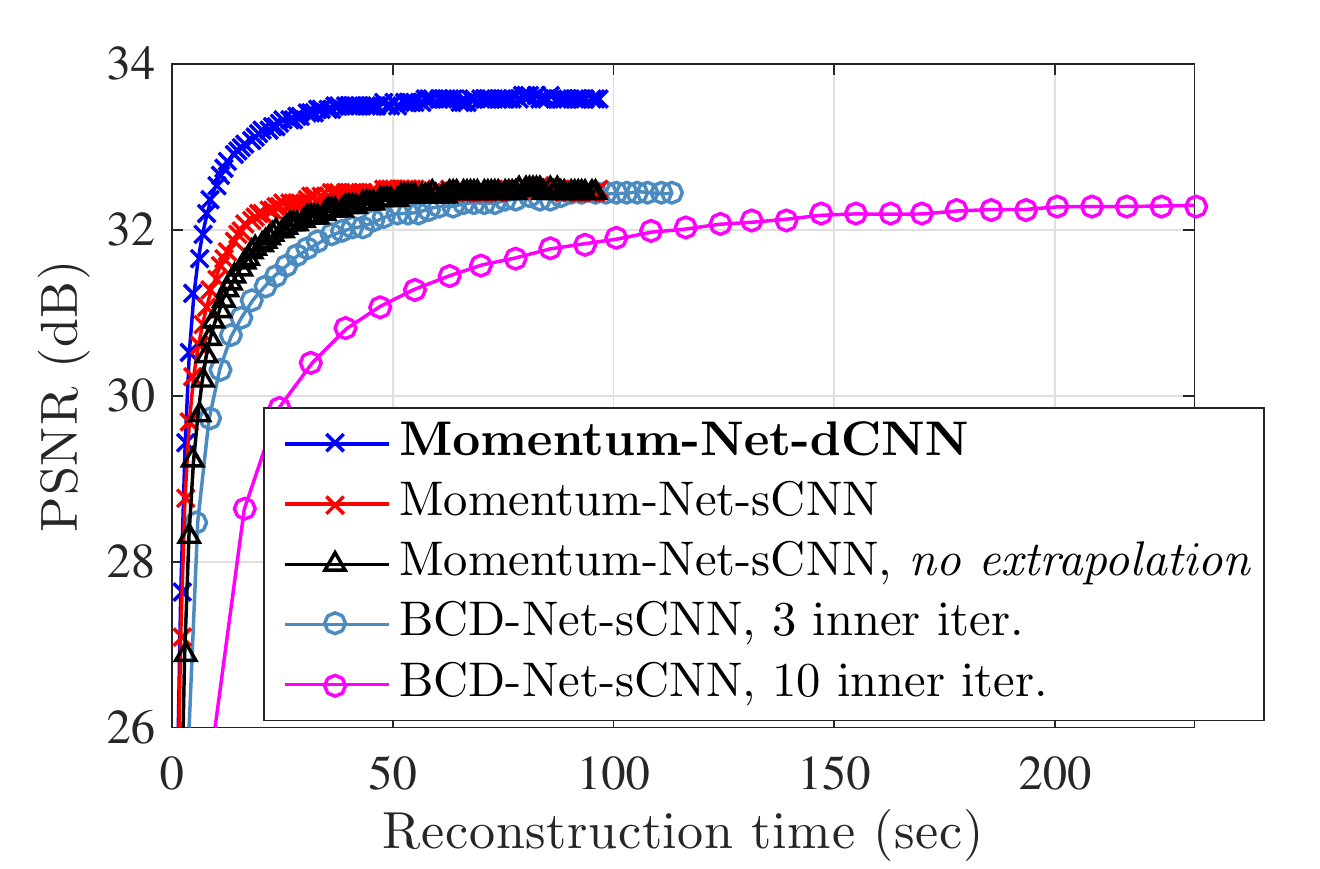} 
\end{tabular}

\vspace{-0.75em}
\caption{PSNR maximization comparisons
between different INNs in LF photography using a focal stack 
(LF photography systems with $C \!=\! 5$ detectors obtain a focal stack of
LFs consisting of $S \!=\! 81$ sub-aperture images; 
averaged PSNR values across three test reconstructed images).
}
\label{fig:timecomp:lf}
\vspace{-0.25em}
\end{figure}

First, compare sCNN results in Figs.~\ref{fig:itercomp:ct}--\ref{fig:timecomp:ct}
and Figs.~\ref{fig:itercomp:lf}--\ref{fig:timecomp:lf}, for 
sparse-view CT and LF photography using a focal stack, respectively. For both applications, the
proposed Momentum-Net using extrapolation significantly improves MBIR speed and accuracy,
compared to the existing soft-refining INNs,
\cite{Aggarwal&Mani&Jacobs:18TMI, Romano&Elad&Milanfar:17SJIS, Buzzard&etal:18SJIS, Chen&Pock:17PAMI, Hammernik&etal:17MRM} 
that correspond to BCD-Net \cite{Chun&Fessler:18IVMSP} or Momentum-Net using \textit{no extrapolation},
and ADMM-Net \cite{Yang&etal:16NIPS, Chan&Wang&Elgendy:17TCI, Buzzard&etal:18SJIS},
and the existing hard-refining INN PDS-Net \cite{Ono:17SPL}.
(Note that 
BCD-Net and Momentum-Net require slightly less computational complexity per INN iteration,
compared to ADMM-Net and PDS-Net, respectively, due to having fewer modules.)
Fig.~\ref{fig:timecomp:ct} shows that to reach the $24$ HU RMSE value in sparse-view CT reconstruction, 
the proposed Momentum-Net decreases MBIR time by $53.3$\% and $62.5$\%,
compared to Momentum-Net without extrapolation and BCD-Net using three inner iterations, respectively.
Fig.~\ref{fig:timecomp:lf} shows that to reach the $32$ dB PSNR value in LF reconstruction from a focal stack, 
the proposed Momentum-Net decreases MBIR time by $36.5$\% and $61.5$\%,
compared to Momentum-Net without extrapolation and BCD-Net using three inner iterations, respectively.
In addition, Figs.~\ref{fig:timecomp:ct} and \ref{fig:timecomp:lf} show that using extrapolation, 
i.e., \R{eq:momnet:exp} with \R{up:Ex:cvx}--\R{eq:mom_coeff},
improves the performance of Momentum-Net versus iterations.

We conjecture that the larger performance gap between soft-refiner Momentum-Net and hard-refiner PDS-Net, 
in Fig.~\ref{fig:itercomp:ct}(a) compared to Fig.~\ref{fig:itercomp:lf}(a),
is because the LF problem needs stronger regularization, 
i.e., a smaller tuned factor $\chi^\star$ in \R{eq:reg:select}, than the CT problem.
Similarly, comparing Fig.~\ref{fig:itercomp:ct}(b) to Fig.~\ref{fig:itercomp:lf}(b) shows that 
the LF problem has small performance gaps between BCD-Net and ADMM-Net.

For both the applications,
using dCNN refiners \R{sys:dcnn} instead of  sCNN refiners \R{sys:auto:res} has a negligible effect on total run time of Momentum-Net,
because reconstruction time of MBIR modules \R{eq:momnet:mbir} (in CPUs) dominates inference time of 
image refining modules \R{eq:momnet:map} (in GPUs).
Compare results between Momentum-Net-sCNN and -dCNN in 
Figs.~\ref{fig:timecomp:ct} \& \ref{fig:timecomp:lf} and Tables~\ref{tab:recon:ct} \& \ref{tab:recon:lf}.

 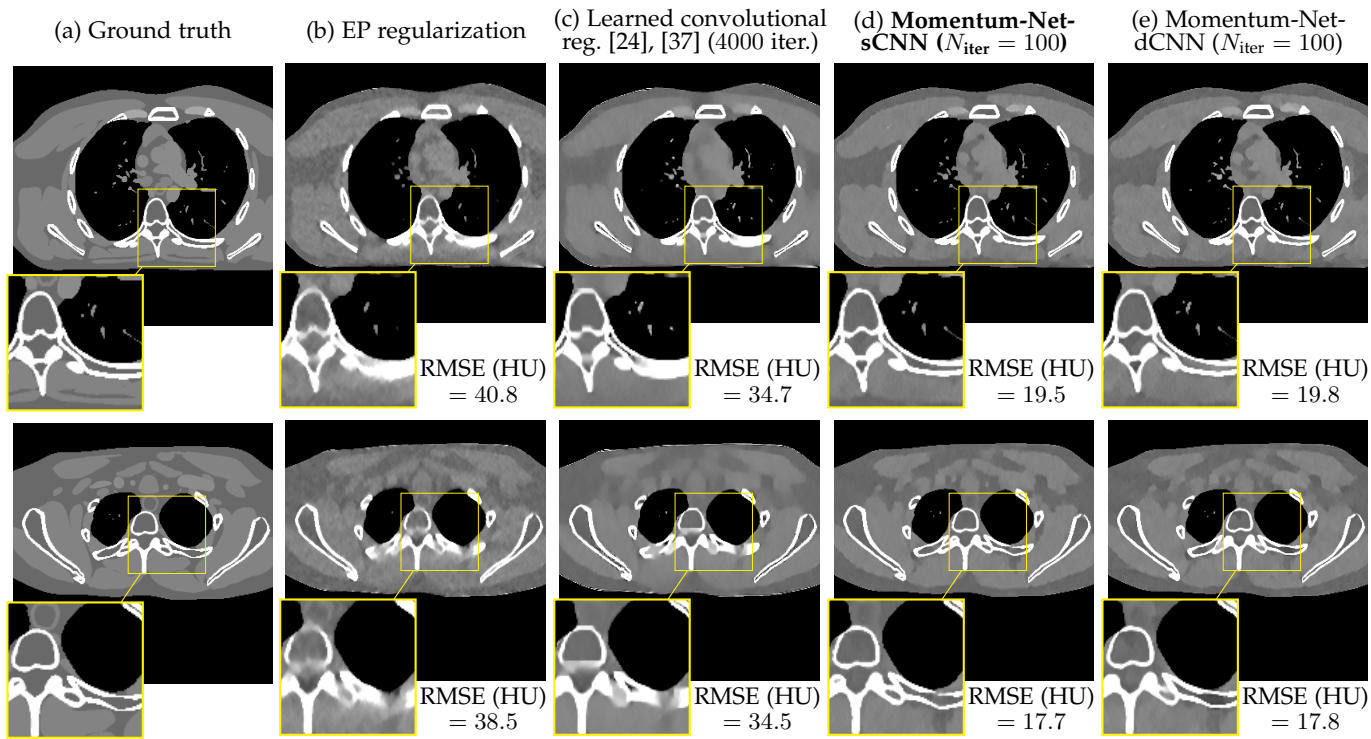
\begin{figure*}[!pt]
% \vspace{-0.75em}
 \centering
 \small\addtolength{\tabcolsep}{-6.5pt}
 \renewcommand{\arraystretch}{0.2}

     \begin{tabular}{ccccc}
     
           {(a) Ground truth} & 
           \specialcell[c]{(b) EP regularization } &
           \specialcell[c]{(c) Learned convolutional \\ reg. \cite{Chun&Fessler:20TIP, Chun&Fessler:18Asilomar} ($4000$ iter.)} & 
           \specialcell[c]{(d) {\bfseries Momentum-Net-} \\ {\bfseries sCNN ($N_{\text{iter}} = 100$)}} &
           \specialcell[c]{(e) Momentum-Net- \\ dCNN ($N_{\text{iter}} = 100$)}
           \\
        
         \begin{tikzpicture}
             \begin{scope}[spy using outlines={rectangle,yellow,magnification=1.75,size=18mm, connect spies}]
                 \node {\includegraphics[viewport={10mm 15mm 125mm 120mm},clip,width=34.5mm,height=34.5mm]{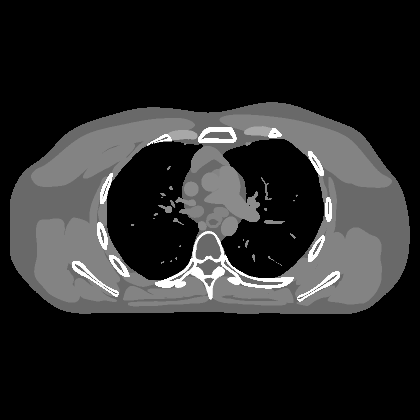}};
 	        \spy on (0.45,-0.42) in node [left] at (0,-1.95);	
             \end{scope}
         \end{tikzpicture} &
        
         \begin{tikzpicture}
             \begin{scope}[spy using outlines={rectangle,yellow,magnification=1.75,size=18mm, connect spies}]
                 \node {\includegraphics[viewport={10mm 15mm 125mm 120mm},clip,width=34.5mm,height=34.5mm]{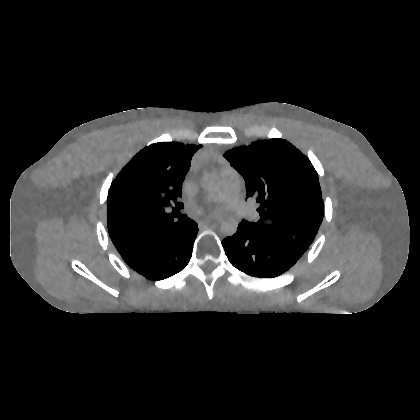}};
 	        \spy on (0.45,-0.42) in node [left] at (0,-1.95);
	        \node [black] at (0.9,-2.5) {\specialcell{$\textmd{RMSE (HU)}$ \\ \\ $= 40.8$}};
             \end{scope}
         \end{tikzpicture} &
        
         \begin{tikzpicture}
             \begin{scope}[spy using outlines={rectangle,yellow,magnification=1.75,size=18mm, connect spies}]
                 \node {\includegraphics[viewport={10mm 15mm 125mm 120mm},clip,width=34.5mm,height=34.5mm]{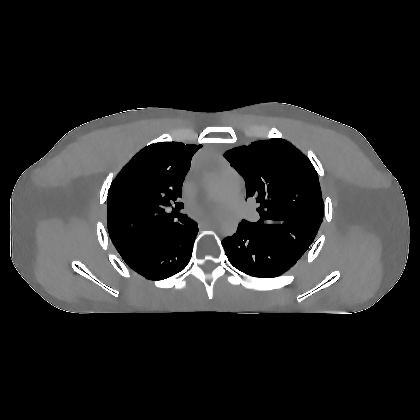}};
 	        \spy on (0.45,-0.42) in node [left] at (0,-1.95);
	        \node [black] at (0.9,-2.5) {\specialcell{$\textmd{RMSE (HU)}$ \\ \\ $= 34.7$}};
             \end{scope}
         \end{tikzpicture} &
        
          \begin{tikzpicture}
             \begin{scope}[spy using outlines={rectangle,yellow,magnification=1.75,size=18mm, connect spies}]
                 \node {\includegraphics[viewport={10mm 15mm 125mm 120mm},clip,width=34.5mm,height=34.5mm]{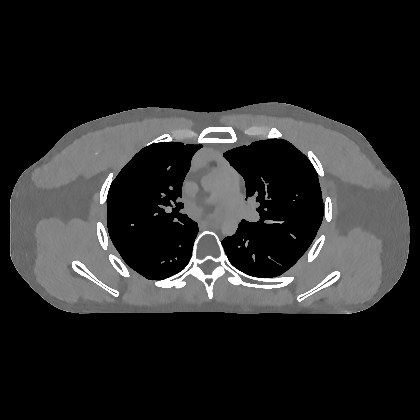}};
 		\spy on (0.45,-0.42) in node [left] at (0,-1.95);
		\node [black] at (0.9,-2.5) {\specialcell{$\textmd{RMSE (HU)}$ \\ \\ $= 19.5$}};
             \end{scope}
         \end{tikzpicture} &
         
         \begin{tikzpicture}
             \begin{scope}[spy using outlines={rectangle,yellow,magnification=1.75,size=18mm, connect spies}]
                 \node {\includegraphics[viewport={10mm 15mm 125mm 120mm},clip,width=34.5mm,height=34.5mm]{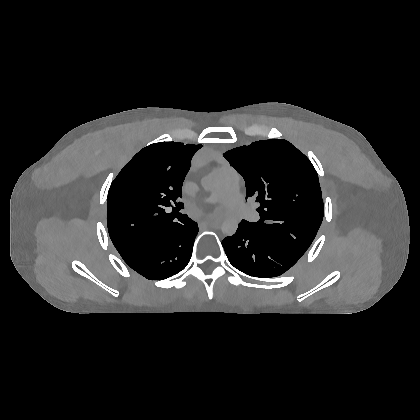}};
 		\spy on (0.45,-0.42) in node [left] at (0,-1.95);
		\node [black] at (0.9,-2.5) {\specialcell{$\textmd{RMSE (HU)}$ \\ \\ $= 19.8$}};
             \end{scope}
         \end{tikzpicture} 
         
	\\
	
	 \begin{tikzpicture}
             \begin{scope}[spy using outlines={rectangle,yellow,magnification=1.75,size=18mm, connect spies}]
                 \node {\includegraphics[viewport={15mm 5mm 130mm 110mm},clip,width=34.5mm,height=34.5mm]{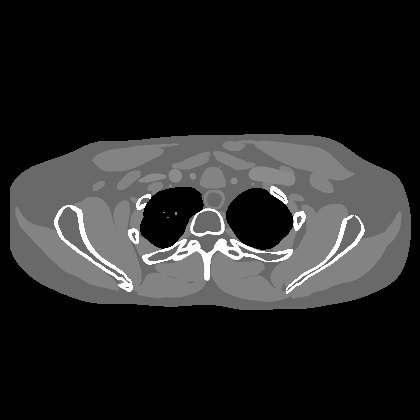}};
 	        \spy on (0.32,0.25) in node [left] at (0,-1.55);	
             \end{scope}
         \end{tikzpicture} &

         \begin{tikzpicture}
             \begin{scope}[spy using outlines={rectangle,yellow,magnification=1.75,size=18mm, connect spies}]
                 \node {\includegraphics[viewport={15mm 5mm 130mm 110mm},clip,width=34.5mm,height=34.5mm]{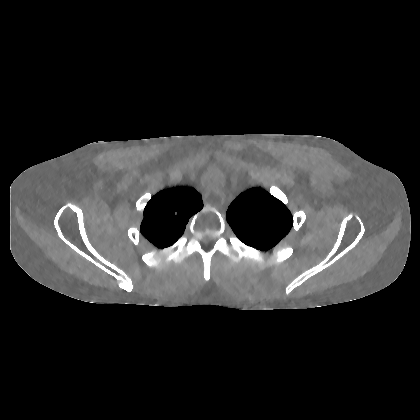}};
 	        \spy on (0.32,0.25) in node [left] at (0,-1.55);
	        \node [black] at (0.9,-2.1) {\specialcell{$\textmd{RMSE (HU)}$ \\ \\ $= 38.5$}};
             \end{scope}
         \end{tikzpicture} &
        
         \begin{tikzpicture}
             \begin{scope}[spy using outlines={rectangle,yellow,magnification=1.75,size=18mm, connect spies}]
                 \node {\includegraphics[viewport={15mm 5mm 130mm 110mm},clip,width=34.5mm,height=34.5mm]{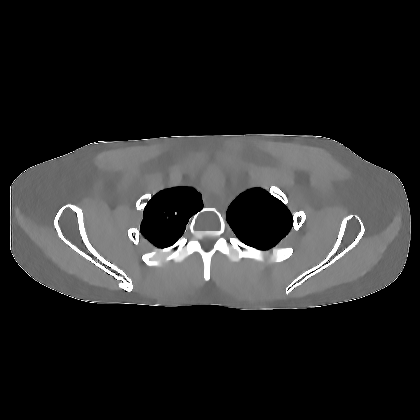}};
 	        \spy on (0.32,0.25) in node [left] at (0,-1.55);
	        \node [black] at (0.9,-2.1) {\specialcell{$\textmd{RMSE (HU)}$ \\ \\ $= 34.5$}};
             \end{scope}
         \end{tikzpicture} &
        
          \begin{tikzpicture}
             \begin{scope}[spy using outlines={rectangle,yellow,magnification=1.75,size=18mm, connect spies}]
                 \node {\includegraphics[viewport={15mm 5mm 130mm 110mm},clip,width=34.5mm,height=34.5mm]{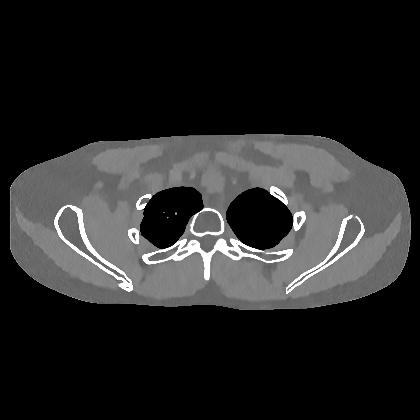}};
 		\spy on (0.32,0.25) in node [left] at (0,-1.55);
		\node [black] at (0.9,-2.1) {\specialcell{$\textmd{RMSE (HU)}$ \\ \\ $= 17.7$}};
             \end{scope}
         \end{tikzpicture} &
         
         \begin{tikzpicture}
             \begin{scope}[spy using outlines={rectangle,yellow,magnification=1.75,size=18mm, connect spies}]
                 \node {\includegraphics[viewport={15mm 5mm 130mm 110mm},clip,width=34.5mm,height=34.5mm]{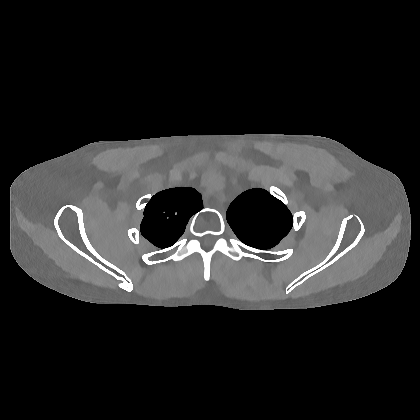}};
 		\spy on (0.32,0.25) in node [left] at (0,-1.55);
		\node [black] at (0.9,-2.1) {\specialcell{$\textmd{RMSE (HU)}$ \\ \\ $= 17.8$}};
             \end{scope}
         \end{tikzpicture} 

     \end{tabular}
    
 \vspace{-0.5em}
 \caption{Comparison of reconstructed images from different MBIR methods in sparse-view CT (fan-beam geometry with $12.5$\% projections views and $10^5$ incident photons; 
 images outside zoom-in boxes are magnified to better show differences; display window $[800, 1200]$ HU). 
 See also Fig.~\ref{fig:recon:anal}.
 }
 \label{fig:recon:ct}
\end{figure*}

 \begin{figure*}[!pt]
% \vspace{-0.75em}
 \centering
 \small\addtolength{\tabcolsep}{-6.5pt}
 \renewcommand{\arraystretch}{0.2}

     \begin{tabular}{cccccc}
     
           \specialcell{(a) Ground truth at the \\ $(5,5)\rth$ angular coord.} & & 
           \specialcell{(b) Error maps of \\ 4D EP reg. \cite{Lien&etal:20NP}} &
           \specialcell{(c) Error maps of \\ Momentum-Net-sCNN } &
           \specialcell{(d) Error maps of \\ {\bfseries Momentum-Net-dCNN}} &
           \\
        
         \begin{tikzpicture}
                 \node {\includegraphics[width=34.5mm,height=34.5mm]{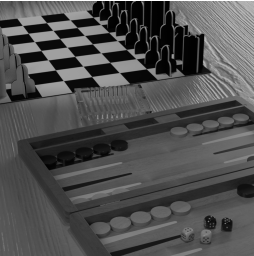}};
         \end{tikzpicture} &
         
         \begin{tikzpicture}
                 \node {\includegraphics[trim=9.45cm 0.1cm 0cm 0cm,clip,height=25mm]{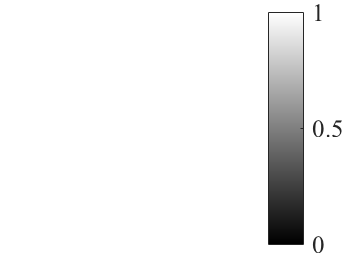}};
         \end{tikzpicture} &
         
         \begin{tikzpicture}
                \node {\includegraphics[clip,width=34.5mm,height=34.5mm]{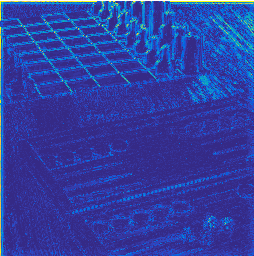}};
         \end{tikzpicture} &
        
         \begin{tikzpicture}
                 \node {\includegraphics[clip,width=34.5mm,height=34.5mm]{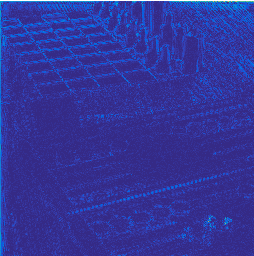}};
         \end{tikzpicture} &
         
         \begin{tikzpicture}
                 \node {\includegraphics[clip,width=34.5mm,height=34.5mm]{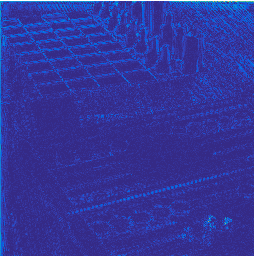}};
         \end{tikzpicture} &
         
         \begin{tikzpicture}
                 \node {\includegraphics[trim=9.1cm 0.1cm 0cm 0cm,clip,height=25mm]{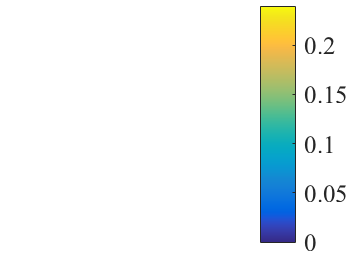}};
         \end{tikzpicture}      
         
         \\
         & &       
         {$\textmd{PSNR (dB)} = 29.9~(32.0)$} & {$\textmd{PSNR (dB)} = 37.7~(35.8)$} & {$\textmd{PSNR (dB)} = 38.2~(37.1)$}
         \\
         
          \begin{tikzpicture}
                 \node {\includegraphics[width=34.5mm,height=34.5mm]{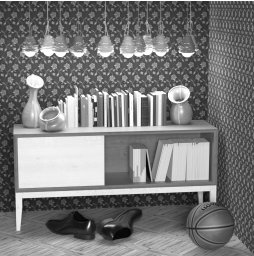}};
         \end{tikzpicture} &
         
         \begin{tikzpicture}
                 \node {\includegraphics[trim=9.45cm 0.1cm 0cm 0cm,clip,height=25mm]{./Fig/lf/colorbar_0-1.png}};
         \end{tikzpicture} &

         \begin{tikzpicture}
               \node {\includegraphics[clip,width=34.5mm,height=34.5mm]{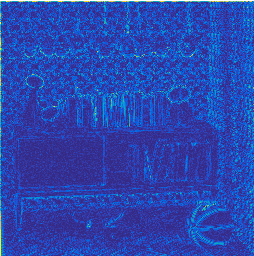}};
         \end{tikzpicture} &
        
         \begin{tikzpicture}
                 \node {\includegraphics[clip,width=34.5mm,height=34.5mm]{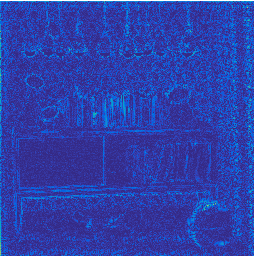}};
         \end{tikzpicture}  &
         
         \begin{tikzpicture}
                 \node {\includegraphics[clip,width=34.5mm,height=34.5mm]{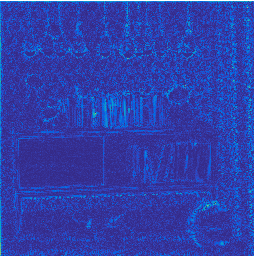}};
         \end{tikzpicture}  &
         
         \begin{tikzpicture}
                 \node {\includegraphics[trim=9.1cm 0.1cm 0cm 0cm,clip,height=25mm]{./Fig/lf/colorbar_0-0p24.png}};
         \end{tikzpicture}      
         
          \\
         & & 
         {$\textmd{PSNR (dB)} = 30.7~(28.1)$} & {$\textmd{PSNR (dB)} = 33.9~(30.7)$} &  {$\textmd{PSNR (dB)} = 34.6~(32.0)$}
         \\
         
         \begin{tikzpicture}
                 \node {\includegraphics[width=34.5mm,height=34.5mm]{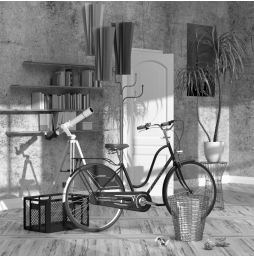}};
         \end{tikzpicture} &
         
         \begin{tikzpicture}
                 \node {\includegraphics[trim=9.45cm 0.1cm 0cm 0cm,clip,height=25mm]{./Fig/lf/colorbar_0-1.png}};
         \end{tikzpicture} &
         
         \begin{tikzpicture}
               \node {\includegraphics[clip,width=34.5mm,height=34.5mm]{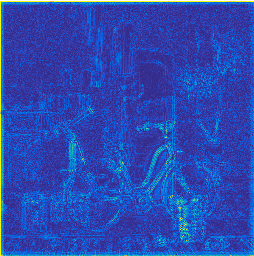}};
         \end{tikzpicture} &
        
         \begin{tikzpicture}
                 \node {\includegraphics[clip,width=34.5mm,height=34.5mm]{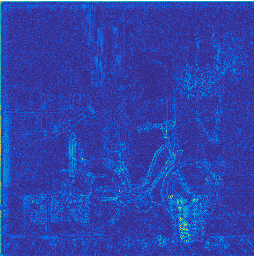}};
         \end{tikzpicture}  &
         
         \begin{tikzpicture}
                 \node {\includegraphics[clip,width=34.5mm,height=34.5mm]{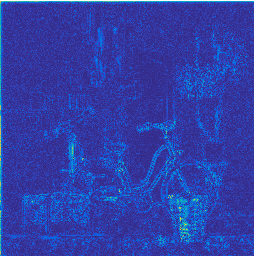}};
         \end{tikzpicture}  &
         
         \begin{tikzpicture}
                 \node {\includegraphics[trim=9.1cm 0.1cm 0cm 0cm,clip,height=25mm]{./Fig/lf/colorbar_0-0p24.png}};
         \end{tikzpicture}      
         
          \\
         & & 
         {$\textmd{PSNR (dB)} = 27.3~(28.1)$} & {$\textmd{PSNR (dB)} = 32.9~(30.9)$} & {$\textmd{PSNR (dB)} = 33.6~(31.7)$}

     \end{tabular}
    
 \vspace{-0.5em}
 \caption{Error map comparisons of reconstructed sub-aperture images from different MBIR methods in LF photography using a focal stack 
 (LF photography systems with $C \!=\! 5$ detectors capture a focal stack of
LFs consisting of $S \!=\! 81$ sub-aperture images; 
 sub-aperture images at the $(5,5)\rth$ angular coordinate;
 the PSNR values in parenthesis were measured from reconstructed LFs).
 See also Fig.~\ref{fig:recon:anal}.
 }
 \label{fig:recon:lf}
\end{figure*}

 \begin{figure*}[!pt]
 \centering
 \small\addtolength{\tabcolsep}{-6.5pt}
 \renewcommand{\arraystretch}{0.2}

     \begin{tabular}{ccccc}
     
           \specialcell{(a) Ground truth \\ depth maps} & 
           \specialcell{(b) Estimated depth \\ from ground truth LF} & 
           \specialcell{(c) Estimated depth \\  from LF reconstructed \vphantom{p} \\ by 4D EP reg. \cite{Lien&etal:20NP}} &
           \specialcell{(d) Estimated depth \\ from LF reconstructed by \\ Momentum-Net-sCNN} &
           \specialcell{(e) Estimated depth \\ from LF recon. by \\ {\bfseries Momentum-Net-dCNN}} 
            \\
        
         \begin{tikzpicture}
             \begin{scope}[spy using outlines={rectangle,yellow,magnification=1.75,size=18mm, connect spies}]
                 \node {\includegraphics[width=34.5mm,height=34.5mm]{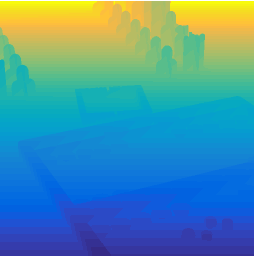}};
             \end{scope}
         \end{tikzpicture} &
         
          \begin{tikzpicture}
             \begin{scope}[spy using outlines={rectangle,yellow,magnification=1.75,size=18mm, connect spies}]
                 \node {\includegraphics[clip,width=34.5mm,height=34.5mm]{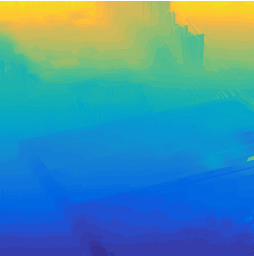}};
             \end{scope}
         \end{tikzpicture} &
	
         \begin{tikzpicture}
             \begin{scope}[spy using outlines={rectangle,yellow,magnification=1.75,size=18mm, connect spies}]
                 \node {\includegraphics[clip,width=34.5mm,height=34.5mm]{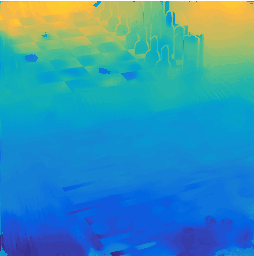}};
             \end{scope}
         \end{tikzpicture} &
        
         \begin{tikzpicture}
             \begin{scope}[spy using outlines={rectangle,yellow,magnification=1.75,size=18mm, connect spies}]
                 \node {\includegraphics[clip,width=34.5mm,height=34.5mm]{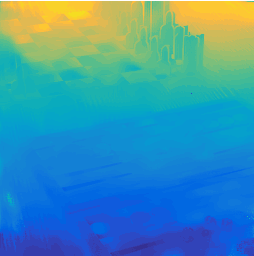}};
             \end{scope}
         \end{tikzpicture} &
         
         \begin{tikzpicture}
             \begin{scope}[spy using outlines={rectangle,yellow,magnification=1.75,size=18mm, connect spies}]
                 \node {\includegraphics[clip,width=34.5mm,height=34.5mm]{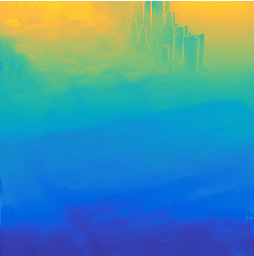}};
             \end{scope}
         \end{tikzpicture} 
         
         \\
         
         \parbox[c]{25mm}{\includegraphics[trim=0.2cm 0.4cm 0.2cm 10.5cm,clip,width=25mm]{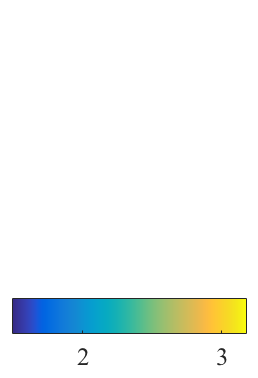}} &
         \parbox[c]{34.5mm}{$\textmd{RMSE (m)} = 4.7 \!\times\! 10^{-2}$ \\ } &
         \parbox[c]{34.5mm}{$\textmd{RMSE (m)} = 13.8 \!\times\! 10^{-2}$ \\ } &
         \parbox[c]{34.5mm}{$\textmd{RMSE (m)} = 8.0 \!\times\! 10^{-2}$ \\ } &
         \parbox[c]{34.5mm}{$\textmd{RMSE (m)} = 5.7 \!\times\! 10^{-2}$ \\ }           
         \\
         
           \begin{tikzpicture}
             \begin{scope}[spy using outlines={rectangle,yellow,magnification=1.75,size=18mm, connect spies}]
                 \node {\includegraphics[width=34.5mm,height=34.5mm]{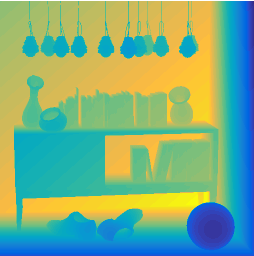}};
             \end{scope}
         \end{tikzpicture} &
         
          \begin{tikzpicture}
             \begin{scope}[spy using outlines={rectangle,yellow,magnification=1.75,size=18mm, connect spies}]
                 \node {\includegraphics[clip,width=34.5mm,height=34.5mm]{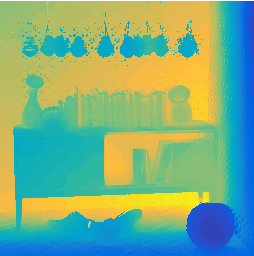}};
             \end{scope}
         \end{tikzpicture} &
	
         \begin{tikzpicture}
             \begin{scope}[spy using outlines={rectangle,yellow,magnification=1.75,size=18mm, connect spies}]
                 \node {\includegraphics[clip,width=34.5mm,height=34.5mm]{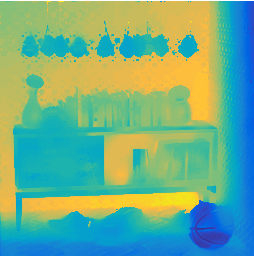}};
             \end{scope}
         \end{tikzpicture} &
        
         \begin{tikzpicture}
             \begin{scope}[spy using outlines={rectangle,yellow,magnification=1.75,size=18mm, connect spies}]
                 \node {\includegraphics[clip,width=34.5mm,height=34.5mm]{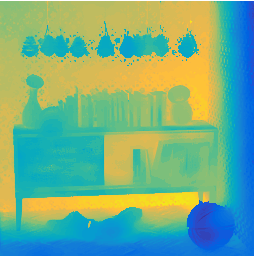}};
             \end{scope}
         \end{tikzpicture} &
         
          \begin{tikzpicture}
             \begin{scope}[spy using outlines={rectangle,yellow,magnification=1.75,size=18mm, connect spies}]
                 \node {\includegraphics[clip,width=34.5mm,height=34.5mm]{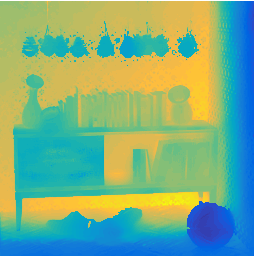}};
             \end{scope}
         \end{tikzpicture} 
         
         \\
         
         \parbox[c]{25mm}{\includegraphics[trim=0.2cm 0.4cm 0.2cm 10.5cm,clip,width=25mm]{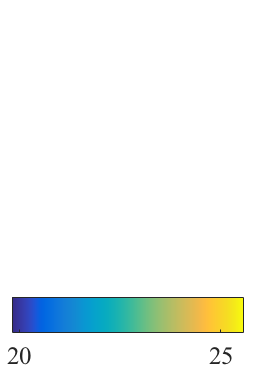}} &
         \parbox[c]{34.5mm}{$\textmd{RMSE (m)} = 30.5 \!\times\! 10^{-2}$ \\ } &
         \parbox[c]{34.5mm}{$\textmd{RMSE (m)} = 39.5 \!\times\! 10^{-2}$ \\ } &
         \parbox[c]{34.5mm}{$\textmd{RMSE (m)} = 34.6 \!\times\! 10^{-2}$ \\ } & 
         \parbox[c]{34.5mm}{$\textmd{RMSE (m)} = 31.9 \!\times\! 10^{-2}$ \\ }                
         \\
         
          \begin{tikzpicture}
             \begin{scope}[spy using outlines={rectangle,yellow,magnification=1.75,size=18mm, connect spies}]
                 \node {\parbox[c]{25mm}{\includegraphics[trim=0.2cm 0.4cm 0.2cm 10.5cm,clip,width=25mm]{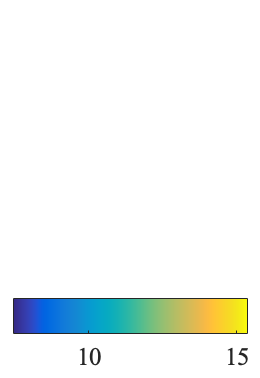}}};     
                 \node[text width=2.1cm] at (0.1,1.5) {Ground truth does not exist.};                   
             \end{scope}
         \end{tikzpicture} &
         
          \begin{tikzpicture}
             \begin{scope}[spy using outlines={rectangle,yellow,magnification=1.75,size=18mm, connect spies}]
                 \node {\includegraphics[clip,width=34.5mm,height=34.5mm]{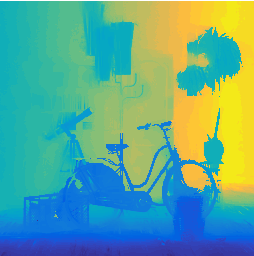}};
             \end{scope}
         \end{tikzpicture} &
	
         \begin{tikzpicture}
             \begin{scope}[spy using outlines={rectangle,yellow,magnification=1.75,size=18mm, connect spies}]
                 \node {\includegraphics[clip,width=34.5mm,height=34.5mm]{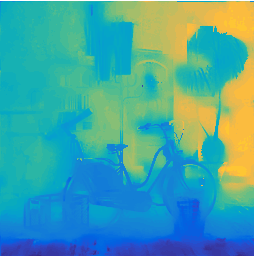}};
             \end{scope}
         \end{tikzpicture} &
        
         \begin{tikzpicture}
             \begin{scope}[spy using outlines={rectangle,yellow,magnification=1.75,size=18mm, connect spies}]
                 \node {\includegraphics[clip,width=34.5mm,height=34.5mm]{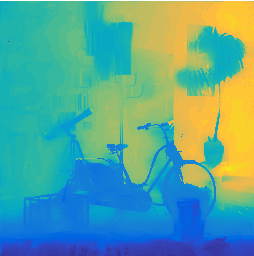}};
             \end{scope}
         \end{tikzpicture} &
         
         \begin{tikzpicture}
             \begin{scope}[spy using outlines={rectangle,yellow,magnification=1.75,size=18mm, connect spies}]
                 \node {\includegraphics[clip,width=34.5mm,height=34.5mm]{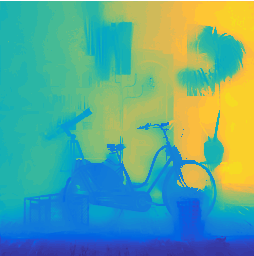}};
             \end{scope}
         \end{tikzpicture} 
      
     \end{tabular}
    
 \vspace{-0.5em}
 \caption{
Comparisons of estimated depths from LFs reconstructed by different MBIR methods in LF photograph using a focal stack 
(LF photography systems with $C \!=\! 5$ detectors capture a focal stack of
LFs consisting of $S \!=\! 81$ sub-aperture images; 
SPO depth estimation \cite{Zhang&etal:16CVIU} was applied to reconstructed LFs in Fig.~\ref{fig:recon:lf}; 
display window in meters).
See also Fig.~\ref{fig:recon:anal}.
 }
 \label{fig:depth}
\end{figure*}

\subsection{Comparisons between different MBIR methods} \label{sec:comp:MBIR}

In sparse-view CT using $12.5$\% of the full projection views,
Fig.~\ref{fig:recon:ct}(b)--(e) and Table~\ref{tab:recon:ct}(b)--(f) show that 
the proposed Momentum-Net 
achieves significantly better reconstruction quality compared to 
the conventional EP MBIR method 
and the state-of-the-art MBIR method within an unsupervised learning setup, 
MBIR model using a learned convolutional regularizer \cite{Chun&Fessler:20TIP, Chun&Fessler:18Asilomar}.
In particular, Momentum-Net recovers 
both low- and high-contrast regions (e.g., soft tissues and bones, respectively)
more accurately than MBIR using a learned convolutional regularizer; see Fig.~\ref{fig:recon:ct}(c)--(e).
In addition, when their shallow convolutional autoencoders need identical computational complexities,
Momentum-Net achieves much faster MBIR compared to MBIR using a learned convolutional regularizer;
see Table~\ref{tab:recon:ct}(c)--(d). 

In LF photography using five focal sensors,
regardless of scene parameters and camera settings,
Momentum-Net consistently achieves significantly more accurate image recovery,
compared to MBIR model using the state-of-the-art non-trained regularizer, 4D EP \cite{Lien&etal:20NP}.
The effectiveness of Momentum-Net is more evident for a scene with less fine details.
See Fig.~\ref{fig:recon:lf}(b)--(d) and Table~\ref{tab:recon:lf}(b)--(d).
Regardless of the scene distances from LF imaging systems, the reconstructed LFs by Momentum-Net significantly improve the depth estimation accuracy over those reconstructed by the state-of-the-art non-trained regularizer, 4D EP \cite{Lien&etal:20NP}.
See Fig.~\ref{fig:depth}(c)--(e) and Table~\ref{tab:depth}(c)--(e).

In general,
Momentum-Net needs more computations per iteration than EP MBIR, 
because its refining NNs use more and larger filters than the small finite-difference filters in EP MBIR,
and EP MBIR algorithms can be often further accelerated by gradient approximations, 
e.g., ordered-subsets methods \cite{Hudson&Larkin:94TMI, Erdogan&Fessler:99PMB}.

\section{Conclusions} \label{sec:conclusion}

Developing rapidly converging INNs is important, 
because \textit{1)} it leads to fast MBIR 
by reducing the computational complexity
in calculating data-fit gradients or 
applying refining NNs, 
and \textit{2)} training INNs with many iterations requires long training time
or it is challenging when refining NNs are fixed across INN iterations.
The proposed Momentum-Net framework is applicable for a wide range of inverse problems, while achieving fast and convergent MBIR.
To achieve fast MBIR, 
Momentum-Net uses momentum in extrapolation modules, 
and noniterative MBIR modules at each iteration via majorizers.
For sparse-view CT and LF photography using a focal stack,
Momentum-Net achieves faster and more accurate MBIR compared to 
the existing soft-refining INNs,
\cite{Aggarwal&Mani&Jacobs:18TMI, Romano&Elad&Milanfar:17SJIS, Buzzard&etal:18SJIS, Chen&Pock:17PAMI, Hammernik&etal:17MRM} 
that correspond to BCD-Net \cite{Chun&Fessler:18IVMSP} or Momentum-Net using \textit{no extrapolation},
and ADMM-Net \cite{Yang&etal:16NIPS, Chan&Wang&Elgendy:17TCI, Buzzard&etal:18SJIS},
and the existing hard-refining INN PDS-Net \cite{Ono:17SPL}.
When an application needs strong regularization strength, e.g., LF photography using limited detectors,
using dCNN refiners with moderate depth significantly improves the MBIR accuracy of Momentum-Net compared to sCNNs,
only marginally increasing total MBIR time.
In addition, Momentum-Net guarantees convergence to a fixed-point 
for general differentiable (non)convex MBIR functions (or data-fit terms) and convex feasible sets, 
under some mild conditions and two asymptotic conditions.
The proposed regularization parameter selection scheme uses the \dquotes{spectral spread} of majorization matrices,
and is useful to consider data-fit variations across training/testing samples.

There are a number of avenues for future work.
First, we expect to further improve performances of Momentum-Net (e.g., MBIR time and accuracy) by using sharper majorizer designs.
Second, we expect to further reduce MBIR time of Momentum-Net with the stochastic gradient perspective (e.g., ordered subset \cite{Hudson&Larkin:94TMI, Erdogan&Fessler:99PMB}).
On the regularization parameter selection side, 
our future work is learning the factor $\chi$ in \R{eq:reg:select} from datasets while training refining NNs.

% use section* for acknowledgment
%\ifCLASSOPTIONcompsoc
%  % The Computer Society usually uses the plural form
%  \section*{Acknowledgments}
%\else
%  % regular IEEE prefers the singular form
%  \section*{Acknowledgment}
%\fi
%
%We thank Xuehang Zheng for providing CT imaging simulation setup.

% Can use something like this to put references on a page
% by themselves when using endfloat and the captionsoff option.
\ifCLASSOPTIONcaptionsoff
  \newpage
\fi

%%%%%%%%%%%%%%%%%%%%%%%%%%%%%%%%%%%%%%%%%%%%%%%%%%%%%%%%%%%%%%%%%%%%%%%%%%%%%%%%
%                                References
%%%%%%%%%%%%%%%%%%%%%%%%%%%%%%%%%%%%%%%%%%%%%%%%%%%%%%%%%%%%%%%%%%%%%%%%%%%%%%%%
\bibliographystyle{IEEEtran}
\bibliography{referenceBibs_Bobby}
%%%%%%%%%%%%%%%%%%%%%%%%%%%%%%%%%%%%%%%%%%%%%%%%%%%%%%%%%%%%%%%%%%%%%%%%%%%%%%%%

%%%%%%%%%%%%%%%%%%%%%%%%%%%%%%%%%%%%%%%%%%%%%%%%%%%%%%%%%%%%%%%%%%%%%%%%%%%%%%%%
%                                Supplement
%%%%%%%%%%%%%%%%%%%%%%%%%%%%%%%%%%%%%%%%%%%%%%%%%%%%%%%%%%%%%%%%%%%%%%%%%%%%%%%%

\onecolumn

\renewcommand{\thefigure}{A.\arabic{figure}}
\renewcommand{\thetable}{A.\arabic{table}}
\renewcommand{\thesection}{A.\arabic{section}}
\renewcommand{\thefootnote}{A.\arabic{footnote}}
\renewcommand{\thetheorem}{A.\arabic{theorem}}
\renewcommand{\theequation}{A.\arabic{equation}}
\renewcommand{\thealgorithm}{A.\arabic{algorithm}} 

\setcounter{section}{0}
\setcounter{equation}{0}
\setcounter{theorem}{0}
\setcounter{figure}{0}
\setcounter{table}{0}
\setcounter{algorithm}{0}
\setcounter{footnote}{0}

\begin{center}
% \Huge
\fontsize{23.4}{23.4}\selectfont 
Momentum-Net: Fast and convergent \\ 
iterative neural network for inverse problems \\ 
(Appendices)
\end{center}

\vspace{0.5in}
This supplementary material for \cite{Chun&etal:18arXiv:momnet:supp} 
\textit{a)}~reviews the Block Proximal Extrapolated Gradient method using a Majorizer (BPEG-M) 
\cite{Chun&Fessler:18TIP:supp, Chun&Fessler:20TIP:supp}, 
\textit{b)}~lists parameters of Momentum-Net, and summarizes selection guidelines or gives default values,
\textit{c)}~compares the convergence properties between Momentum-Net and BCD-Net, and
\textit{d)}~provides mathematical proofs or detailed descriptions to support several arguments in
the main manuscript. 
We use the prefix \dquotes{A} for the numbers
in section, theorem, equation, figure, table, and footnote in the supplement.

\section{BPEG-M: Review}
\label{sec:bpgm}

This section explains \textit{block multi-(non)convex} optimization problems,
and summarizes the state-of-the-art method for block multi-(non)convex optimization, 
BPEG-M \cite{Chun&Fessler:18TIP:supp, Chun&Fessler:20TIP:supp}, 
along with its convergence guarantees.

\subsection{Block multi-(non)convex optimization} \label{sec:bpgm:block}

In a block optimization problem, the variables of the underlying optimization problem are treated either as a single block or multiple disjoint blocks. 
In \textit{block multi-(non)convex} optimization, we consider the following problem:
\be{
\label{sys:multiConvx}
\min_{u} \, F(u_1,\ldots,u_B) \triangleq f(u_1,\ldots,u_B) + \sum_{b=1}^B r_b (u_b)
%\tag{P1}
}
where variable $u$ is decomposed into $B$ blocks $u_1,\ldots,u_B$ ($\{ u_b \in \bbR^{n_b} : b=1,\ldots,B \}$), 
$f$ is assumed to be (continuously) differentiable,
but functions $\{ r_b : b = 1,\ldots,B \}$ are not necessarily differentiable. 
The function $r_b$ can incorporate the constraint $u_b \in \cU_b$, by allowing $r_b$'s to be extended-valued, e.g., $r_b (u_b) = \infty$ if $u_b \notin \cU_b$, for $b=1,\ldots,B$.
It is standard to assume that both $f$ and $\{ r_b \}$ are proper and closed, and the sets $\{ \cU_b \}$ are closed.
We consider either that \R{sys:multiConvx} has block-wise convexity (but \R{sys:multiConvx} is jointly nonconvex in general) \cite{Xu&Yin:13SIAM:supp, Chun&Fessler:18TIP:supp}
or that $f$, $\{ r_b \}$, or $\{ \cU_b \}$ are not necessarily convex \cite{Xu&Yin:17JSC:supp, Chun&Fessler:20TIP:supp}.
Importantly, $r_b$ can include (non)convex and nonsmooth $\ell^p$ (quasi-)norm, $p \in [0, 1]$.
The next section introduces our optimization framework that solves \R{sys:multiConvx}.

The following sections review BPEG-M \cite{Chun&Fessler:18TIP:supp, Chun&Fessler:20TIP:supp}, the state-of-the-art optimization framework for solving block multi-(non)convex problems, when used with sufficiently sharp majorizers.
BPEG-M uses block-wise extrapolation, majorization, and proximal mapping.
By using a more general Lipschitz continuity (see Definition~\ref{d:QM}) for block-wise gradients, BPEG-M is particularly useful for rapidly calculating majorizers involved with large-scale problems, and successfully applied to some large-scale machine learning and computational imaging problems; see \cite{Chun&Fessler:18TIP:supp, Chun&Fessler:20TIP:supp, Chun&Fessler:18Asilomar:supp} and references therein.

%\subsection{Preliminaries}
%
%The BPEG-M framework considers $M$-Lipschitz continuity (see Definition~\ref{d:QM}) of the gradient.
%If the gradient of a function is $M$-Lipschitz continuous, 
%then one obtains the following quadratic majorizer \cite{Lange&Hunter&Yang:00JCGS, Jacobson&Fessler:07TIP} at a given point $v$ without assuming convexity:
%
%\lem{[Quadratic majorization via $M$-Lipschitz continuous gradients~\!\mbox{\cite{Chun&Fessler:20TIP}}] \label{l:QM}
%Let $f(u) : \bbR^n \rightarrow \bbR$. If $\nabla f$ is $M$-Lipschitz continuous, then 
%$f(u) \leq f(v) + \ip{\nabla_u f(v)}{u-v} + \frac{1}{2} \nm{u - v}_M^2$, $\forall u,v \in \bbR^n$.
%}

\subsection{BPEG-M}

This section summarizes the BPEG-M framework.
Using Definition~\ref{d:QM} and Lemma~\ref{l:QM}, the proposed method, BPEG-M, is given as follows.
To solve \R{sys:multiConvx}, we minimize majorizers of $F$ cyclically over each block $u_1,\ldots,u_B$, while fixing the remaining blocks at their previously updated variables. Let $u_b^{(i+1)}$ be the value of $u_b$ after its $i\rth$ update, and define
\bes{
%\label{eq:def:f_block}
f_b^{(i+1)}(u_b) \triangleq f \Big( u_1^{(i+1)}, \ldots, u_{b-1}^{(i+1)}, u_b, u_{b+1}^{(i)}, \ldots, u_{B}^{(i)} \Big),
}
for all $b,i$.
At the $b\rth$ block of the $i\rth$ iteration, we apply Lemma~\ref{l:QM} to functional $f_b^{(i+1)}(u_b)$ with a $M^{(i+1)}$-Lipschitz continuous gradient at the extrapolated point $\acute{u}_b^{(i+1)}$, and minimize a majorized function.
In other words, we consider the updates
\begingroup
%\setlength{\thinmuskip}{1.5mu}
%\setlength{\medmuskip}{2mu plus 1mu minus 2mu}
%\setlength{\thickmuskip}{2.5mu plus 2.5mu}
%\fontsize{9.5pt}{11.4pt}\selectfont
\allowdisplaybreaks
\ea{
u_b^{(i+1)} 
&= \argmin_{ u_b } \, \ip{ \nabla f_b^{(i+1)} (\acute{u}_b^{(i+1)}) }{ u_b - \acute{u}_b^{(i+1)} } 
+ \frac{1}{2} \nm{ u_b - \acute{u}_b^{(i+1)} }_{\widetilde{M}_b^{(i+1)}}^2 
+ r_b (u_b)
\nn 
\\
&= \mathrm{Prox}_{r_b}^{\widetilde{M}_b^{(i+1)}} \!\! \bigg( \! \underbrace{\acute{u}_b^{(i+1)} - \left( \! \widetilde{M}_b^{(i+1)} \! \right)^{\!\!-1} \nabla f_b^{(i+1)} (\acute{u}_b^{(i+1)}) }_{\text{\emph{extrapolated gradient step using a majorizer} of $f_b^{(i+1)}$}} \! \bigg), 
\label{up:x}
}
\endgroup
where 
\be{
\label{up:xacute}
\acute{u}_b^{(i+1)} = u_{b}^{(i)} + E_b^{(i+1)} \big( u_b^{(i)} - u_b^{(i-1)} \big),
}
the proximal operator is defined by \R{eq:d:prox},
$\nabla f_b^{(i+1)} (\acute{u}_b^{(i+1)})$ is the block-partial gradient of $f$ at $\acute{u}_b^{(i+1)}$, 
a \textit{scaled majorization matrix} is given by
\be{
\label{up:Mtilde}
\widetilde{M}_b^{(i+1)} = \lambda_b \cdot M_b^{(i+1)} \succ 0, \qquad \lambda_b \geq 1,
}
and $M_b^{(i+1)} \!\in\! \bbR^{n_b \times n_b}$ is a symmetric positive definite \textit{majorization matrix} of $\nabla f_b^{(i+1)}(u_b)$.
In \R{up:xacute}, the $\bbR^{n_b \times n_b}$ matrix $E_b^{(i+1)} \succeq 0$ is an \textit{extrapolation matrix} that accelerates convergence in solving block multi-convex problems \cite{Chun&Fessler:18TIP:supp}.
We design it to satisfy conditions \R{cond:Eb:cvx} or \R{cond:Eb:ncvx} below.
In \R{up:Mtilde}, $\{ \lambda_b = 1 : \forall b \}$ and $\{ \lambda_b > 1 : \forall b \}$, for block multi-convex and block multi-nonconvex problems, respectively.

For some $f_b^{(i+1)}$ having sharp majorizers, we expect that extrapolation \R{up:xacute} has no benefits in accelerating convergence, and use $\{ E_b^{(i+1)} = 0 : \forall i \}$.
Other than the blocks having sharp majorizers, one can apply some increasing momentum coefficient formula \cite{Nesterov:13MP:supp, Beck&Teboulle:09SIAM:supp} to the corresponding extrapolation matrices.
The choice in \cite{Chun&Fessler:18TIP:supp, Chun&Fessler:20TIP:supp, Xu&Yin:13SIAM:supp} accelerated BPEG-M for some machine learning and data science applications.
Algorithm~\ref{alg:bpgm} summarizes these updates.

\begin{algorithm}[pt!]
\caption{BPEG-M \cite{Chun&Fessler:18TIP:supp, Chun&Fessler:20TIP:supp}}
\label{alg:bpgm}

\begin{algorithmic}
\REQUIRE $\{ u_b^{(0)} = u_b^{(-1)} : \forall b \}$, $\{ w_{b}^{(i)} \in [0, 1], \forall b,i \}$, $i=0$

\WHILE{a stopping criterion is not satisfied}

\FOR{$b = 1,\ldots,B$}

\STATE \begingroup
\fontsize{9.5pt}{11.4pt}\selectfont
Calculate $\widetilde{M}_b^{(i+1)}$ by \R{up:Mtilde}, and $E_b^{(i+1)}$ to satisfy \R{cond:Eb:cvx} or \R{cond:Eb:ncvx}
\endgroup
\STATE $\displaystyle \acute{u}_b^{(i+1)} = u_{b}^{(i)} + E_b^{(i+1)} \big( u_b^{(i)} - u_b^{(i-1)} \big)$
\STATE $\displaystyle u_b^{(i+1)} = \mathrm{Prox}_{r_b}^{\widetilde{M}_b^{(i+1)}} \! \bigg( \! \acute{u}_b^{(i+1)} - \Big( \! \widetilde{M}_b^{(i+1)} \! \Big)^{\!\!-1} \!\! \nabla f_b^{(i+1)} (\acute{u}_b^{(i+1)}) \! \bigg)$

\ENDFOR

\STATE $i = i+1$

\ENDWHILE

\end{algorithmic}
\end{algorithm}

\subsection{Convergence results} \label{sec:bpgm:analysis}

This section summarizes convergence results of Algorithm~\ref{alg:bpgm} under the following assumptions:

\bulls{
%[\setlength{\topsep}{1pt}]

\item {\em Assumption~S.1)}
In \R{sys:multiConvx}, $F$ is proper and lower bounded in $\dom(F) \triangleq \{ u : F(u) < \infty \}$.
In addition,
\bulls{[\setlength{\topsep}{1pt}]

\item[] {for \emph{block multi-convex}~\R{sys:multiConvx}}, $f$ is differentiable and \R{sys:multiConvx} has a Nash point or block-coordinate minimizer\footnote{
Given a feasible set $\cU$, a point $u^\star \in \dom(F) \cup \cU$ is a critical point (or stationary point) of $F$ if the directional derivative $d^T \nabla F(u^\star) \geq 0$ for any feasible direction $d$ at $u^\star$.
If $u^\star$ is an interior point of $\cU$, then the condition is equivalent to $0 \in \partial F(u^\star)$.
} (see its definition in \cite[(2.3)--(2.4)]{Xu&Yin:13SIAM:supp});

\item[] {for \emph{block multi-nonconvex}~\R{sys:multiConvx}}, $f$ is continuously differentiable, $r_b$ is lower semicontinuous\footnote{
$F$ is lower semicontinuous at point $u_0$ if $\liminf_{u \rightarrow u_0} F(u) \geq F(u_0)$.
}, $\forall b$, and \R{sys:multiConvx} has a critical point $u^\star$ that satisfies $0 \in \partial F(u^\star)$.
}

\item {\em Assumption~S.2)} $\nabla f_b^{(i+1)} (u_b)$ is $M$-Lipschitz continuous with respect to $u_b$, i.e.,
\begingroup
\setlength\abovedisplayskip{0.5\baselineskip}
\setlength\belowdisplayskip{0.5\baselineskip}
\bes{
%\label{eq:QMbound}
\nm{ \nabla f_b^{(i+1)} (u) - \nabla f_b^{(i+1)} (v)  }_{\left( \! M_b^{(i+1)} \! \right)^{\!-1}} \leq \nm{u - v}_{M_b^{(i+1)}},
}
\endgroup
for $u,v \in \bbR^{n_b}$, where $M_b^{(i+1)}$ is a bounded majorization matrix.

\item {\em Assumption~S.3)} The extrapolation matrices $E_b^{(i+1)} \succeq 0$ satisfy that
\begingroup
\setlength\abovedisplayskip{0.2\baselineskip}
\setlength\belowdisplayskip{0.2\baselineskip}
\ea{
\label{cond:Eb:cvx}
\mbox{for \emph{block multi-convex}~\R{sys:multiConvx},~} &
\big( E_b^{(i+1)} \big)^{\!T} M_b^{(i+1)} E_b^{(i+1)} \preceq \delta^2 \cdot M_b^{(i)};
\\
\label{cond:Eb:ncvx}
\mbox{for \emph{block multi-nonconvex}~\R{sys:multiConvx},~} &
\big( E_b^{(i+1)} \big)^{\!T} M_b^{(i+1)} E_b^{(i+1)} \preceq \frac{\delta^2 (\lambda_b - 1)^2}{4 (\lambda_b + 1)^2} \cdot M_b^{(i)},
}
\endgroup
with $\delta < 1$, $\forall b,i$.
}

\thm{[\emph{Block multi-convex}~\R{sys:multiConvx}: A limit point is a Nash point \mbox{\cite{Chun&Fessler:18TIP:supp}}]
\label{t:Nash_convg}
Under Assumptions~S.1--S.3, let $\{ u^{(i+1)} : i \geq 0\}$ be the sequence generated by Algorithm~\ref{alg:bpgm}. Then any limit point of $\{ u^{(i+1)} : i \geq 0 \}$ is a Nash point of \R{sys:multiConvx}.
}

\thm{[\emph{Block multi-nonconvex}~\R{sys:multiConvx}: A limit point is a critical point \mbox{\cite{Chun&Fessler:20TIP:supp}}]
\label{t:subseqConv}
Under Assumptions~S.1--S.3, let $\{ u^{(i+1)} : i \geq 0\}$ be the sequence generated by Algorithm~\ref{alg:bpgm}.
Then any limit point of $\{ u^{(i+1)} : i \geq 0 \}$ is a critical point of \R{sys:multiConvx}. 
}

\rem{\label{r:convg}
Theorems \ref{t:Nash_convg}--\ref{t:subseqConv} imply that, if there exists a critical point for \R{sys:multiConvx}, i.e., $0 \in \partial F(u^\star)$, then any limit point of $\{ u^{(i+1)} : i \geq 0 \}$ is a critical point.
One can further show global convergence under some conditions: if $\{ u^{(i+1)} : i \geq 0 \}$ is bounded and the critical points are isolated, then $\{ u^{(i+1)} : i \geq 0 \}$ converges to a critical point \cite[Rem.~3.4]{Chun&Fessler:18TIP:supp}, \cite[Cor.~2.4]{Xu&Yin:13SIAM:supp}.
}

\subsection{Application of BPEG-M to solving block multi-(non)convex problem \R{sys:recov&caol}}
\label{sec:bpgm:caol}

For update \R{eq:bpgm:zk}, we do not use extrapolation, i.e., \R{up:xacute}, 
since the corresponding majorization matrices are sharp, so one obtains tight majorization bounds in Lemma~\ref{l:QM}. 
See, for example, \cite[\S\Romnum{5}-B]{Chun&Fessler:20TIP:supp}.
For updates \R{eq:bpgm:zk} and \R{eq:bpgm:x}, we rewrite $\sum_{k=1}^K \| h_k \conv x - \zeta_k \|_2^2$ as $\| x - \sum_{k=1}^K \mathrm{flip}(h_k^*)  \conv \zeta_k \|_2^2$ 
by using the TF condition in \S\ref{sec:mbir:learn_reg} \cite[\S\Romnum{6}]{Chun&Fessler:20TIP:supp}, \cite{Chun&Fessler:18Asilomar:supp}.

\section{Empirical measures related to the convergence of Momentum-Net using sCNN refiners} \label{sec:momnet:assume:scnn}

This section provides empirical measures related to Assumption~4 for Momentum-Net using single-hidden layer autoencoders \R{sys:auto:res}; 
see Fig.~\ref{fig:momnet:assume:scnn} below. 
We estimated the sequence $\{ \epsilon^{(i)} \!:\! i \!=\! 2,\ldots,N_{\text{lyr}} \}$ in Definition~\ref{d:pair-map}, 
 the sequence $\{ \Delta^{(i)} \!:\! i \!=\! 2,\ldots,N_{\text{lyr}} \}$ in Definition~\ref{d:bcm}, 
 and the Lipschitz constants $\{ \kappa^{(i)} \!:\! i \!=\! 1,\ldots,N_{\text{lyr}} \}$ of refining NNs $\{ \cR_{\theta^{(i)}} \!:\! \forall i \}$, 
 based on a hundred sets of randomly selected training samples related with the corresponding bounds of the measures, 
 e.g., $u$ and $v$ in \R{cond:eps} are training input to $\cR_{\theta^{(i+1)}}$ and $\cR_{\theta^{(i)}}$ in \R{eq:momnet:map}, respectively.

 \begin{figure*}[!ht]
% \vspace{-0.5pc}
 \centering
 \small\addtolength{\tabcolsep}{-7.5pt}
 \renewcommand{\arraystretch}{1}

     \begin{tabular}{ccc}
     \multicolumn{3}{c}{\small (a) Sparse-view CT: Condition numbers of data-fit majorizers have \textit{mild} variations.} 
     \\
     \small{(a1) $\{ \Delta^{(i)} : i \geq 2 \}$} &
     \small{(a2) $\{ \epsilon^{(i)} : i \geq 2 \}$} &
     \small{(a3) $\{ \kappa^{(i)} : i \geq 1 \}$} 
     \\
     \includegraphics[scale=0.55, trim=0.2em 0.2em 1.1em 1em, clip]{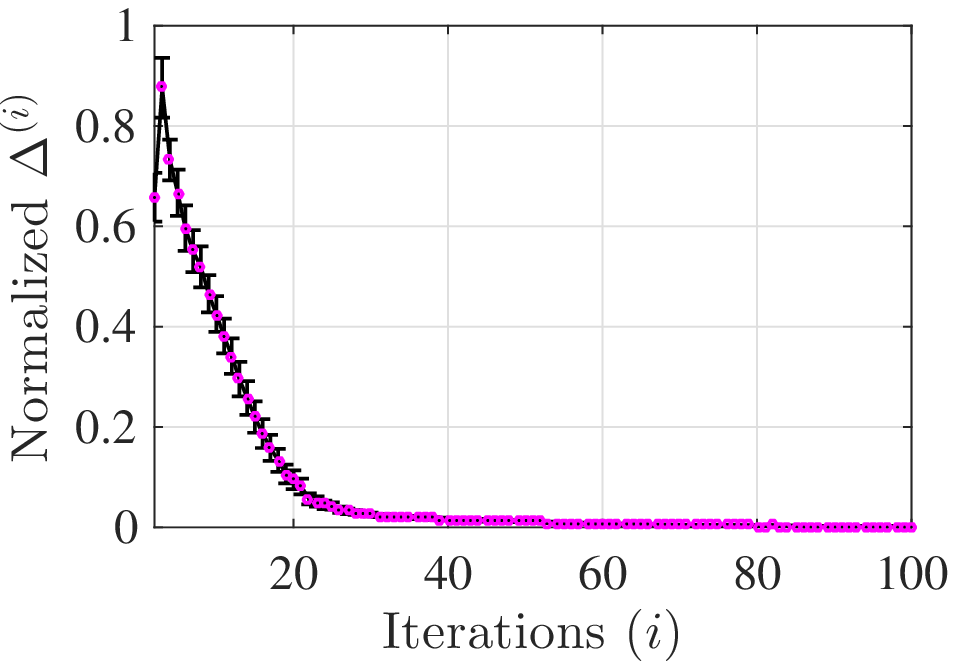} &
     \includegraphics[scale=0.55, trim=0.2em 0.2em 1.1em 1em, clip]{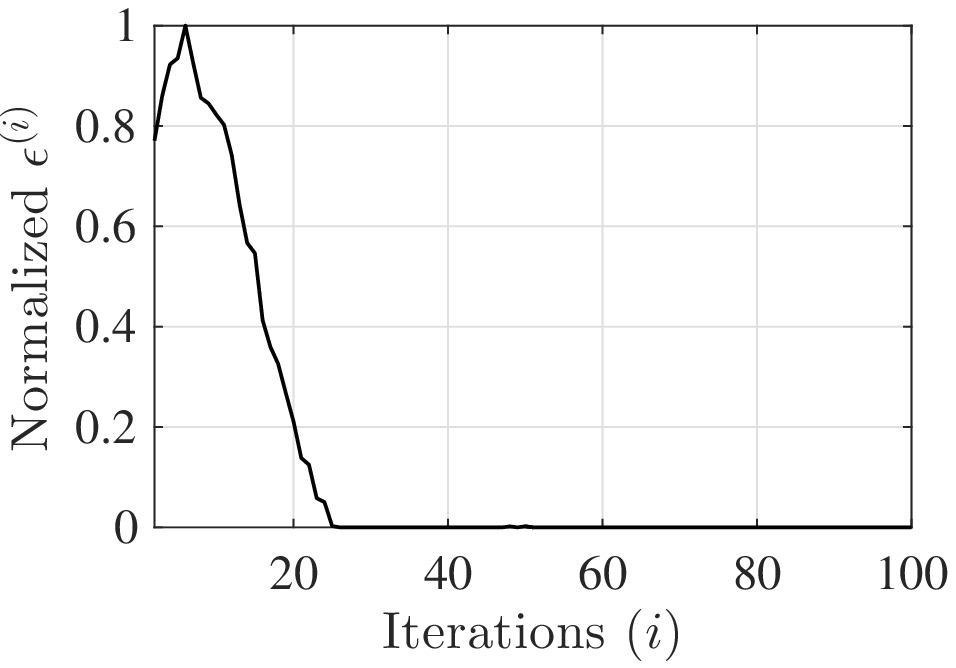} &
     \includegraphics[scale=0.55, trim=0.2em 0.2em 1.1em 1em, clip]{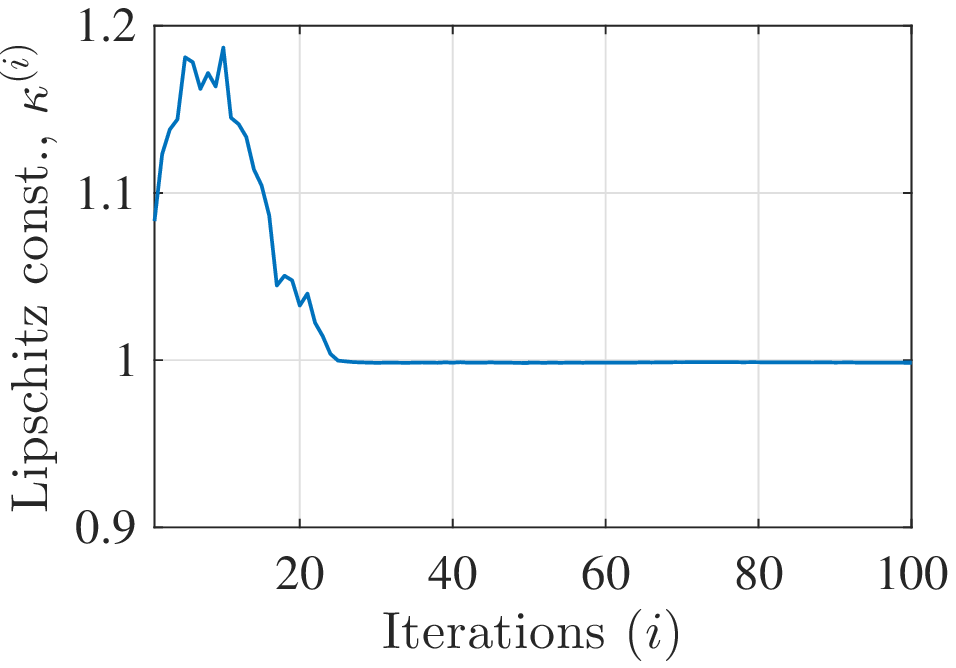} 
     \\
     \multicolumn{3}{c}{\small (b) LF photography using focal stack: Condition numbers of data-fit majorizers have \textit{strong} variations.}
     \\
     \small{(b1) $\{ \Delta^{(i)} : i \geq 2 \}$} &
     \small{(b2) $\{ \epsilon^{(i)} : i \geq 2 \}$} &
     \small{(b3) $\{ \kappa^{(i)} : i \geq 1 \}$} 
     \\
     \includegraphics[scale=0.55, trim=0.2em 0.2em 1.1em 1em, clip]{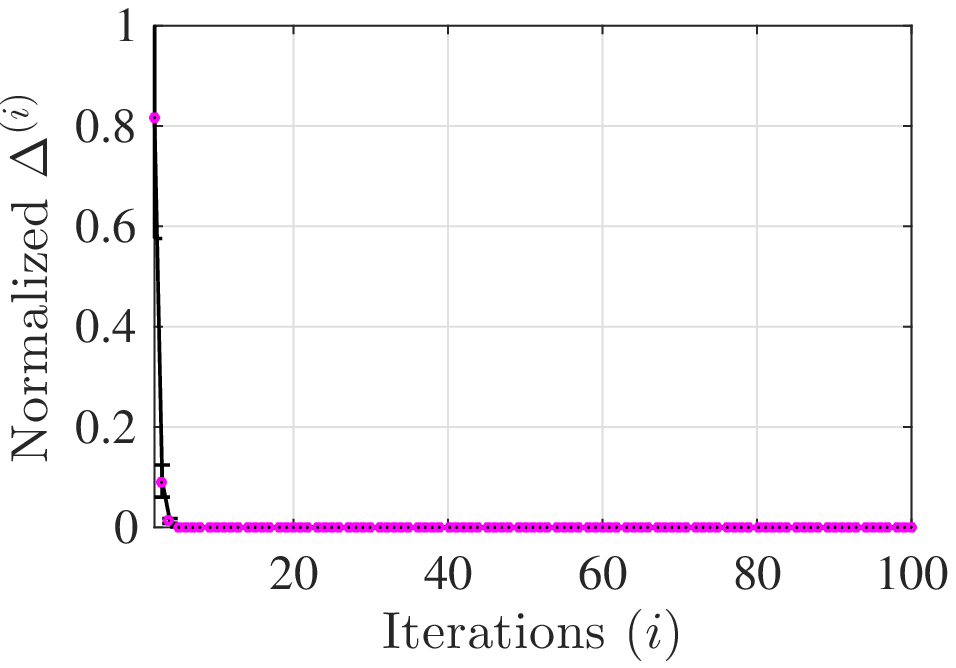} &
     \includegraphics[scale=0.55, trim=0.2em 0.2em 1.1em 1em, clip]{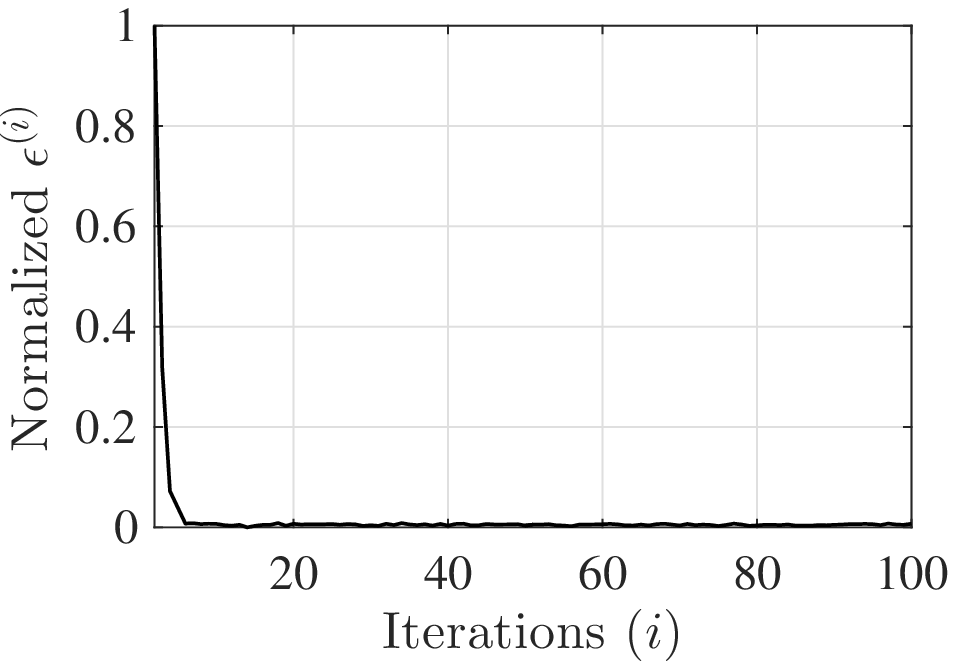} &
	\includegraphics[scale=0.55, trim=0.2em 0.2em 1.1em 1em, clip]{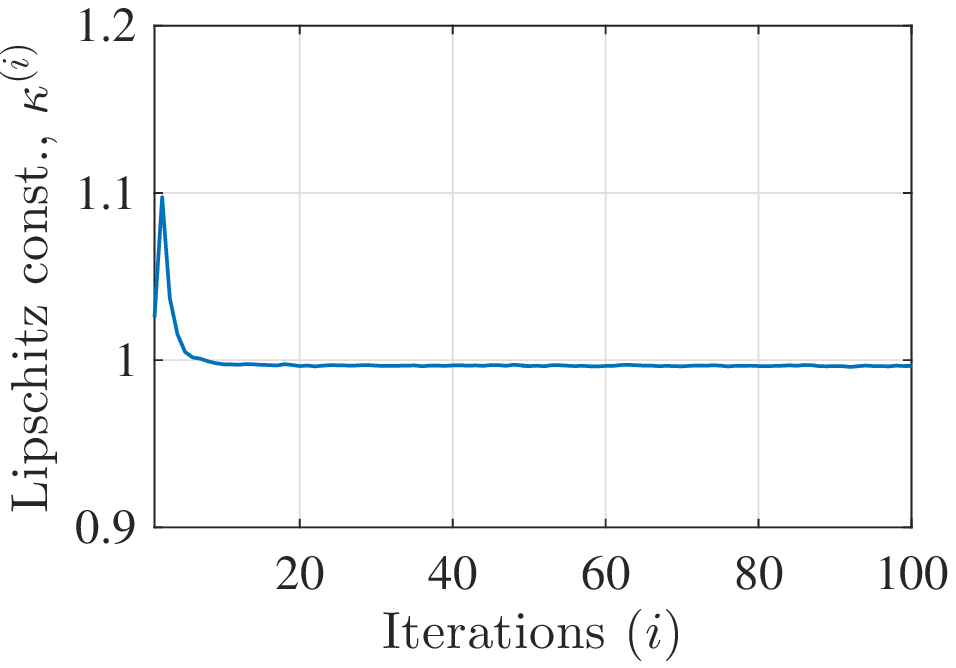}
     \end{tabular}
     
\vspace{-0.75em}
 \caption{
 Empirical measures related to Assumption~4 for guaranteeing convergence of Momentum-Net using sCNN refiners 
 (for details, see \R{sys:auto:res} and \S\ref{sec:exp:INN:param}), in different applications.
 (a)~The sparse-view CT reconstruction experiment used fan-beam geometry with $12.5$\% projections views. 
 (b)~The LF photography experiment used five detectors and reconstructed LFs consisting of $9 \!\times\! 9$ sub-aperture images.
 (a1,~b1)~For both the applications, we observed that $\Delta^{(i)} \rightarrow 0$. 
 This implies that the $z^{(i+1)}$-updates in \R{eq:momnet:map} satisfy the asymptotic block-coordinate minimizer condition in Assumption~4.
 (Magenta dots denote the mean values and black vertical error bars denote standard deviations.)
 (a2)~Momentum-Net trained from training data-fits, where their majorization matrices have \emph{mild} condition number variations,
 shows that $\epsilon^{(i)} \rightarrow 0$.
 This implies that paired NNs $( \cR_{\theta^{(i+1)}},  \cR_{\theta^{(i)}} )$ in \R{eq:momnet:map} are asymptotically nonexpansive.
 (b2)~Momentum-Net trained from training training data-fits, where their majorization matrices have \emph{mild} condition number variations, 
 shows that $\epsilon^{(i)}$ becomes close to zero, but does not converge to zero in one hundred iterations.
 (a3,~b3)~The NNs, $\cR_{\theta^{(i+1)}}$ in \R{eq:momnet:map}, become nonexpansive, 
 i.e., its Lipschitz constant $\kappa^{(i)}$ becomes less than $1$, as $i$ increases.
 }
 \label{fig:momnet:assume:scnn}
\vspace{-0.5pc}
 \end{figure*}

\section{Probabilistic justification for the asymptotic block-coordinate minimizer condition in Assumption~4} 
\label{sec:l:delta:prob}

This section introduces a useful result for an asymptotic block-coordinate minimizer $z^{(i+1)}$:
the following lemma provides a \emph{probabilistic} bound for $\| x^{(i)} - z^{(i+1)} \|_2^2$ in \R{cond:delta}, 
given a subgaussian vector $z^{(i+1)} - z^{(i)}$ with independent and zero-mean entries.

\lem{[Probabilistic bounds for $\| x^{(i)} - z^{(i+1)} \|_2^2$]
\label{l:delta:prob}
Assume that $z^{(i+1)} - z^{(i)}$ is a zero-mean subgaussian vector of which entries are independent and zero-mean subgaussian variables.
Then, each bound in \R{cond:delta} holds with probability at least 
\bes{
1 - \exp \!\! \left( \frac{ - \left( \| z^{(i+1)} - z^{(i)} \|_2^2 + \Delta^{(i+1)} \right)^2 }{8\rho \cdot \sigma^{(i+1)} \cdot  \|  \cR_{\theta^{(i+1)}} (x^{(i)}) - x^{(i)} \|_2^2} \right),
}
where $\sigma^{(i+1)}$ is a subgaussian parameter for $z^{(i+1)} - z^{(i)}$, 
and a random variable is subgaussian with parameter $\sigma$ if  $\bbP \{ | \cdot | \geq t \} \leq 2 \exp( - \frac{t^2}{2\sigma} )$ for $t \geq 0$.
}
\prf{
First, observe that
\begingroup
\allowdisplaybreaks
\ea{
\left\| x^{(i)} - z^{(i+1)} \right\|_2^2 
&= \left\| x^{(i)} - z^{(i)} - ( z^{(i+1)} - z^{(i)} ) \right\|_2^2
\nn \\
&=  \left\| x^{(i)} - z^{(i)} \right\|_2^2 + \left\| z^{(i+1)} - z^{(i)} \right\|_2^2 - 2 \ip{x^{(i)} - z^{(i)}}{z^{(i+1)} - z^{(i)}}
\nn \\
&=  \left\| x^{(i)} - z^{(i)} \right\|_2^2 + \left\| z^{(i+1)} - z^{(i)} \right\|_2^2 - 2 \ip{ z^{(i+1)} - z^{(i)} + \rho (x^{(i)} - \cR_{\theta^{(i+1)}} (x^{(i)}) ) }{ z^{(i+1)} - z^{(i)} }
\label{eq:l:delta:prob:1} 
\\
&=  \left\| x^{(i)} - z^{(i)} \right\|_2^2 - \left\| z^{(i+1)} - z^{(i)} \right\|_2^2 + 2 \rho \ip{ \cR_{\theta^{(i+1)}} (x^{(i)}) - x^{(i)} }{ z^{(i+1)} - z^{(i)} }
\label{eq:l:delta:prob:2} 
}
\endgroup
where the inequality \R{eq:l:delta:prob:1} holds by $x^{(i)} = \rho x^{(i)} - \rho \cR_{\theta^{(i+1)}} + z^{(i+1)}$ via \R{eq:momnet:map}.
We now obtain a probablistic bound for the third quantity in \R{eq:l:delta:prob:2} via a concentration inequality. 
The concentration inequality on the sum of independent zero-mean subgaussian variables (e.g., \cite[Thm.~7.27]{Foucart&Rauhut:book:supp}) yields that for any $t^{(i+1)} \geq 0$
\be{
\label{eq:l:delta:prob:3} 
\bbP \left\{  \ip{ \cR_{\theta^{(i+1)}} (x^{(i)}) - x^{(i)}}{ z^{(i+1)} - z^{(i)} } \geq t^{(i+1)} \right\}
\leq \exp \left( - \frac{ ( t^{(i+1)} )^2 }{ 2 \sigma^{(i+1)} \|  \cR_{\theta^{(i+1)}} (x^{(i)}) - x^{(i)} \|_2^2 } \right)
}
where $\sigma^{(i+1)}$ is given as in Lemma~\ref{l:delta:prob}.
Applying the result \R{eq:l:delta:prob:3} with $t^{(i+1)} = \frac{1}{2\rho} ( \| z^{(i+1)} - z^{(i)} \|_2^2 + \Delta^{(i+1)} )$ to the bound \R{eq:l:delta:prob:2} completes the proofs.
}

Lemma~\ref{l:delta:prob} implies that, 
given sufficiently large $\Delta^{(i+1)}$, 
or sufficiently small $\sigma^{(i+1)}$ (e.g., variance for a Gaussian random vector $z^{(i+1)} - z^{(i)}$) 
or $ \|  \cR_{\theta^{(i+1)}} (x^{(i)}) - x^{(i)} \|_2^2$, 
bound \R{cond:delta} is satisfied with high probability, for each $i$.
In particular, 
$\Delta^{(i+1)}$ can be large for the first several iterations;
if paired operators $( \cR_{\theta^{(i+1)}}, \cR_{\theta^{(i)}} )$ in \R{eq:momnet:map} map their input images to similar output images 
(e.g., the trained NNs $\cR_{\theta^{(i+1)}}$ and $\cR_{\theta^{(i)}}$ have good refining capabilities for $x^{(i)}$ and $x^{(i-1)}$), 
then $\sigma^{(i+1)}$ is small;
if the regularization parameter $\gamma$ in \R{eq:momnet:mbir} is sufficiently large, then $ \|  \cR_{\theta^{(i+1)}} (x^{(i)}) - x^{(i)} \|_2^2$ is small.

\section{Proofs of Proposition~\ref{p:momnet:sum}}
\label{sec:p:momnet:sum}

First, we show that $\sum_{i=0}^{\infty} \| x^{(i+1)} - x^{(i)} \|_2^2 < \infty$ for convex and nonconvex $F(x;y,z^{(i+1)})$ cases. 
\bulls{

\item \underline{\emph{Convex}~$F(x;y,z^{(i+1)})$ case}:
Using Assumption 2 and $\{ \widetilde{M}^{(i+1)} \!=\! M^{(i+1)} \!:\! \forall i \}$ for the convex case via \R{up:MFtilde}, we obtain the following results for any $\cX$:
%\begingroup
%\allowdisplaybreaks
\ea{
&~ F \Big( x^{(i)}; y,  z^{(i)} \Big) - F \Big( x^{(i+1)}; y,  z^{(i+1)} \Big) + \gamma \Delta^{(i+1)}
\nn \\
&\geq F \Big( x^{(i)}; y,  z^{(i+1)} \Big) - F \Big( x^{(i+1)}; y,  z^{(i+1)} \Big) 
\label{eq:p:momnet-convg-cnvx:1}
\\
&\geq \frac{1}{2} \nm{ x^{(i+1)} - \acute{x}^{(i+1)}  }_{M^{(i+1)}}^2 + \left( \acute{x}^{(i+1)} - x^{(i)} \right)^T M^{(i+1)} \left( x^{(i+1)} - \acute{x}^{(i+1)} \right)
\label{eq:p:momnet-convg-cnvx:2} 
\\
& = \frac{1}{2} \nm{ x^{(i+1)} -  x^{(i)} }_{M^{(i+1)}}^2 - \frac{1}{2} \nm{ E^{(i+1)} \left( x^{(i)} -  x^{(i-1)} \right) }_{M^{(i+1)}}^2
\label{eq:p:momnet-convg-cnvx:3} 
\\
&\geq \frac{1}{2} \nm{ x^{(i+1)} -  x^{(i)} }_{M^{(i+1)}}^2 - \frac{\delta^2}{2} \nm{ x^{(i)} -  x^{(i-1)} }_{M^{(i)}}^2  
\label{eq:p:momnet-convg-cnvx:4}
}
%\endgroup
where the inequality \R{eq:p:momnet-convg-cnvx:1} uses the condition \R{cond:delta} in Assumption~4,
the inequality \R{eq:p:momnet-convg-cnvx:2} is obtained by using the results in \cite[Lem.~S.1]{Chun&Fessler:18TIP:supp},
the equality \R{eq:p:momnet-convg-cnvx:3} uses the extrapolation formula \R{eq:momnet:exp} and the symmetry of $M^{(i+1)}$,
the inequality \R{eq:p:momnet-convg-cnvx:4} holds by Assumption~3.

Summing the inequality of $F(x^{(i)}; y,z^{(i)}) - F(x^{(i+1)}; y, z^{(i+1)}) + \gamma \Delta^{(i+1)}$ in \R{eq:p:momnet-convg-cnvx:4} over $i=0,\ldots,N_{\text{lyr}}-1$, we obtain
\begingroup
\setlength{\thinmuskip}{1.5mu}
\setlength{\medmuskip}{2mu plus 1mu minus 2mu}
\setlength{\thickmuskip}{2.5mu plus 2.5mu}
\allowdisplaybreaks
\ea{
F \Big( x^{(0)}; y, z^{(0)} \Big)  - F \Big( x^{(N_{\text{lyr}})}; y, z^{(N_{\text{lyr}})} \Big) 
+ \gamma \sum_{i=0}^{N_{\text{lyr}}-1} \Delta^{(i+1)}
&\geq \sum_{i=0}^{N_{\text{lyr}}-1} \frac{1}{2} \nm{ x^{(i+1)} -  x^{(i)} }_{M^{(i+1)}}^2 - \frac{\delta^2}{2} \nm{ x^{(i)} -  x^{(i-1)} }_{M^{(i)}}^2 
\nn \\
&\geq \sum_{i=0}^{N_{\text{lyr}}-1} \frac{1-\delta^2}{2}  \nm{ x^{(i+1)} -  x^{(i)} }_{M^{(i+1)}}^2 
\nn \\
&\geq \sum_{i=0}^{N_{\text{lyr}}-1} \frac{m_{F,\min} (1-\delta^2)}{2} \nm{ x^{(i+1)} -  x^{(i)} }_2^2 
\label{eq:p:momnet-convg-cnvx:7}
}
\endgroup
where the inequality \R{eq:p:momnet-convg-cnvx:7} holds by Assumption 2.
Due to the lower boundedness of $F(x;y,z)$ in Assumption~1 and the summability of $\{ \Delta^{(i+1)} \geq 0 : \forall i \}$ in Assumption~4,
taking $N_{\text{lyr}} \rightarrow \infty$ gives
\be{
\label{eq:p:momnet:convg:sqSum:u}
\sum_{i=0}^{\infty} \nm{ x^{(i+1)} - x^{(i)}}_2^2 < \infty.
}

\item \underline{\emph{Nonconvex}~$F(x;y,z^{(i+1)})$ case}:
Using Assumption 2, we obtain the following results without assuming that $F(x;y,z^{(i+1)})$ is convex:
\begingroup
\allowdisplaybreaks
\ea{
&~  F \Big( x^{(i)}; y,  z^{(i)} \Big) - F \Big( x^{(i+1)}; y,  z^{(i+1)} \Big) + \gamma \Delta^{(i+1)}
\nn \\
&\geq F \Big( x^{(i)}; y,  z^{(i+1)} \Big) - F \Big( x^{(i+1)}; y,  z^{(i+1)} \Big) 
\label{eq:p:momnet-convg-ncnvx:0} 
\\
&\geq \frac{\lambda - 1}{4} \nm{ x^{(i+1)} - x^{(i)} }_{M^{(i+1)}}^2 - \frac{ (\lambda + 1)^2 }{\lambda - 1} \nm{ x^{(i)} - \acute{x}_b^{(i+1)} }_{M^{(i+1)}}^2 
\label{eq:p:momnet-convg-ncnvx:1} 
\\
&= \frac{\lambda-1}{4} \nm{ x^{(i+1)} -  x^{(i)} }_{M^{(i+1)}}^2 - \frac{(\lambda+1)^2}{\lambda-1} \nm{ E^{(i+1)} \left( x^{(i)} -  x^{(i-1)} \right) }_{M^{(i+1)}}^2
\label{eq:p:momnet-convg-ncnvx:2} 
\\
&\geq \frac{\lambda-1}{4} \left( \nm{ x^{(i+1)} -  x^{(i)} }_{M^{(i+1)}}^2 - \delta^2 \nm{ x^{(i)} -  x^{(i-1)} }_{M^{(i)}}^2 \right)
\label{eq:p:momnet-convg-ncnvx:3} 
}
\endgroup
where the inequality \R{eq:p:momnet-convg-ncnvx:0} uses the condition \R{cond:delta} in Assumption~4,
the inequality \R{eq:p:momnet-convg-ncnvx:1} use the results in \cite[\S S.3]{Chun&Fessler:20TIP:supp},
the equality \R{eq:p:momnet-convg-ncnvx:2} holds by \R{eq:momnet:exp},
the inequality \R{eq:p:momnet-convg-ncnvx:3} is obtained by Assumption~3.

Summing the inequality of $F(x^{(i)}; y,z^{(i)}) - F(x^{(i+1)}; y, z^{(i+1)}) + \gamma \Delta^{(i+1)}$ in \R{eq:p:momnet-convg-ncnvx:3} over $i=0,\ldots,N_{\text{lyr}}-1$, we obtain
\begingroup
\setlength{\thinmuskip}{1.5mu}
\setlength{\medmuskip}{2mu plus 1mu minus 2mu}
\setlength{\thickmuskip}{2.5mu plus 2.5mu}
\allowdisplaybreaks
\eas{
F \Big( x^{(0)}; y, z^{(0)} \Big)  - F \Big( x^{(N_{\text{lyr}})}; y, z^{(N_{\text{lyr}})} \Big) 
+ \gamma \cdot \sum_{i=0}^{N_{\text{lyr}}-1} \Delta^{(i+1)} 
&\geq \sum_{i=0}^{N_{\text{lyr}}-1} \frac{\lambda-1}{4} \left( \nm{ x^{(i+1)} -  x^{(i)} }_{M^{(i+1)}}^2 - \delta^2 \nm{ x^{(i)} -  x^{(i-1)} }_{M^{(i)}}^2 \right)
\nn \\
&\geq \sum_{i=0}^{N_{\text{lyr}}-1} \frac{(\lambda-1)(1-\delta^2)}{2} \nm{ x^{(i+1)} -  x^{(i)} }_{M^{(i+1)}}^2
\nn \\
&\geq \sum_{i=0}^{N_{\text{lyr}}-1} \frac{m_{F,\min} (\lambda-1) (1-\delta^2)}{2} \nm{ x^{(i+1)} - x^{(i)} }_2^2,
}
\endgroup
where we follow the arguments in obtaining \R{eq:p:momnet-convg-cnvx:7} above.
Again, using the lower boundedness of $F(x;y,z)$ and the summability of $\{ \Delta^{(i+1)} \geq 0 : \forall i \}$,
taking $N_{\text{lyr}} \rightarrow \infty$ gives the result \R{eq:p:momnet:convg:sqSum:u} for nonconvex $F(x;y,z^{(i+1)})$. 
}

Second, we show that $\sum_{i=0}^{\infty} \| z^{(i+1)} - z^{(i)} \|_2^2 < \infty$. Observe
\ea{
\nm{z^{(i+1)} - z^{(i)}}_2^2
&= \left\| (1-\rho) \Big( x^{(i)} - x^{(i-1)} \Big) + \rho \Big( \cR_{\theta^{(i+1)}} (x^{(i)}) - \cR_{\theta^{(i)}} (x^{(i-1)}) \Big) \right\|_2^2
\nn \\
&\leq (1-\rho) \nm{ x^{(i)} - x^{(i-1)} }_2^2 + \rho \nm{ \cR_{\theta^{(i+1)}} (x^{(i)}) - \cR_{\theta^{(i)}} (x^{(i-1)}) }_2^2
\nn \\
&\leq \nm{ x^{(i)} - x^{(i-1)} }_2^2 + \rho \epsilon^{(i+1)}
\label{eq:p:momnet-convg:mapdiff}
}
where the first equality uses the image mapping formula in \R{eq:momnet:map},
the first inequality holds by applying Jensen's inequality to the (convex) squared $\ell^2$-norm,
the second inequality is obtained by using the asymptotically non-expansiveness of the paired operators $( \cR_{\theta^{(i+1)}},  \cR_{\theta^{(i)}} )$ in Assumption~4. 
Summing the inequality of $\| z^{(i+1)} - z^{(i)} \|_2^2$ in \R{eq:p:momnet-convg:mapdiff} over $i=0,\ldots,N_{\text{lyr}}-1$, we obtain
\ea{
\label{eq:p:momnet:sqSum:u0}
\sum_{i=0}^{N_{\text{lyr}}-1}  \nm{z^{(i+1)} - z^{(i)}}_2^2 
&\leq \sum_{i=0}^{N_{\text{lyr}}-2} \nm{ x^{(i+1)} - x^{(i)} }_2^2 + \rho \sum_{i=0}^{N_{\text{lyr}}-1} \epsilon^{(i+1)},
}
where we used $x^{(0)} = x^{(-1)}$ as given in Algorithm~\ref{alg:momnet}.
By taking $N_{\text{lyr}} \rightarrow \infty$ in \R{eq:p:momnet:sqSum:u0}, using result \R{eq:p:momnet:convg:sqSum:u}, and the summability of the sequence $\{ \epsilon^{(i+1)} : i \geq 0 \}$, we obtain
\be{
\label{eq:p:momnet:convg:sqSum:u0}
\sum_{i=0}^{\infty} \nm{ z^{(i+1)} - z^{(i)}}_2^2 < \infty.
}
Combining the results in \R{eq:p:momnet:convg:sqSum:u} and \R{eq:p:momnet:convg:sqSum:u0} completes the proofs.

\section{Proofs of Theorem~\ref{t:momnet:convg}}
\label{sec:t:momnet:convg}

Let $\bar{x}$ be a limit point of $\{ x^{(i)} : i \geq 0 \}$ and $\{ x^{(i_j)} \}$ be the subsequence converging to $\bar{x}$.
Let $\bar{z}$ be a limit point of $\{ z^{(i)} : i \geq 0 \}$ and $\{ z^{(i_j)} \}$ be the subsequence converging to $\bar{z}$.
The closedness of $\cX$ implies that $\bar{x} \in \cX$.
Using the results in Proposition~\ref{p:momnet:sum}, 
$\{ x^{(i_j+1)} \}$ and $\{ z^{(i_j+1)} \}$ also converge to $\bar{x}$ and $\bar{z}$, respectively.
Taking another subsequence if necessary, the subsequence $\{ M^{(i_j+1)} \}$ converges to some $\bar{M}$, since $M^{(i+1)}$ is bounded by Assumption~2.
The subsequences $\{ \theta^{(i_j+1)} \}$ converge to some $\bar{\theta}$, since $x^{(i_j+1)} \rightarrow \bar{x}$, $z^{(i_j+1)} \rightarrow \bar{z}$, and $\{ \theta^{(i+1)} \}$ is bounded via Assumption~4.

Next, we show that the convex proximal minimization \R{eq:t:momnet:convg:cvx:1} below is continuous in the sense that the output point $x^{(i_j+1)}$ continuously depends on the input points $\acute{x}^{(i_j+1)}$ and $z^{(i_j+1)}$, and majorization matrix $\widetilde{M}^{(i_j+1)}$:
\ea{
x^{(i_j+1)} &= \argmin_{ x \in \cX } \, \ip{\nabla F(\acute{x}^{(i_j+1)}; y, z^{(i_j+1)})}{x - \acute{x}^{(i_j+1)}} + \frac{1}{2} \nm{x - \acute{x}^{(i_j+1)}}_{\widetilde{M}^{(i_j+1)}}^2 
\label{eq:t:momnet:convg:cvx:1}
\\
&= \mathrm{Prox}_{\bbI_{\cX}}^{\widetilde{M}^{(i_j+1)}} \! \Big( \acute{x}^{(i_j+1)} - \big( \widetilde{M}^{(i_j+1)} \big)^{-1} \nabla F (\acute{x}^{(i_j+1)}; y, z^{(i_j+1)}) \Big).
\nn
%\label{eq:t:momnet:convg:cvx:2}
}
where the proximal mapping operator $\mathrm{Prox}_{\bbI_{\cX}}^{\widetilde{M}^{(i_j+1)}} \! (\cdot)$ is given as in \R{eq:d:prox}.
We consider the two cases of majorization matrices $\{ M^{(i+1)} \}$ given in Theorem~\ref{t:momnet:convg}:
\bulls{
\item \underline{For a sequence of diagonal majorization matrices, i.e., $\{ M^{(i+1)} : i \geq 0 \}$},
one can obtain the continuity of the convex proximal minimization \R{eq:t:momnet:convg:cvx:1} with respect to $\acute{x}^{(i_j+1)}$, $z^{(i_j+1)}$, and $\widetilde{M}^{(i_j+1)}$, 
by extending the existing results in \cite[Thm.~2.26]{Rockafellar&Wets:book:supp}, \cite{Rockafellar:76SIAMCO:supp} with the separability of \R{eq:t:momnet:convg:cvx:1} to element-wise optimization problems.

\item \underline{For a fixed general majorization matrix, i.e., $M = M^{(i+1)}$, $\forall i$}, we obtain that the convex proximal minimization \R{eq:t:momnet:convg:cvx:M:1} below is continuous with respect to the input points $\acute{x}^{(i_j+1)}$ and $z^{(i_j+1)}$:
\ea{
x^{(i_j+1)} 
&= \argmin_{ x \in \cX } \, \ip{\nabla F(\acute{x}^{(i_j+1)}; y, z^{(i_j+1)})}{x - \acute{x}^{(i_j+1)}} + \frac{1}{2} \nm{x - \acute{x}^{(i_j+1)}}_{\widetilde{M}}^2
\label{eq:t:momnet:convg:cvx:M:1} 
\\ 
&= \mathrm{Prox}_{\bbI_{\cX}}^{\widetilde{M}}  \big( \acute{x}^{(i_j+1)} - \widetilde{M}^{-1} \nabla F (\acute{x}^{(i_j+1)}; y, z^{(i_j+1)}) \big)
\nn
\\
&= \big( \mathrm{Id} + \widetilde{M}^{-1} \hat{\partial} \bbI_{\cX} \big)^{-1} \big( \acute{x}^{(i_j+1)} - \widetilde{M}^{-1} \nabla F (\acute{x}^{(i_j+1)}; y, z^{(i_j+1)}) \big)
\label{eq:t:momnet:convg:cvx:M:2}
}
where $\hat{\partial} f(x)$ is the subdifferential of $f$ at $x$ and $\mathrm{Id}$ denotes the identity operator, 
and the proximal mapping of $\bbI_{\cX}$ relative to $\| \cdot \|_{\widetilde{M}}$ is uniquely determined by 
the resolvent of the operator $\widetilde{M}^{-1} \hat{\partial} \bbI_{\cX}$ in \R{eq:t:momnet:convg:cvx:M:2}.

First, we obtain that the operator $\widetilde{M}^{-1} \hat{\partial} \bbI_{\cX}$ is monotone.
For a convex extended-valued function $f_e: \bbR^N \rightarrow \bbR \cup \{ \infty \}$, observe that $\widetilde{M}^{-1} \hat{\partial} f_e$ is a monotone operator:
\be{
\ip{ \widetilde{M}^{-1} \hat{\partial} f_e (u) - \widetilde{M}^{-1} \hat{\partial} f_e (v) }{ u - v }
= \ip{ \underbrace{ \widetilde{M}^{-1} \widetilde{M} }_{= I} \hat{\partial} f_e (\widetilde{M} \tilde{u}) - \underbrace{ \widetilde{M}^{-1} \widetilde{M} }_{=I} \hat{\partial} f_e (\widetilde{M} \tilde{v}) }{ \widetilde{M} \tilde{u} - \widetilde{M} \tilde{v} } 
\geq 0, \qquad \forall u, v,
\label{eq:t:momnet:convg:monotone}
}
where the equality uses the variable change $\{ u = \widetilde{M} \tilde{u}, v = \widetilde{M} \tilde{v} \}$,  
a chain rule of the subdifferential of a composition of a convex extended-valued function 
and an affine mapping \cite[\S7]{Mordukhovich&Nam:17Opt:supp}, and the symmetry of $\widetilde{M}$,
and the inequality holds because the subdifferential of convex extended-valued function is a monotone operator \cite[\S4.2]{Ryu&Boyd:16ACM:supp}.
Because characteristic function of a convex set is extended-valued function, 
the result in \R{eq:t:momnet:convg:monotone} implies that the operator $\widetilde{M}^{-1} \hat{\partial} \bbI_{\cX}$ is monotone.
Second, note that the resolvent of a monotone operator $\widetilde{M}^{-1} \hat{\partial} \bbI_{\cX}$ (with a parameter $1$), i.e., 
$( \mathrm{Id} + \widetilde{M}^{-1} \hat{\partial} \bbI_{\cX} )^{-1}$ in \R{eq:t:momnet:convg:cvx:M:2}, 
is nonexpansive \cite[\S6]{Rockafellar&Wets:book:supp} and thus continuous.
We now obtain that the convex proximal minimization \R{eq:t:momnet:convg:cvx:M:1} is continuous with respect to the input points $\acute{x}^{(i_j+1)}$ and $z^{(i_j+1)}$, 
because the proximal mapping operator $( \mathrm{Id} + \widetilde{M}^{-1} \hat{\partial} \bbI_{\cX} )^{-1}$ in \R{eq:t:momnet:convg:cvx:M:2}, the affine mapping $\widetilde{M}^{-1}$, and $\nabla F(x;y,z)$ are continuous with respect to their input points.
}

For the two cases above, 
using the fact that $x^{(i_j+1)} \rightarrow \bar{x}$, $\acute{x}^{(i_j+1)} \rightarrow \bar{x}$, $z^{(i_j+1)} \rightarrow \bar{z}$, and $M^{(i_j+1)} \rightarrow \bar{M}$ (or $\bar{M} = M$ for the $\{ M^{(i+1)} = M \}$ case) as $j \rightarrow \infty$, \R{eq:t:momnet:convg:cvx:1} becomes
\be{
%\label{eq:t:momnet:convg:cvx:3}
\bar{x} = \argmin_{ x \in \cX } \, \ip{\nabla F(\bar{x}; y,\bar{z})}{x - \bar{x}} + \frac{1}{2} \nm{x - \bar{x}}_{\bar{M}}^2 .
}
Thus, $\bar{x}$ satisfies the first-order optimality condition of $\min_{x \in \cX} F(x;y,\bar{z})$:
\bes{
\ip{\nabla F (\bar{x};y, \bar{z}) }{ x - \bar{x} } \geq 0, \qquad \mbox{for any $x \in \cX$},
}
and this completes the proof of the first result.

Next, note that the result in Proposition~\ref{p:momnet:sum} imply
\be{
\label{t:momnet:convg:cvx:convg_to0}
\nm{ \cA_{\cR_{\theta^{(i+1)}}}^{M^{(i+1)}} \!\! \left( \left[ \arraycolsep=1.5pt \begin{array}{c} x^{(i)} \\ x^{(i-1)} \end{array} \right] \right) - \left[ \arraycolsep=1.5pt \begin{array}{c} x^{(i)} \\ x^{(i-1)} \end{array} \right] }_2 \rightarrow 0.
}
Additionally, note that a function $\cA_{\cR_{\theta^{(i+1)}}}^{M^{(i+1)}} \!\! - I$ is continuous.
To see this, observe that the convex proximal mapping in \R{eq:momnet:mbir} is continuous (see the obtained results above), and $\cR_{\theta^{(i+1)}}$ is continuous (see Assumption~4).
Combining \R{t:momnet:convg:cvx:convg_to0}, 
the convergence of $\{ M^{(i_j+1)}, \cR_{\theta^{(i_j+1)}} \}$, 
and the continuity of $\cA_{\cR_{\theta^{(i+1)}}}^{M^{(i+1)}} \!\! - I$, 
we obtain $[\bar{x}^T \!, \bar{x}^T]^T \!=\! \cA_{\cR_{\bar{\theta}}}^{\bar{M}} ( [\bar{x}^T \!, \bar{x}^T]^T )$, and this completes the proofs of the second result.

\section{Proofs of Corollary~\ref{c:momnet:convg:whole}}
\label{sec:c:momnet:convg:whole}

To prove the first result, we use proof by contradiction. 
Suppose that $\mathrm{dist} (x^{(i)}, \cS) \nrightarrow 0$.
Then there exists $\epsilon > 0$ and a subsequence $\{ x^{(i_j)} \}$ such that $\mathrm{dist} (x^{(i_j)}, \cS) \geq \epsilon$, $\forall j$.
However, the boundedness assumption of $\{ x^{(i_j)} \}$ in Corollary~\ref{c:momnet:convg:whole} implies that there must exist a limit point $\bar{x} \in \cS$ via Theorem~\ref{t:momnet:convg}.
This is a contradiction, and gives the first result (via the result in Proposition~\ref{p:momnet:sum}).
Under the isolation point assumption in Corollary~\ref{c:momnet:convg:whole}, using the obtained results, $\| x^{(i+1)} - x^{(i)} \|_2 \rightarrow 0$ (via Proposition~\ref{p:momnet:sum}) and $\mathrm{dist} (x^{(i+1)}, \cS) \rightarrow 0$, and the following the proofs in \cite[Cor.~2.4]{Xu&Yin:13SIAM:supp}, we obtain the second result.

\section{Momentum-Net vs.~BCD-Net}
\label{sec:momnet-vs-bcdnet}

This section compares the convergence properties of Momentum-Net (Algorithm~\ref{alg:momnet})  
and BCD-Net (Algorithm~\ref{alg:bcdnet}).
We first show that for convex $f(x;y)$ and $\cX$, 
the sequence of reconstructed images generated by BCD-Net converges:

\prop{[Sequence convergence]
\label{p:bcdnet:cauchy}
In Algorithm~\ref{alg:bcdnet}, let $f(x;y)$ be convex and subdifferentiable, and $\cX$ be convex.
Assume that the paired operators $( \cR_{\theta^{(i+1)}},  \cR_{\theta^{(i)}} )$ are asymptotically contractive, i.e.,
\bes{
\left\|   \cR_{\theta^{(i+1)}} (u) -  \cR_{\theta^{(i)}}  (v) \right\|_2 <  \nm{u - v}_2 + \epsilon^{(i+1)},
}
with $\sum_{i=0}^{\infty} \epsilon^{(i+1)} < \infty$ and $\{ \epsilon^{(i+i)} \in [0, \infty) : \forall i \}$, $\forall u,v, i$.
Then, the sequence $\{ x^{(i+1)} : i \geq 0 \}$ generated by Algorithm~\ref{alg:bcdnet} is convergent.
}
\prf{
We rewrite the updates in Algorithm~\ref{alg:bcdnet} as follows:
\eas{
x^{(i+1)} 
& = \argmin_{x \in \cX} ~ f(x;y) + \frac{\gamma}{2} \nm{ x -  \cR_{\theta^{(i+1)}} ( x^{(i)} ) }_2^2 
= \mathrm{Prox}_{f + \bbI_{\cX}}^{\gamma I} \! \big(  \cR_{\theta^{(i+1)}} ( x^{(i)} ) \big)
\\
&=  \big( \mathrm{Id} + \gamma^{-1} \hat{\partial} ( f(x;y) + \bbI_{\cX} ) \big)^{-1} \big(  \cR_{\theta^{(i+1)}} ( x^{(i)} ) \big)
\\
& =: \cA^{(i+1)} (x^{(i)}).
}
We first show that the paired operators $\{ \cA^{(i+1)}, \cA^{(i)} \}$ is asymptotically contractive:
\ea{
&~ \nm{\cA^{(i+1)} (u) - \cA^{(i)} (v) }_2
\nn \\
&= \left\| \big( \mathrm{Id} + \gamma^{-1} \hat{\partial} ( f(x;y) + \bbI_{\cX} ) \big)^{-1} \big(  \cR_{\theta^{(i+1)}} ( u ) \big) - \big( \mathrm{Id} + \gamma^{-1} \hat{\partial} ( f(x;y) + \bbI_{\cX} ) \big)^{-1} \big(  \cR_{\theta^{(i)}} ( v ) \big) \right\|_2 
\nn \\
&\leq \nm{  \cR_{\theta^{(i+1)}} ( u ) -  \cR_{\theta^{(i)}} ( v ) }_2
\label{eq:p:bcdnet:nonexp}
\\
&\leq L' \nm{ u - v }_2 + \epsilon^{(i+1)} \nm{ u - v }_2,
\label{eq:p:bcdnet:lipschitz}
}
$\forall u,v$, 
where the inequality \R{eq:p:bcdnet:nonexp} holds because 
the subdifferential of the convex extended-valued function $f(x;y) + \bbI_{\cX}$
(the characteristic function of a convex set $\cX$, $\bbI_{\cX}$, is convex, and the sum of the two convex functions, $f(x;y) + \bbI_{\cX}$, is convex) 
is a monotone operator \cite[\S4.2]{Ryu&Boyd:16ACM:supp}, 
and the resolvent of a monotone relation with a positive parameter, 
i.e., $( \mathrm{Id} + \gamma^{-1} \hat{\partial} ( f(x;y) + \bbI_{\cX} ) )^{-1}$ with $\gamma^{-1} > 0$,
is nonexpansive \cite[\S6]{Ryu&Boyd:16ACM:supp},
and the inequality \R{eq:p:bcdnet:lipschitz} holds by $L' < 1$ via the contractiveness of the paired operators $( \cR_{\theta^{(i+1)}},  \cR_{\theta^{(i)}} )$, $\forall i$.
Note that the inequality \R{eq:p:bcdnet:nonexp} does not hold for nonconvex $f(x;y)$ and/or $\cX$.
Considering that $L' < 1$, we show that the sequence $\{ x^{(i+1)} : i \geq 0 \}$ is Cauchy sequence:
\begingroup
\allowdisplaybreaks
\eas{
\nm{ x^{(i+l)} - x^{(i)} }_2 
&= \nm{ ( x^{(i+l)} - x^{(i+l-1)}  ) + \ldots + ( x^{(i+1)} - x^{(i)} ) }_2
\\
&\leq \nm{ x^{(i+l)} - x^{(i+l-1)}  } + \ldots + \nm{ x^{(i+1)} - x^{(i)} }_2
\\
&\leq \big( {L'}^{l-1} + \ldots + 1 \big) \nm{ x^{(i+1)} - x^{(i)}  }_2 + \big( \epsilon^{(i+l)} + \ldots + \epsilon^{(i+1)} \big)
\\
&\leq \frac{1}{1 - {L'}} \nm{ x^{(i+1)} - x^{(i)}  }_2 + \sum_{i'=1}^{l} \epsilon^{(i+i')}
}
\endgroup
where the second inequality uses the result in \R{eq:p:bcdnet:lipschitz}. 
Since the sequence $\{ x^{(i+1)} : i \geq 0 \}$ is Cauchy sequence, $\{ x^{(i+1)} : i \geq 0 \}$ is convergent, and this completes the proofs.
}

In terms of guaranteeing convergence, 
BCD-Net has three theoretical or practical limitations compared to Momentum-Net:

\bulls{
%[\setlength{\topsep}{1pt}]

\item Different from Momentum-Net, BCD-Net assumes the asymptotic contractive condition for the paired operators $\{  \cR_{\theta^{(i+1)}},  \cR_{\theta^{(i)}} \}$. 
When image mapping operators in \R{eq:bcdnet:map} are identical across iterations, i.e., $\{ \cR_{\theta} = \cR_{\theta^{(i+1)}} : i \!\geq\! 0 \}$, then $\cR_{\theta}$ is assumed to be contractive.
On the other hand, a mapping operator (identical across iterations) of Momentum-Net only needs to be nonexpansive.
Note, however, that when $f(x;y) = \frac{1}{2} \| y - Ax \|_W^2$ with $A^H W A \succ 0$ (e.g., Example~\ref{eg:bcdnet:denoising}), 
BCD-Net can guarantee the sequence convergence with the asymptotically nonexpansive paired operators $( \cR_{\theta^{(i+1)}},  \cR_{\theta^{(i)}} )$ 
(see Definition~\ref{d:pair-map}) \cite{Chun&etal:19MICCAI:supp}.

\item When one applies an iterative solver to \R{eq:bcdnet:recon}, 
there always exist some numerical errors and these obstruct the sequence convergence guarantee in Proposition~\ref{p:bcdnet:cauchy}. To guarantee sufficiently small numerical errors from iterative methods solving \R{eq:bcdnet:recon} (so that one can find a critical point solution for the MBIR problem \R{eq:bcdnet:recon}), one needs to use sufficiently many inner iterations that can substantially slow down entire MBIR.

\item BCD-Net does not guarantee the sequence convergence for nonconvex data-fit $f(x;y)$, whereas Momentum-Net guarantees convergence to a fixed-point for both convex $f(x;y)$ and nonconvex $f(x;y)$.
}

\section{For the sCNN architecture \R{sys:auto:res}, connection between convolutional training loss \R{sys:auto:res:train} and its patch-based training loss} \label{sec:l:train}

This section shows that given the sCNN architecture \R{sys:auto:res},
the convolutional training loss in \R{sys:auto:res:train} has three advantages over 
the patch-based training loss in \cite{Chun&Fessler:18IVMSP:supp, Chun&etal:19MICCAI:supp}
that may use all the extracted overlapping patches of size $R$:
\bulls{
\item The corresponding patch-based loss does not model the patch aggregation process that is inherently modeled in \R{sys:auto:res}.
\item It is an upper bound of the convolutional loss \R{sys:auto:res:train}.
\item It requires about $R$ times more memory than \R{sys:auto:res:train}.
}
We prove the benefits of \R{sys:auto:res:train} using the following lemma.

\lem{\label{l:train}
The loss function \R{sys:auto:res:train} for training the residual convolutional autoencoder in \R{sys:auto:res} is bounded by the patch-based loss function:
\be{
\label{eq:l:train}
\frac{1}{2L} \sum_{s=1}^S \Big\| \hat{x}_{s}^{(i)} -  \frac{1}{R} \sum_{k=1}^K d_k \circledast \cT_{\alpha_k} ( e_k \circledast x_{s}^{(i)} ) \Big\|_2^2 
\leq  \frac{1}{2LR} \sum_{s=1}^S \big\| \widehat{X}_{s}^{(i)}  - D \cT_{\tilde{\alpha}} ( E X_{s}^{(i)} ) \big\|_{\mathrm{F}}^2,
}
where the residual is defined by $\hat{x}_{s}^{(i)} \triangleq x_{s} - x_{s}^{(i)}$, 
$\{ x_{s}, x_{s}^{(i)} \}$ are given as in \R{sys:auto:res:train}, 
$\widehat{X}_{s} \in \bbR^{R \times V_{s}}$ and $X_{s} \in \bbR^{R \times V_{s}}$ are 
the $l\rth$ training data matrices whose columns are $V_{s}$ vectorized patches extracted from 
the images $\hat{x}_{s}$ and $x_{s}$ (with the circulant boundary condition and the \dquotes{stride} parameter $1$), respectively,
$D \triangleq [d_1, \ldots, d_K] \in \bbC^{R \times K}$ is a decoding filter matrix,
and $E \triangleq [e_1^*, \ldots, e_K^*]^H \in \bbC^{K \times R}$ is an encoding filter matrix.
Here, the definition of soft-thresholding operator in \R{eq:soft-threshold} is generalized by
\be{
\label{eq:soft-threshold:gen}
( \cT_{\tilde{\alpha}} (u) )_k \triangleq \left\{ \begin{array}{cc} u_k - \alpha_k \cdot \sgn (u_k), & | u_k | > \alpha_k, \\ 0, & \mbox{otherwise}, \end{array} \right.
}
for $K = 1,\ldots,K$, where $\tilde{\alpha} = [\alpha_1,\ldots,\alpha_K]^T$.
See other related notations in \R{sys:auto:res}.
}

\prf{
First, we have the following reformulation \cite[\S S.1]{Chun&Fessler:20TIP:supp}:
\be{
\label{eq:l:train:reform}
\left[ \begin{array}{c} e_1 \conv u \\ \vdots \\ e_K \conv u \end{array} \right] = \underbrace{ P' \left[ \begin{array}{c} E P_1 \\ \vdots \\ E P_N \end{array} \right] }_{\mbox{$\triangleq \widetilde{E}$}} u, \quad \forall u,
}
where $P' \in \bbC^{KN \times KN}$ is a permutation matrix, $E$ is defined in Lemma~\R{l:train}, and $P_n \in \bbC^{R \times N}$ is the $n\rth$ patch extraction operator for $n = 1,\ldots,N$.
Considering that $\frac{1}{R} \sum_{k=1}^K \mathsf{flip}(e_k^*) \circledast ( e_k \circledast u ) = \widetilde{E}^H \widetilde{E} u$ via the definition of $\widetilde{E}$ in \R{eq:l:train:reform} (see also the reformulation technique in \cite[\S S.1]{Chun&Fessler:20TIP:supp}), 
we obtain the following reformulation result:
\ea{
\label{eq:reform:idt}
\frac{1}{R} \sum_{k=1}^K \mathsf{flip}(e_k^*) \circledast \cT_{\alpha_k} ( e_k \circledast x_{s}^{(i)} ) =  \frac{1}{R} \sum_{n=1}^N P_n^H E^H \cT_{\widetilde{\alpha}} \big( E P_n x_{s}^{(i)} \big) 
}
where the soft-thresholding operators $\{ \cT_{\alpha_k}(\cdot) : \forall k \}$ and $\cT_{\tilde{\alpha}}(\cdot)$ are defined in \R{eq:soft-threshold:gen} and
we use the permutation invariance of the thresholding operator $\cT_\alpha (\cdot)$, i.e., $\cT_{\alpha} ( P (\cdot) ) =  P \cdot \cT_{\alpha} ( \cdot )$ for any $\alpha$.
Finally, we obtain the result in \R{eq:l:train} as follows:
\ea{
\label{eq:reform:dist:1}
\frac{1}{2L} \sum_{s=1}^S \Big\| \hat{x}_{s}^{(i)} -  \frac{1}{R} \sum_{k=1}^K d_k \circledast \cT_{\alpha_k} ( e_k \circledast x_{s}^{(i)} ) \Big\|_2^2 
&= \frac{1}{2L}  \sum_{s=1}^S \Big\| \hat{x}_{s}^{(i)} -  \frac{1}{R} \sum_{n=1}^N P_n^H D \cT_{\tilde{\alpha}} \big( E P_n x_{s}^{(i)} \big)  \Big\|_2^2
\\
\label{eq:reform:dist:2} 
& = \frac{1}{2L}  \sum_{s=1}^S \Big\| \frac{1}{R} \sum_{n=1}^N P_n^H P_n \hat{x}_{s}^{(i)} -  \frac{1}{R} \sum_{n=1}^N P_n^H D \cT_{\tilde{\alpha}} \big( E P_n x_{s}^{(i)} \big)  \Big\|_2^2
\\
& =  \frac{1}{2LR^2} \sum_{s=1}^S \bigg\| \sum_{n=1}^N P_n^H \Big( \hat{x}_{l,n}^{(i)} - D \cT_{\tilde{\alpha}} \big( E x_{l,n}^{(i)} \big) \Big) \bigg\|_2^2
\nn
\\
\label{eq:reform:dist:3} 
& \leq  \frac{1}{2LR} \sum_{s=1}^S \sum_{n=1}^N \Big\| \hat{x}_{l,n}^{(i)} - D \cT_{\tilde{\alpha}} \big( E x_{l,n}^{(i)} \big) \Big\|_2^2  
\\
&=  \frac{1}{2LR} \sum_{s=1}^S \nm{ \widehat{X}_{s}^{(i)} - D \cT_{\tilde{\alpha}} \left( E X_{s}^{(i)} \right)  }_{\mathrm{F}}^2,
\nn
}
where $D$ is defined in Lemma~\ref{l:train},
$\{ \hat{x}_{l,n}^{(i)} = P_n \hat{x}_{s}^{(i)} \in \bbC^R, x_{l,n}^{(i)} = P_n x_{s}^{(i)} \in \bbC^R: n = 1,\ldots,N \} $ is a set of extracted patches, 
the training matrices $\{ \widehat{X}_{s}^{(i)}, X_{s}^{(i)} \}$ are defined by $\widehat{X}_{s}^{(i)} \triangleq [ \hat{x}_{l,n}^{(i)}, \ldots, \hat{x}_{l,N}^{(i)} ]$ 
and $X_{s}^{(i)} \triangleq [ x_{l,1}^{(i)}, \ldots, x_{l,N}^{(i)} ]$. 
Here, the equality \R{eq:reform:dist:1} uses the result in \R{eq:reform:idt}, 
the equality \R{eq:reform:dist:2} holds by $\sum_{n=1}^N P_n^H P_n = R \cdot I$ (for the circulant boundary condition in Lemma~\ref{l:train}), 
and the inequality \R{eq:reform:dist:3} holds by $\widetilde{P} \widetilde{P}^H \preceq R \cdot I$ with $\widetilde{P} \triangleq [P_1^H \cdots P_N^H]^H$.
}

Lemma~\ref{l:train} reveals that
when the patch-based training approach extract all the $R$-size overlapping patches,
\textit{1)} the corresponding patch-based loss is an upper bound of the convolutional loss \R{sys:auto:res:train};
\textit{2)} it requires about $R$-times larger memory than \R{sys:auto:res:train} 
because $V_{s} \approx R N_{s}$ for $x \in \bbR^{N_{s}}$ and the boundary condition described in Lemma~\ref{l:train}, $\forall l$;
and \textit{3)} it misses modeling the patch aggregation process that is inherently modeled in \R{sys:auto:res}
--  see that the patch aggregation operator $\sum_{n=1}^N P_n^H (\cdot)_n$ is removed in the inequality \R{eq:reform:dist:3} in the proof of Lemma~\ref{l:train}.
In addition, different from the patch-based training approach \cite{Chun&Fessler:18IVMSP:supp, Chun&etal:19MICCAI:supp}, 
i.e., training with the function on the right-hand side in \R{eq:l:train}, 
one can use different sizes of filters $\{ e_k, d_k : \forall k \}$ 
in the convolutional training loss,
i.e., the function on the left-hand side in \R{eq:l:train}.

\section{Details of experimental setup}

\subsection{Majorization matrix designs for quadratic data-fit}

For (real-valued) quadratic data-fit $f(x;y)$ in the form of $\frac{1}{2} \| y -  A x \|_W^2$, 
if a majorization matrix $M$ exists such that $A^H W A \preceq M$, 
it is straightforward to verify that the gradient of quadratic data-fit $f(x;y)$ satisfies the $M$-Lipchitz continuity in Definition~\ref{d:QM}, i.e.,
\bes{
\| \nabla f (u;y) - \nabla f (v;y) \|_{M^{-1}} 
=  \| A^H W A u - A^H W A v \|_{M^{-1}}
\leq \| u - v \|_{M}^2, \quad \forall u,v \in \bbR^N.
}
because the assumption $A^T W A \preceq M \Leftrightarrow M^{-1/2} A^T W A M^{-1/2} \preceq I$ implies that 
the eigenspectrum of $M^{-1/2} A^T W A M^{-1/2}$ lies in the interval $[0, 1]$, 
and gives the following result:
\bes{
\big( M^{-1/2} A^T W A M^{-1/2} \big)^2 \preceq I 
\Leftrightarrow ( A^T W A  ) M^{-1} ( A^T W A ) \preceq M.
}

Next, we review a useful lemma in designing majorization matrices for a wide class of quadratic data-fit $f(x;y)$:

\lem{[\!\!\protect{\cite[Lem.~S.3]{Chun&Fessler:18TIP:supp}}]
\label{l:diag(|At|W|A|1)}
For a (possibly complex-valued) matrix $A$ and a diagonal matrix $W$ with non-negative entries, $A^H W A  \preceq \diag( | A^H | W | A | 1 )$, where $| A |$ denotes the matrix consisting of the absolute values of the elements of $A$.
}

\subsection{Parameters for MBIR optimization models: Sparse-view CT reconstruction} \label{sec:param:mbir:ct}

For MBIR model using EP regularization, 
we combined a EP regularizer $\sum_{n =1}^{N} \sum_{n' \in \cN_{n}} \iota_{n} \iota_{n'} \varphi(x_n - x_{n'})$
and the data-fit $f(x;y)$ in \S\ref{sec:appl:ct}, 
where $\cN_n$ is the set of indices of the neighborhood, 
$\iota_n$ and $\iota_{n'}$ are parameters that encourage uniform noise \cite{Cho&Fessler:15TMI:supp},
and $\varphi(\cdot)$ is the Lange penalty function, i.e., $\varphi(t) = \delta^2 (  | t/\delta | - \log(1+| t/\delta |) )$, with $\delta \!=\! 10$ in HU.
We chose the regularization parameter (e.g., $\gamma$ in \R{sys:mbir:z}) as $2^{15.5}$.
We ran the relaxed linearized augmented Lagrangian method \cite{Nien&Fessler:16:TMI:supp}
with $100$ iterations and $12$ ordered-subsets,
and initialized the EP MBIR algorithms with a conventional FBP method using a Hanning window.

For MBIR model using a learned convolutional regularizer \cite[(P2)]{Chun&Fessler:18Asilomar:supp}, 
we trained convolutional regularizer with filters of $\{ h_k \in \bbR^R : R \!=\! K \!=\! 7^2 \}$ 
via CAOL \cite{Chun&Fessler:20TIP:supp} in an unsupervised training manner;
see training details in \cite{Chun&Fessler:20TIP:supp}.
The regularization parameters (e.g., $\gamma$ in \R{sys:recov&caol}) were selected
by applying the \dquotes{spectral spread} based selection scheme in \S\ref{sec:reg:sel}
with the tuned factor $\chi^\star \!=\! 167.64$.
We selected the spatial-strength-controlling hard-thresholding parameter 
(i.e., $\alpha'$ in \cite[(P2)]{Chun&Fessler:18Asilomar:supp}) as follows:
for Test samples~$\#1$--$2$, we chose it is as $10^{-10}$ and $6^{-11}$, respectively.
We initialized the MBIR model using a learned regularizer with the EP MBIR results obtained above.
We terminated the iterations if the relative error stopping criterion (e.g., \cite[(44)]{Chun&Fessler:18TIP:supp}) is met before reaching the maximum number of iterations.
We set the tolerance value as $10^{-13}$ and the maximum number of iterations to $4 \!\times\! 10^3$.

\subsection{Parameters for MBIR optimization models: LF photography using a focal stack} \label{sec:param:mbir:lf}

For MBIR model using 4D EP regularization \cite{Lien&etal:20NP:supp}, 
we combined a 4D EP regularizer $\sum_{n =1}^{N} \sum_{n' \in \cN_{n}} \varphi(x_n - x_{n'})$
and the data-fit $f(x;y)$ in \S\ref{sec:appl:lf}, 
where $\cN_n$ is the set of indices of the 4D neighborhood,
and $\varphi(\cdot)$ is the hyperbola penalty function, i.e., $\varphi(t) = \delta^2 ( \sqrt{1+ | t/\delta |^2} - 1 )$.
We selected the hyperbola function parameter $\delta$ and 
regularization parameter (e.g., $\gamma$ in \R{sys:mbir:z}) as follows: 
for Test samples~$\#1$--$3$, we chose them as 
$\{ \delta \!=\! 10^{-4}, \gamma \!=\! 10^3 \}$, 
$\{ \delta \!=\! 10^{-1}, \gamma \!=\! 10^7 \}$, 
and $\{ \delta \!=\! 10^{-1}, \gamma \!=\! 5 \!\times\! 10^3 \}$, respectively.
We ran the conjugate gradient method with $100$ iterations,
and initialized the 4D EP MBIR algorithms with $A^T y$ rescaled in the interval $[0, 1]$.

\subsection{Reconstruction accuracy and depth estimation accuracy of different MBIR methods} 

Tables~\ref{tab:recon:ct}--\ref{tab:depth} below provide reconstruction accuracy numerics of different MBIR methods in sparse-view CT reconstruction 
and LF photography using a focal stack,
and reports the SPO depth estimation \cite{Zhang&etal:16CVIU:supp} accuracy numerics on reconstructed LFs from different MBIR methods.

\begin{table}[h!]	

\centering
\renewcommand{\arraystretch}{1.1}
	
\caption{RMSE (HU) of different CT MBIR methods 
\protect\linebreak
(fan-beam geometry with $12.5\%$ projections views and $10^5$ incident photons)}
\vspace{-0.5em}	
\label{tab:recon:ct}
	
\begin{tabular}{C{1cm}|C{1cm}C{1cm}C{3.3cm}C{2.8cm}C{3cm}C{2.8cm}}
\hline \hline
&
(a) FBP  & 
(b) EP reg. & 
\protect{(c) Learned convolutional reg. \cite{Chun&Fessler:20TIP:supp, Chun&Fessler:18Asilomar:supp}} & 
\specialcell[c]{(d) Momentum-Net- \\ sCNN } &
\specialcell[c]{(e) {\bfseries Momentum-Net-} \\ {\bfseries sCNN} w/ larger width} &
\specialcell[c]{(f) Momentum-Net- \\ dCNN} 
\\ 
\hline
Test $\#1$ & 
82.8 & 40.8 & 35.2  & 19.9 & \textbf{19.5} & 19.8 
\\ 
Test $\#2$ & 
74.9 & 38.5 & 34.5 & 18.4 & \textbf{17.7} & 17.8 
\\ 
\hline \hline 
\end{tabular}

\smallskip
\begin{myquote}{0.5in}
(c)'s convolutional regularizer uses $\{ R \!=\! K \!=\! 7^2 \}$
\\
(d)'s refining sCNNs are in the form of residual single-hidden layer convolutional autoencoder \R{sys:auto:res} with $\{ R \!=\! K \!=\! 7^2 \}$.
\\
(e)'s refining sCNNs are in the form of residual single-hidden layer convolutional autoencoder \R{sys:auto:res} with $\{ R \!=\! 7^2, K \!=\! 9^2 \}$. 
This setup gives results in Fig.~\ref{fig:recon:ct}(d), as described in \S\ref{sec:exp:INN:param}.
\\
(f)'s refining dCNNs are in the form of residual multi-hidden layer CNN \R{sys:dcnn} with $\{ L \!=\! 4, R \!=\! 3^2, K \!=\! 64 \}$.
\end{myquote}
\vspace{-1pc}

%\vspace{-1pc}
\end{table}

\begin{table}[h!]	

\centering
\renewcommand{\arraystretch}{1.1}
	
\caption{PSNR (dB) of different LF MBIR methods 
\protect\linebreak
(LF photography systems with $C \!=\! 5$ detectors obtain a focal stack of
LFs consisting of $S \!=\! 81$ sub-aperture images)}
\vspace{-0.5em}	
\label{tab:recon:lf}
	
\begin{tabular}{C{1cm}|C{1.3cm}C{2.3cm}C{3.5cm}C{3.5cm}}
\hline \hline
&
(a) $A^T y$  & 
(b) 4D EP reg.~\cite{Lien&etal:20NP:supp} & 
\specialcell[c]{(c) Momentum-Net-sCNN} &
\specialcell[c]{(d) {\bfseries Momentum-Net-dCNN}}
\\ 
\hline
Test $\#1$ & 
16.4 & 32.0 & 35.8 & \textbf{37.1}  
\\ 
Test $\#2$ & 
21.1 & 28.1 & 30.7 & \textbf{32.0}
\\
Test $\#3$ & 
21.6 & 28.1 & 30.9 & \textbf{31.7}
\\ 
\hline \hline
\end{tabular}

\smallskip
\begin{myquote}{0.5in}
(c)'s refining sCNNs are in the form of residual single-hidden layer convolutional autoencoder with $\{ R \!=\! 5^2, K \!=\! 32 \}$.
\\
(d)'s refining dCNNs are in the form of residual multi-hidden layer CNN \R{sys:dcnn} with $\{ L \!=\! 6, R \!=\! 3^2, K \!=\! 16 \}$.
\\
Momentum-Nets use refining CNNs in an epipolar-domain; see details in \S\ref{sec:exp:INN:param}.
\end{myquote}

\vspace{-1pc}
\end{table}

\begin{table}[h!]	

\centering
\renewcommand{\arraystretch}{1.1}
	
\caption{RMSE (in $10^{-2}$, m) of estimated depth from reconstructed LFs with different LF MBIR methods 
\protect\linebreak
(LF photography systems with $C \!=\! 5$ detectors obtain a focal stack of
LFs consisting of $S \!=\! 81$ sub-aperture images)}
\vspace{-0.5em}	
\label{tab:depth}
	
\begin{tabular}{C{1cm}|C{2.2cm}C{2.7cm}C{2.6cm}C{3.4cm}C{3.4cm}}
\hline \hline
&
\specialcell[c]{(a) Ground truth \\ LF}  &
\specialcell[c]{(b) Reconstructed LF \\ by $A^T y$}  & 
\specialcell[c]{(c) Reconstructed LF \\ by 4D EP reg.~\cite{Lien&etal:20NP:supp}} & 
\specialcell[c]{(d) Reconstructed LF \\ by Momentum-Net-sCNN} & 
\specialcell[c]{(e) Reconstructed LF \\ by {\bfseries Momentum-Net-dCNN}}
\\ 
\hline
Test $\#1$ & 
4.7  &  41.0  &  13.8  & 8.0 & \textbf{5.7} 
\\
Test $\#2$ & 
30.5  &  117.6  &  39.5  & 34.6 & \textbf{31.9}
\\  
Test $\#3$ & 
n/a$^\dagger$  &  n/a$^\dagger$  &  n/a$^\dagger$  &  n/a$^\dagger$ & n/a$^\dagger$
\\  
\hline \hline
\end{tabular}

\smallskip
\begin{myquote}{0.5in}
\protect{SPO depth estimation \cite{Zhang&etal:16CVIU:supp} was applied to reconstructed LFs.}
\\
(d)'s refining sCNNs are in the form of residual single-hidden layer convolutional autoencoder with $\{ R \!=\! 5^2, K \!=\! 32 \}$.
\\
(e)'s refining dCNNs are in the form of residual multi-hidden layer CNN \R{sys:dcnn} with $\{ L \!=\! 6, R \!=\! 3^2, K \!=\! 16 \}$.
\\
Momentum-Nets use refining CNNs in an epipolar-domain; see details in \S\ref{sec:exp:INN:param}.
\\
$^\dagger$The ground truth depth map for Test sample $\#3$ does not exist in the LF dataset \cite{Honauer&etal:16ACCV:supp}.
\end{myquote}

%\vspace{-1pc}
\end{table}

\subsection{Reconstructed images and estimated depths with noniterative analytical methods}

\begin{figure*}[!h]
% \vspace{-0.75em}
 \centering
 \small\addtolength{\tabcolsep}{-6.5pt}
 \renewcommand{\arraystretch}{0.2}

     \begin{tabular}{c}
     
        	 \specialcell[c]{Fig.~\ref{fig:recon:ct}: FBP}
	 \\
	 
          \begin{tikzpicture}
             \begin{scope}[spy using outlines={rectangle,yellow,magnification=1.75,size=18mm, connect spies}]
                 \node {\includegraphics[viewport={10mm 15mm 125mm 120mm},clip,width=34.5mm,height=34.5mm]{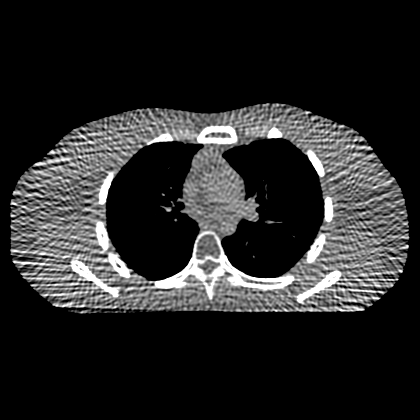}};
 	        \spy on (0.45,-0.42) in node [left] at (0,-1.95);	
	        \node [black] at (0.9,-2.5) {\specialcell{$\textmd{RMSE (HU)}$ \\ \\ $= 82.8$}};
             \end{scope}
         \end{tikzpicture}          
	\\
         
          \begin{tikzpicture}
             \begin{scope}[spy using outlines={rectangle,yellow,magnification=1.75,size=18mm, connect spies}]
                 \node {\includegraphics[viewport={15mm 5mm 130mm 110mm},clip,width=34.5mm,height=34.5mm]{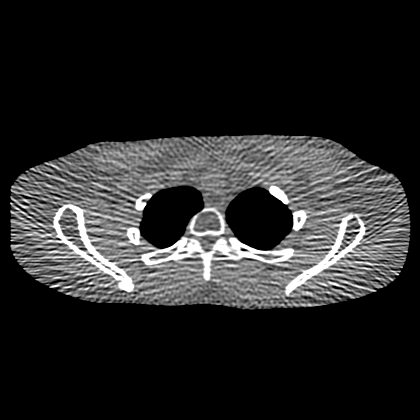}};
 	        \spy on (0.32,0.25) in node [left] at (0,-1.55);	
	        \node [black] at (0.9,-2.1) {\specialcell{$\textmd{RMSE (HU)}$ \\ \\ $= 74.9$}};
             \end{scope}
         \end{tikzpicture}

     \end{tabular}
     \hspace{1pc}
     \begin{tabular}{c}
     
           \specialcell[c]{\vphantom{$A^T y$} \\ Fig.~\ref{fig:recon:lf}: Error maps of $A^T y$} 
           \\
         
          \begin{tikzpicture}
                 \node {\includegraphics[clip,width=34.5mm,height=34.5mm]{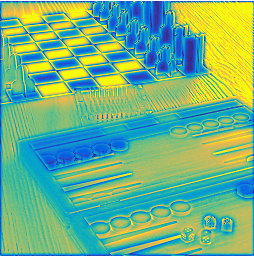}};
         \end{tikzpicture}  
         \\
 
         {$\textmd{PSNR (dB)} = 16.5~(16.4)$} 
	\\
         
          \begin{tikzpicture}
                \node {\includegraphics[clip,width=34.5mm,height=34.5mm]{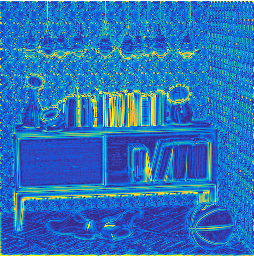}};
         \end{tikzpicture} 
          \\

         {$\textmd{PSNR (dB)} = 22.6~(21.1)$} 
         \\
                  
          \begin{tikzpicture}
                \node {\includegraphics[clip,width=34.5mm,height=34.5mm]{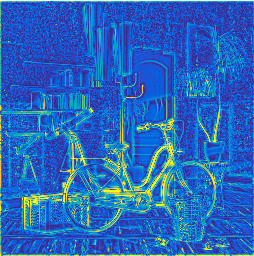}};
         \end{tikzpicture} 
         
         \\

         {$\textmd{PSNR (dB)} = 23.4~(21.6)$} 

     \end{tabular}
     \hspace{1pc}
     \begin{tabular}{c}
     
           \specialcell{Fig.~\ref{fig:depth}: Estimated depth \\ from LF recon.~by $A^T y$}
           \\
        
	\begin{tikzpicture}
             \begin{scope}[spy using outlines={rectangle,yellow,magnification=1.75,size=18mm, connect spies}]
                 \node {\includegraphics[clip,width=34.5mm,height=34.5mm]{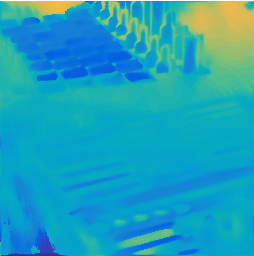}};
             \end{scope}
         \end{tikzpicture} 
 	\\         
        \parbox[c]{34.5mm}{$\textmd{RMSE (m)} = 41.0 \!\times\! 10^{-2}$} 
         \\
         
	\begin{tikzpicture}
             \begin{scope}[spy using outlines={rectangle,yellow,magnification=1.75,size=18mm, connect spies}]
                 \node {\includegraphics[clip,width=34.5mm,height=34.5mm]{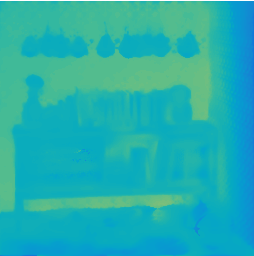}};
             \end{scope}
         \end{tikzpicture}
         \\
          \parbox[c]{34.5mm}{$\textmd{RMSE (m)} = 117.6 \!\times\! 10^{-2}$} 
         \\
         
         \begin{tikzpicture}
             \begin{scope}[spy using outlines={rectangle,yellow,magnification=1.75,size=18mm, connect spies}]
                 \node {\includegraphics[clip,width=34.5mm,height=34.5mm]{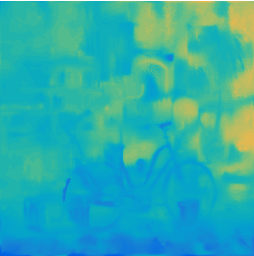}};
             \end{scope}
         \end{tikzpicture}
	\\
         {\small n/a} 
	      
     \end{tabular}
   
 \vspace{-0.5em}
 \caption{Reconstructed images from analytical back-projection methods. We used such results in the first two columns to initialize INN methods.
 }
 \label{fig:recon:anal}
\end{figure*}

This section provides reconstructed images by an analytical back-projection method in sparse-view CT reconstruction 
and LF photography using a focal stack (see the first two columns in Fig.~\ref{fig:recon:anal}),
and estimated depths from reconstructed LFs via the SPO depth estimation method \cite{Zhang&etal:16CVIU:supp} (see the third column in Fig.~\ref{fig:recon:anal}(c)).
Results in Fig.~\ref{fig:recon:anal} below are supplementary to Fig.~\ref{fig:recon:ct}, Fig.~\ref{fig:recon:lf}, and Fig. \ref{fig:depth},
and the first two columns visualize initial input images to INN methods.

\section{How to choose parameters of image refining modules in soft-refining INNs?}
\label{sec:discuss:param}

In soft-refining INNs using iterative-wise refining NNs, 
one does not need to greatly increase parameter dimensions of refining NNs \cite{Lim&etal:20TMI:supp, Chun&etal:19MICCAI:supp}.
The natural question then arises, \dquotes{How one can choose between sCNN \R{sys:auto:res} and dCNN \R{sys:dcnn} refiners, 
and select their parameters ($R$, $K$, and $L$)?}
The first answer to this question depends on some understanding of data-fit $f(x;y)$ in MBIR problem \R{sys:mbir},
e.g., the regularization strength $\gamma$ and the condition number variations across training data-fit majorizers.
(An additional criteria could be general understandings between sample size/diversity and parameter dimension of NNs.)

For example, the sparse-view CT system in \S\ref{sec:appl:ct} needs moderate regularization strength ($\chi^\star \!=\! 167.64$) 
and the majorization matrices of its training data-fits have mild condition number variations (the standard deviation is $1.1$).
training data-fits have mild parameter variations across samples.
Comparing results between Momentum-Net-sCNN and -dCNN in Fig.~\ref{fig:timecomp:ct} and Table~\ref{tab:recon:ct}
demonstrates that sCNN \R{sys:auto:res} seems suffice.
Table~\ref{tab:recon:ct}(d)--(e) shows that one can further improve the refining accuracy of sCNN \R{sys:auto:res} by increasing its width, i.e., $K$.
The LF photography system using a limited focal stack in \S\ref{sec:appl:lf} needs a large $\gamma$ value ($\chi^\star \!=\! 1.5$),
and the majorization matrices of its training data-fits have large condition number variations (the standard deviation is $2245.5$).
Comparing results between Momentum-Net-sCNN and -dCNN in Fig.~\ref{fig:timecomp:lf} and Table~\ref{tab:recon:lf}
demonstrates that dCNN \R{sys:dcnn} yields higher PSNR than sCNN \R{sys:auto:res}.
For dCNN \R{sys:dcnn}, we observed increasing its depth, i.e., $L$, up to a certain number 
is more effective than increasing its width, i.e., $K$, as briefly discussed in \S\ref{sec:exp:INN:param}.

For choosing the relaxation parameter $\rho$ in \R{eq:momnet:map}, 
we also suggest considering the regularization strength in \R{eq:momnet:mbir}.
For an application that needs moderate regularization strength,
e.g., sparse-view CT in \S\ref{sec:appl:ct},
we suggest setting $\rho$ to $0.5$ so as to mix information between input and output of refining NNs,
rather than $1 - \varepsilon$ that does not mix input and output.
For an application that needs strong regularization,
e.g., LF photography using a limited focal stack in \S\ref{sec:appl:lf},
we suggest using $\rho \!=\! 1 - \varepsilon$ than $\rho \!=\! 0.5$.
Results in the next section validate this suggestion.

\subsection{Performance of Momentum-Net with different relaxation parameters $\rho$ in \R{eq:momnet:map}} 
\label{sec:rho}

\begin{figure}[!th]
%\vspace{-0.5pc}
 \centering
% \small\addtolength{\tabcolsep}{-7.5pt}
 \renewcommand{\arraystretch}{1}

     \begin{tabular}{cc}
     {\small (a) Sparse-view CT} &
     {\small (b) Light-field photography using focal stack} 
     \\
     \includegraphics[scale=0.55, trim=0.2em 0.2em 1.1em 1em, clip]{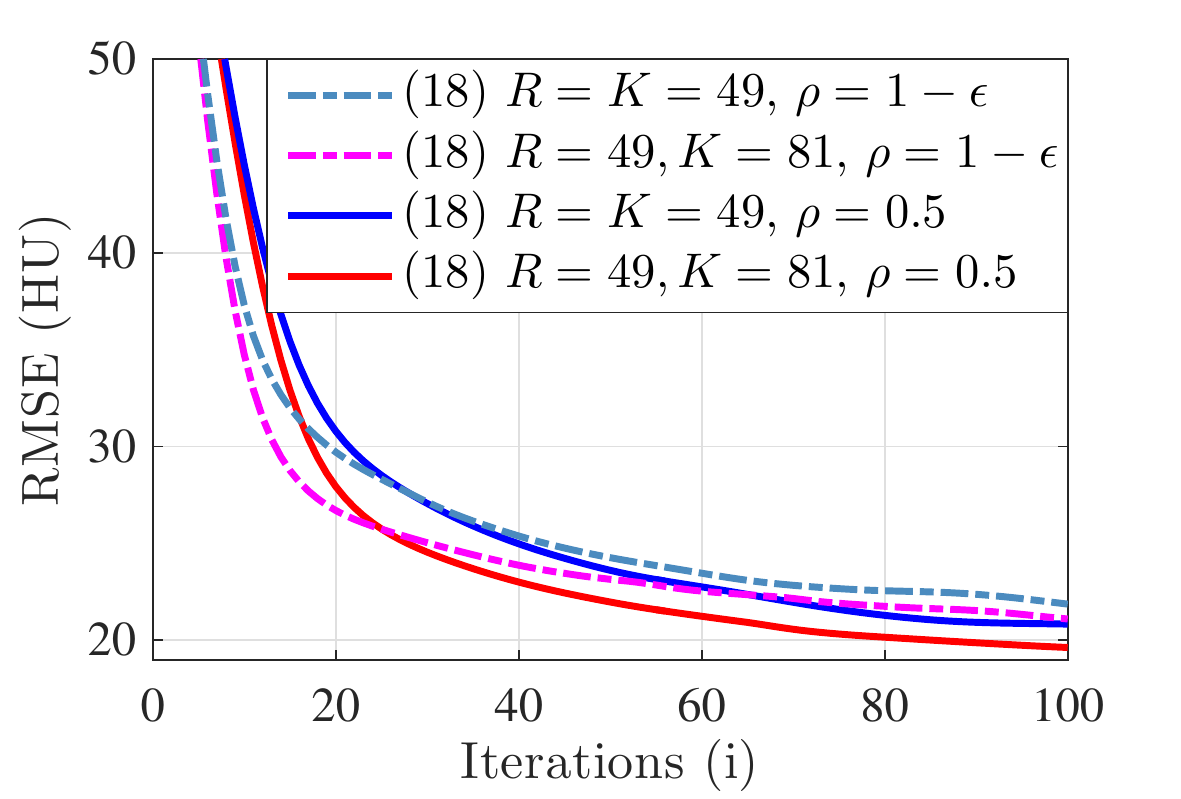} &
     \includegraphics[scale=0.55, trim=0.2em 0.2em 1.1em 1em, clip]{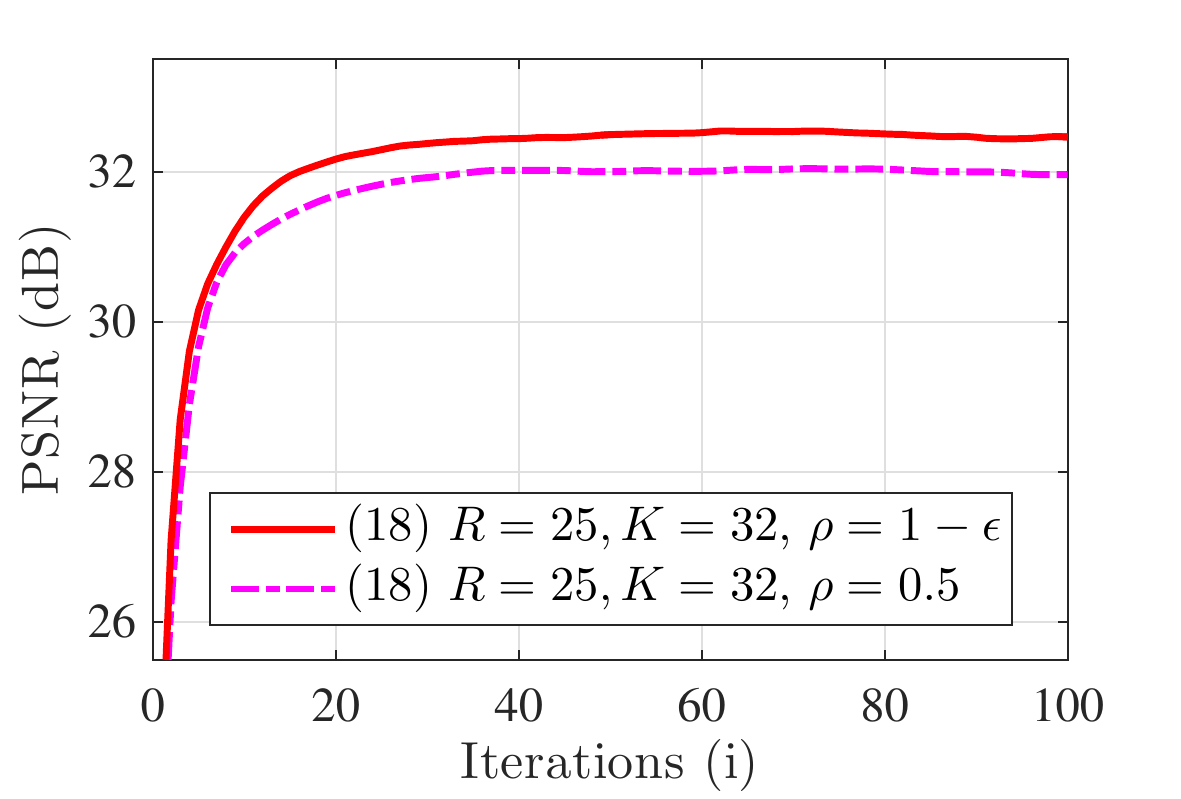}
     \end{tabular}
     
\vspace{-0.5em}
\caption{Convergence behavior of Momentum-Net-sCNN with different relaxation parameters, $\rho = 0.5$ and $\rho =  1 - \varepsilon$. 
For both applications (see their imaging setups in \S\ref{sec:exp:imaging}), PyTorch ver.~0.3.1 was used.}
\label{fig:convg:rho}
\vspace{-0.5pc}
\end{figure}

Fig.~\ref{fig:convg:rho} below compares the performances of Momentum-Net-sCNN with different $\rho$ values.
The results in Fig.~\ref{fig:convg:rho} support the $\rho$ selection guideline in \S\ref{sec:exp:INN:test}.
One can maximize the MBIR accuracy of Momentum-Net by properly selecting $\rho$.

Note that $\rho \in (0,1)$ controls strength of inference from refining NNs in \R{eq:momnet:map}, but does not affect the convergence guarantee of Momentum-Net. 
Fig.~\ref{fig:convg:rho} illustrates that Momentum-Net appears to converge regardless of $\rho$ values.

\section{Parameters of Momentum-Net} 
\label{sec:param}

Table~\ref{tab:param} below lists parameters of Momentum-Net, and summarizes selection guidelines or default values.
Similar to BCD-Net/ADMM-Net, 
the main tuning jobs to maximize the performance of Momentum-Net
include selecting architectures of refining NNs $\{ \cR_{\theta^{(i)}} \!:\! \forall i \}$ in \R{eq:momnet:map},
and choosing a regularization parameter $\gamma$ in \R{eq:momnet:mbir} by tuning $\chi$ in \S\ref{sec:reg:sel}.
One can simplify the tuning process by using the selection guidelines in \S\ref{sec:discuss:param} for selecting architectures of $\{ \cR_{\theta^{(i)}} \!:\! \forall i \}$, and
training $\chi$ in \S\ref{sec:reg:sel}.
Note that one designs majorization matrices $\{ M^{(i)} \!:\! \forall i \}$ rather than tuning them:
majorization matrices can be analytically designed, e.g., Lemma~\ref{l:diag(|At|W|A|1)} as used in \S\ref{sec:exp:INN:param};
one can algorithmically design them \cite{Mcgaffin&Fessler:15arXiv:supp}.
Tighter majorization matrices are expected to further accelerate the convergence of Momentum-Net \cite{Chun&Fessler:18TIP:supp, Chun&Fessler:20TIP:supp}.

\begin{table}[ht!]	
% \vspace{-0.5pc}
\centering
\renewcommand{\arraystretch}{1.1}
	
\caption{Guidelines for choosing parameters of Momentum-Net}
\vspace{-0.5em}	
\label{tab:param}
	
\begin{tabular}{C{1.5cm}|C{1.1cm}|L{5cm}}
\hline \hline
Param. & Module & Guidelines or default values
\\
\hline
$\{ \cR_{\theta^{(i)}} \!:\! \forall i \}$ & \R{eq:momnet:map} & 
Trainable by \S\ref{sec:train:NN}. 
For selecting their architecture/param., see guideline \S\ref{sec:discuss:param}.
\\
$\rho \in (0,1)$ & \R{eq:momnet:map} & Use regularization strength $\gamma$; see guideline in \S\ref{sec:discuss:param}.
\\
$\delta < 1$ in \R{up:Ex:cvx}--\R{up:Ex:ncvx} & \R{eq:momnet:exp} & $1 \!-\! \varepsilon$
\\
$\{ M^{(i)} \!:\! \forall i \}$ & \R{eq:momnet:mbir} & Designed off-line. For large-scale inverse problems with quadratic data-fit, use Lemma~\ref{l:diag(|At|W|A|1)}. 
\\
$\lambda \geq 1$ in \R{up:MFtilde} & \R{eq:momnet:mbir} & 
For convex $F(x;y,z^{(i+1)})$, $\lambda \!=\! 1$;
\hspace{1.5pc}for nonconvex $F(x;y,z^{(i+1)})$, $\lambda \!=\! 1 \!+\! \varepsilon$.
\\
$\gamma \!>\!  0$ & \R{eq:momnet:mbir} & Chosen by tuning/training $\chi$ in \S\ref{sec:reg:sel}
\\ 
\hline \hline 
\end{tabular}

\smallskip
\begin{myquote}{0.5in}
All INN methods also must select a number of INN iterations, $N_{\text{iter}}$.
One could determine it by using the convergence behavior of iteration-wise refiners in Fig.~\ref{fig:momnet:convg}.
\end{myquote}
\vspace{-0.5pc}

\end{table}

% Can use something like this to put references on a page
% by themselves when using endfloat and the captionsoff option.
\ifCLASSOPTIONcaptionsoff
  \newpage
\fi

%%%%%%%%%%%%%%%%%%%%%%%%%%%%%%%%%%%%%%%%%%%%%%%%%%%%%%%%%%%%%%%%%%%%%%%%%%%%%%%%
%                                References
%%%%%%%%%%%%%%%%%%%%%%%%%%%%%%%%%%%%%%%%%%%%%%%%%%%%%%%%%%%%%%%%%%%%%%%%%%%%%%%%
\bibliographystyleSupp{IEEEtran}
\bibliographySupp{referencesSupp_Bobby}
%%%%%%%%%%%%%%%%%%%%%%%%%%%%%%%%%%%%%%%%%%%%%%%%%%%%%%%%%%%%%%%%%%%%%%%%%%%%%%%%

\end{document}